\documentclass[10pt,letterpaper,oneside,twocolumns,romanappendices]{IEEEtran}

\usepackage{amsmath, amsfonts, amssymb, amsthm, amsbsy, amscd, bm} 
\usepackage{mathrsfs}
\usepackage{cite}
\usepackage{array}
\usepackage{rotating}

\usepackage{enumerate}
\usepackage{hyperref}
\usepackage{subfigure}

\usepackage{comment,array}
\usepackage{graphicx}
\usepackage[table]{xcolor}
\usepackage{mwe}
\usepackage[percent]{overpic}

\makeatletter
\def\hlinewd#1{%
  \noalign{\ifnum0=`}\fi\hrule \@height #1 \futurelet
   \reserved@a\@xhline}
\makeatother

\DeclareMathOperator{\Res}{Res}
\definecolor{mygray1}{rgb}{0.2,.2,.2}
\definecolor{mygray2}{rgb}{.5,.5,.5}
\definecolor{color_CNN}{rgb}{0.9,1,1}
\definecolor{color_Decoupled}{rgb}{1, 0.9, 0.8}
\definecolor{color_Combined}{rgb}{1, 1, 0.9}
\definecolor{color_Tikhonov}{rgb}{1, 0.95, 0.9}
\definecolor{color_Shrinkage}{rgb}{0.95, 0.85, 0.9}
\definecolor{color_TV}{rgb}{0.9, 0.9, 1}
\definecolor{color_STAT}{rgb}{0.9, 1, 0.9}

\makeatletter
\def\hlinewd#1{%
  \noalign{\ifnum0=`}\fi\hrule \@height #1 \futurelet
   \reserved@a\@xhline}
\makeatother

\newcommand{\mcrot}[4]{\multicolumn{#1}{#2}{\rlap{\rotatebox{#3}{#4}~}}} 
\newcommand*{\twoelementtable}[3][l]%
{%
    \renewcommand{\arraystretch}{0.8}%
    \begin{tabular}[t]{@{}#1@{}}%
        #2\tabularnewline
        #3%
    \end{tabular}%
}

\definecolor{mygreen1}{rgb}{0, .8, 0}
\definecolor{mygreen2}{rgb}{0, .4, 0}

\graphicspath{{figures/}}

\begin{document}
%
\title{Convolutional Deblurring for Natural Imaging}

\author{Mahdi~S.~Hosseini,~\IEEEmembership{Member,~IEEE,}~and~Konstantinos~N.~Plataniotis,~\IEEEmembership{Fellow,~IEEE}
\thanks{Copyright (c) 2019 IEEE. Personal use of this material is permitted. However, permission to use this material for any other purposes must be obtained from the IEEE by sending a request to pubs-permissions@ieee.org. 

M. S. Hosseini and K. N. Plataniotis are with The Edward S. Rogers Sr. Department of Electrical and Computer Engineering, University of Toronto, Toronto, ON M5S 3G4, Canada e-mail: \url{mahdi.hosseini@mail.utoronto.ca}.

This paper has supplementary downloadable material available at \url{http://ieeexplore.ieee.org} and \url{https://github.com/mahdihosseini/1Shot-MaxPol}, provided by the author. The material includes source codes for 1ShotMaxPol implementation and additional visual results on deblurring for comparison. Contact \url{mahdi.hosseini@mail.utoronto.ca} for further questions about this work}}%

\markboth{For Publication in IEEE Transaction on Image Processing 2019}%
{Shell \MakeLowercase{\textit{et al.}}: Bare Demo of IEEEtran.cls for IEEE Journals}

\maketitle

\begin{abstract}
In this paper, we propose a novel design of image deblurring in the form of one-shot convolution filtering that can directly convolve with naturally blurred images for restoration. The problem of optical blurring is a common disadvantage to many imaging applications that suffer from optical imperfections. Despite numerous deconvolution methods that blindly estimate blurring in either inclusive or exclusive forms, they are practically challenging due to high computational cost and low image reconstruction quality. Both conditions of high accuracy and high speed are prerequisites for high-throughput imaging platforms in digital archiving. In such platforms, deblurring is required after image acquisition before being stored, previewed, or processed for high-level interpretation. Therefore, on-the-fly correction of such images is important to avoid possible time delays, mitigate computational expenses, and increase image perception quality. We bridge this gap by synthesizing a deconvolution kernel as a linear combination of Finite Impulse Response (FIR) even-derivative filters that can be directly convolved with blurry input images to boost the frequency fall-off of the Point Spread Function (PSF) associated with the optical blur. We employ a Gaussian low-pass filter to decouple the image denoising problem for image edge deblurring. Furthermore, we propose a blind approach to estimate the PSF statistics for two Gaussian and Laplacian models that are common in many imaging pipelines. Thorough experiments are designed to test and validate the efficiency of the proposed method using $2054$ naturally blurred images across six imaging applications and seven state-of-the-art deconvolution methods.

\end{abstract}

\begin{IEEEkeywords}
Image deconvolution, point spread function (PSF), optical blur, generalized Gaussian, MaxPol derivatives
\end{IEEEkeywords}

%
\IEEEpeerreviewmaketitle

\section{Introduction}
\IEEEPARstart{B}{lurring} in many imaging modalities is caused by inadequate optical configuration in image acquisition. In an imperfect optical system, a ray of light passing through the optical setup will spread over the image domain instead of converting to a single end point. This spreading effect is known as the point spread function (PSF) and characterizes the  response (a.k.a impulse response) of the optical system \cite{mahajan1998optical, joshi2008psf}. The corresponding observation model is usually expressed by a linear convolution
\begin{align}
f_{\textit{B}}(x) = f_{\textit{L}}(x) \ast h_{\textit{PSF}}(x) + \eta(x),
\label{eq_psf_model_1}
\end{align}
where $f_{\textit{B}}$ is the blurry observation {(sampled image)}, $f_{\textit{L}}$ is the latent sharp image {(image to be recovered)}, $h_{\textit{PSF}}$ is the associated PSF kernel, and  $\eta$ is the noise contamination artifact. The problem of deblurring (a.k.a deconvolution) refers to the restoration of the latent image from its blurry observation, which is inherently an ill-posed problem. When the PSF is given, this is known as the \textit{``non-blind''} image deconvolution problem; otherwise, it is called the \textit{``blind''} approach. In either case, {image deconvolution} has been explored for a long time to address low-level blurring deficiencies, such as lens aberrations \cite{mahajan1998optical, joshi2008psf, keller2006objective, bishop2012light}, turbid medium \cite{hufnagel1964modulation, fried1966optical, narasimhan2002vision, bonettini2008scaled, bertero2009image, zhu2013removing, zhu2015fast}, out-of-focus \cite{cannon1976blind, savakis1993blur, elder1998local}, and motion artifacts \cite{fergus2006removing, cho2009fast, whyte2014deblurring, sun2015learning}.

The energy fall-off of the high frequency band is a common in the PSF - it suppresses sharp edges and leads to blurry observations. Aberrations (including turbid medium and out-of-focus) are mainly identified by a symmetric PSF which preserves the image geometry and is known to be a common problem in many imaging modalities. It is known that no matter how well the system is in focus, including no motion artifacts, the aberrations are still barriers to generating high quality images. One way to improve the image perception quality is to deploy more sophisticated optical hardware such high numerical aperture lens \cite{mahajan1998optical}. However, this is not a cost effective approach for applications such as consumer cameras. A more viable approach would be to integrate fast deblurring algorithms such as unsharp masking techniques in order to maintain real-time acquisition problem \cite{polesel2000image, deng2011generalized}. Although such masks require no more processing cost than one-shot filter convolution, they do not necessarily comply with the inverse response of the PSF for fall-off correction, and hence produce sub-par image quality with over-sharpening artifacts. 


In this paper, we focus on the proper correction of the fall-off frequency in the PSF by casting image deconvolution as a one-shot convolution filter problem. Our method is divided into two main steps. We first find that the \textit{a priori} PSF model can be inferred by a scale-space analysis of the blurred image in the Fourier domain. After making an assumption about the frequency fall-off of natural images, we blindly estimate the statistics of the PSF for two different models of the Gaussian and Laplacian as variants of the generalized Gaussian distribution. In the second step, we provide a closed-form solution to the inverse PSF for deblurring by fitting a series of polynomials in the frequency domain and then obtaining its equivalent representation in the spatial domain as a linear combination of FIR derivative filters. In doing so, we avoid producing ringing artifacts in the restored image while optimally preserving edge information on both fine and coarse resolutions. We show that the proposed deblurring method is capable of addressing diverse PSF models produced by various imaging {modalities such as consumer cameras, narrow-angle planetary observation cameras, etc}. 

\subsection{Related Works}
An overview of existing image deconvolution methods for image recovery divided into seven categories is listed in Table \ref{table_deblurring_methods}. Here, we describe each category and analyze their strengths and weaknesses.

\begin{table*}[htp]
\renewcommand{\arraystretch}{1.2}
\caption{List of existing image deconvolution methodologies, listed in seven categories: statistical prior modeling; Tikhonov regularization; total variation regularized minimization approaches; sparse modeling; iterative shrinkage; including combined approach of different regularization; and recently with emerging CNN models.}
\label{table_deblurring_methods}
\centering
\scriptsize
\begin{tabular}{lp{.4cm}p{.2cm}cccccccl}
\hlinewd{1.5pt}
Author & Year &
\mcrot{1}{c}{45}{\textcolor{mygray1}{Blind (B)/NonBlind (N)}} &
\mcrot{1}{c}{45}{\textcolor{mygray1}{Statistical priors}} &
\mcrot{1}{c}{45}{\textcolor{mygray1}{Tikhonvov regulation}} &
\mcrot{1}{c}{45}{\textcolor{mygray1}{Total variation}} &
\mcrot{1}{c}{45}{\textcolor{mygray1}{Sparse modeling}} &
\mcrot{1}{c}{45}{\textcolor{mygray1}{Iterative shrinkage}} &
\mcrot{1}{c}{45}{\textcolor{mygray1}{Combined regulation}} &
\mcrot{1}{c}{45}{\textcolor{mygray1}{Deep learning design}} &
\hspace{1in}Method description\\ \hlinewd{1.5pt}
%
%
\rowcolor{color_CNN}
Tao \cite{tao2018scale} & $2018$ & N &&&&&&&$\textcolor{mygray1}{\bullet}$& Coarse to fine scale analysis using residual CNN encoder/decoder
\\ \hlinewd{.75pt}
\rowcolor{color_CNN}
Wang \cite{wang2018training} & $2018$ & N &&&$\textcolor{mygray1}{\bullet}$&$\textcolor{mygray1}{\bullet}$&&&$\textcolor{mygray1}{\bullet}$& 
Image is pre-deconvolved via Wiener method and fed into a CNN model to predict sharp residual
\\ \hlinewd{.75pt}
\rowcolor{color_CNN}
Schuler \cite{schuler2013machine, schuler2016learning} & $2016$ & B &&&&&&&$\textcolor{mygray1}{\bullet}$& An end-to-end deep network is trained to estimate blur kernel for image deconvolution
\\ \hlinewd{.75pt}
\rowcolor{color_CNN}
Sun \cite{sun2015learning} & $2015$ & N &$\textcolor{mygray1}{\bullet}$&&&&&&$\textcolor{mygray1}{\bullet}$ & Learn non-uniform motion blur via CNN model and feed into Gaussian mixture model minimization
\\ \hlinewd{.75pt}
\rowcolor{color_CNN}
Hardis \cite{hradivs2015convolutional} & $2015$ & B &&&&&&&$\textcolor{mygray1}{\bullet}$ & Design a CNN with 10/15 convolution layers to deconvolve text images from their blurry observation
\\ \hlinewd{.75pt}
\rowcolor{color_CNN}
Xu \cite{xu2014deep} & $2014$ & N &&&&&&&$\textcolor{mygray1}{\bullet}$ & Design a deconvolution CNN with two hidden layers assembled with separable kernels
\\ \hlinewd{.75pt}
%
%
\rowcolor{color_Decoupled}
Zhang \cite{zhang2017learning} & $2017$ & N &&&&&&$\textcolor{mygray1}{\bullet}$&$\textcolor{mygray1}{\bullet}$ & Decoupled minimization, image denoising prior by pre-trained CNN model
\\ \hlinewd{.75pt}
\rowcolor{color_Decoupled}
Chan \cite{chan2017plug} & $2017$ & N &$\textcolor{mygray1}{\bullet}$&&&&&$\textcolor{mygray1}{\bullet}$&& Decoupled minimization, image denoising prior by transform recursive edge-preserving filters \cite{gastal2011domain}
\\ \hlinewd{.75pt}
\rowcolor{color_Decoupled}
Romano \cite{romano2017little} & $2017$ & N &$\textcolor{mygray1}{\bullet}$&&&&&$\textcolor{mygray1}{\bullet}$&& Decoupled minimization, image denoising prior by adaptive Laplacian-based regularization
\\ \hlinewd{.75pt}
\rowcolor{color_Decoupled}
Danielyan\cite{danielyan2012bm3d} & $2012$ & N &&&&$\textcolor{mygray1}{\bullet}$&&$\textcolor{mygray1}{\bullet}$&& Decoupled minimization, Learn an overcomplete dictionary for sparse representation
\\ \hlinewd{.75pt}
\rowcolor{color_Decoupled}
Zoran \cite{zoran2011learning} & $2011$ & N &$\textcolor{mygray1}{\bullet}$&&&&&$\textcolor{mygray1}{\bullet}$&& Decoupled minimization, image prior by maximizing the expected patch log-likelihood distribution
\\ \hlinewd{.75pt}
%
\rowcolor{color_Combined}
Anwar \cite{anwar2018image} & $2018$ & B &&$\textcolor{mygray1}{\bullet}$&&$\textcolor{mygray1}{\bullet}$&&$\textcolor{mygray1}{\bullet}$& & sparsity prior using band-pass filter responses and incorporate it into a quadratic framework for recovery
\\ \hlinewd{.75pt}
\rowcolor{color_Combined}
Li \cite{li2018learning} & $2018$ & B &&$\textcolor{mygray1}{\bullet}$&&$\textcolor{mygray1}{\bullet}$&&$\textcolor{mygray1}{\bullet}$&$\textcolor{mygray1}{\bullet}$ & Trained CNN model to classify blur vs clean as image prior for regularized minimization formulation
\\ \hlinewd{.75pt}
\rowcolor{color_Combined}
Simoes \cite{simoes2016framework} & $2016$ & NB &&&&&&$\textcolor{mygray1}{\bullet}$&& Diagonalizing unknown convolution operator using FFT and solving via ADMM
\\ \hlinewd{.75pt}
\rowcolor{color_Combined}
Kim \cite{kim2015generalized} & $2015$ & B &&&$\textcolor{mygray1}{\bullet}$&&&$\textcolor{mygray1}{\bullet}$&& Encode temporal/spatial coherency of dynamic scene using optical-flow/TV regularized minimization
\\ \hlinewd{.75pt}
\rowcolor{color_Combined}
Liu \cite{liu2014blind} & $2014$ & B &&&$\textcolor{mygray1}{\bullet}$&$\textcolor{mygray1}{\bullet}$&&$\textcolor{mygray1}{\bullet}$ && Estimate blur from image spectral property and feed into a  regularized TV/eigenvalue minimization
\\ \hlinewd{.75pt}
\rowcolor{color_Combined}
Mosleh \cite{mosleh2014image} & $2014$ & N &&$\textcolor{mygray1}{\bullet}$&&&$\textcolor{mygray1}{\bullet}$&$\textcolor{mygray1}{\bullet}$ && Encode ringing artifacts using Gabor wavelets and fit into a regularized minimization for cancelation
\\ \hlinewd{.75pt}
\rowcolor{color_Combined}
Pan \cite{pan2014deblurring} & $2014$ & N &&&$\textcolor{mygray1}{\bullet}$&$\textcolor{mygray1}{\bullet}$&&$\textcolor{mygray1}{\bullet}$&& Text image deblurring regularized by sparse encoding of spatial/gradient domains
\\ \hlinewd{.75pt}
\rowcolor{color_Combined}
Pan\cite{pan2013fast} & $2013$ & B &&&&$\textcolor{mygray1}{\bullet}$&&$\textcolor{mygray1}{\bullet}$&& Estimates the kernel and the deblurred image from a combined sparse regularization framework
\\ \hlinewd{.75pt}
\rowcolor{color_Combined}
Kim \cite{kim2013dynamic} & $2013$ & B &&$\textcolor{mygray1}{\bullet}$&$\textcolor{mygray1}{\bullet}$&&&$\textcolor{mygray1}{\bullet}$&& Dynamic image deconvolution using TV/Tikhonov/temporal-sparsity regularized minimization
\\ \hlinewd{.75pt}
\rowcolor{color_Combined}
Shen \cite{shen2012spatially} & $2012$ & B &&$\textcolor{mygray1}{\bullet}$&$\textcolor{mygray1}{\bullet}$&&&$\textcolor{mygray1}{\bullet}$&& TV/Tikhonov regularized minimization for image deconvolution
\\ \hlinewd{.75pt}
\rowcolor{color_Combined}
Sroubek \cite{sroubek2012robust} & $2012$ & B &&&&$\textcolor{mygray1}{\bullet}$&&$\textcolor{mygray1}{\bullet}$&& $\ell_1$-regularized minimization for image deconvolution
\\ \hlinewd{.75pt}
\rowcolor{color_Combined}
Dong \cite{dong2011image} & $2011$ & NB &&&&$\textcolor{mygray1}{\bullet}$&&$\textcolor{mygray1}{\bullet}$&& Learn adaptive bases and use in adaptive regularized minimization for sparse reconstruction
\\ \hlinewd{.75pt}
\rowcolor{color_Combined}
Zhang \cite{zhang2011sparse, zhang2011close} & $2011$ & B &&$\textcolor{mygray1}{\bullet}$&&$\textcolor{mygray1}{\bullet}$&&$\textcolor{mygray1}{\bullet}$&& Sparse regulation  of images via KSVD library for deconvolution and apply to facial recognition
\\ \hlinewd{.75pt}
%
%
\rowcolor{color_TV}
Bai \cite{bai2018graph} & $2018$ & B &&$\textcolor{mygray1}{\bullet}$&$\textcolor{mygray1}{\bullet}$&&&&& Both kernel/image recovered via combined regularization using reweighted graph TV priors
\\ \hlinewd{.75pt}
\rowcolor{color_TV}
Lou \cite{lou2015weighted} & $2015$ & N &&&$\textcolor{mygray1}{\bullet}$&&&$\textcolor{mygray1}{\bullet}$&& Weighted differences of TV regularizers in $\ell_1$/$\ell_2$ norms and solved by split variable technique
\\ \hlinewd{.75pt}
\rowcolor{color_TV}
Zhang \cite{zhang2014image} & $2014$ & N &$\textcolor{mygray1}{\bullet}$&&$\textcolor{mygray1}{\bullet}$&&&$\textcolor{mygray1}{\bullet}$&& Local/non-local similarities defined by TV$^1$/TV$^2$ and regulated by combined minimization
\\ \hlinewd{.75pt}
\rowcolor{color_TV}
Xu \cite{xu2012depth} & $2012$ & B &&&$\textcolor{mygray1}{\bullet}$&&&&& Regulate motion by difference of depth map and deconvolve via non-convex TV minimization
\\ \hlinewd{.75pt}
\rowcolor{color_TV}
Chan \cite{chan2011augmented} & $2011$ & N &&&$\textcolor{mygray1}{\bullet}$&&&&& Deconvolve image/videos using spatial/temporal TV regularization solved by split varying technique
\\ \hlinewd{.75pt}
\rowcolor{color_TV}
Afonso \cite{afonso2010fast} & $2010$ & N &&&$\textcolor{mygray1}{\bullet}$&&&&& Deconvolve image using TV regularization solved by split varying technique
\\ \hlinewd{.75pt}
%
%
\rowcolor{color_Tikhonov}
Li \cite{li2018pure} & $2018$ & B && $\textcolor{mygray1}{\bullet}$ &&&& $\textcolor{mygray1}{\bullet}$ && Non-iterative deconvolution via combination of Wiener filters, solution by a system linear equations
\\ \hlinewd{.75pt}
\rowcolor{color_Tikhonov} 
Bertero \cite{bertero2010discrepancy} & $2010$ & N && $\textcolor{mygray1}{\bullet}$ &&&&&& Generalized Kullback-Leiblar divergence function to regularize Poisson images
\\ \hlinewd{.75pt}
\rowcolor{color_Tikhonov}
Cho \cite{cho2009fast} & $2009$ & B &&$\textcolor{mygray1}{\bullet}$&$\textcolor{mygray1}{\bullet}$&&&&& Separate recovery of motion kernel and image from residual image using Tikhonov regularization
\\ \hlinewd{.75pt}
\rowcolor{color_Tikhonov}
Wiener \cite{wiener1949extrapolation, gonzalez2012digital} & $1949$ & N &&$\textcolor{mygray1}{\bullet}$&&&&&& Regulate image spectrum in Fourier domain with inverse kernel response
\\ \hlinewd{.75pt}
%
%
\rowcolor{color_Shrinkage}
Xiao \cite{xiao2016learning} & $2016$ & B &&&&&$\textcolor{mygray1}{\bullet}$&$\textcolor{mygray1}{\bullet}$&& Regulate motion blur and image by half-quadratic minimization and solve by iterative shrinkage
\\ \hlinewd{.75pt}
\rowcolor{color_Shrinkage}
Zuo \cite{zuo2013generalized} & $2013$ & N &&&&$\textcolor{mygray1}{\bullet}$&$\textcolor{mygray1}{\bullet}$&&& Extends the soft-threshold for non-convex sparse coding using generalized iterated shrinkage
\\ \hlinewd{.75pt}
\rowcolor{color_Shrinkage}
Krishnan \cite{krishnan2011blind} & $2011$ & B &&&&$\textcolor{mygray1}{\bullet}$&$\textcolor{mygray1}{\bullet}$&&& Relaxed prior by spherical section property used in regularized minimization
\\ \hlinewd{.75pt}
\rowcolor{color_Shrinkage}
Dabov \cite{dabov2008image} & $2008$ & N &&$\textcolor{mygray1}{\bullet}$&&&$\textcolor{mygray1}{\bullet}$&&& Two steps of denoising in using hard-thresholding in Fourier and deblurring using Wiener filter
\\ \hlinewd{.75pt}
\rowcolor{color_Shrinkage}
Neelamani \cite{neelamani2004forward} & $2004$ & N &&&&&$\textcolor{mygray1}{\bullet}$&&& Apply transform domain shrinkage in both Fourier and wavelet domains solved by mean-square error
\\ \hlinewd{.75pt}
%
%
\rowcolor{color_STAT}
Whyte \cite{whyte2014deblurring} & $2014$ & N &$\textcolor{mygray1}{\bullet}$&&$\textcolor{mygray1}{\bullet}$&&&$\textcolor{mygray1}{\bullet}$&& Encodes motion and saturated pixels in a combined regularization of Richardson-Lucy and TV
\\ \hlinewd{.75pt}
\rowcolor{color_STAT}
Schmidt \cite{schmidt2013discriminative} & $2013$ & N &$\textcolor{mygray1}{\bullet}$&&&&&&& A discriminative cascaded model is proposed based on generalized half-quadratic regularization
\\ \hlinewd{.75pt}
\rowcolor{color_STAT}
Dong \cite{dong2013nonlocally} & $2013$ & B &$\textcolor{mygray1}{\bullet}$&&&$\textcolor{mygray1}{\bullet}$&&&& Sparse coding of noise in a minimization problem recast by a maximum a posterior algorithm
\\ \hlinewd{.75pt}
\rowcolor{color_STAT}
Sun \cite{sun2013edge} & $2013$ & B &$\textcolor{mygray1}{\bullet}$&&&$\textcolor{mygray1}{\bullet}$&&$\textcolor{mygray1}{\bullet}$&& Estimate blur kernel using patch based statistical prior in a combined regularized minimization
\\ \hlinewd{.75pt}
\rowcolor{color_STAT}
Bishop \cite{bishop2012light} & $2012$ & B &$\textcolor{mygray1}{\bullet}$&&$\textcolor{mygray1}{\bullet}$ &&&&& Estimate depth map via TV regularizer to recover blur kernel and employ in Bayesian deconvolution
\\ \hlinewd{.75pt}
\rowcolor{color_STAT}
Whyte \cite{whyte2012non} & $2012$ & B &$\textcolor{mygray1}{\bullet}$&&$\textcolor{mygray1}{\bullet}$&&&&& Models camera motion geometry and feed into MAP estimation using hyper-Laplacian priors
\\ \hlinewd{.75pt}
\rowcolor{color_STAT}
Amizic \cite{amizic2012sparse} & $2012$ & B &$\textcolor{mygray1}{\bullet}$&&&$\textcolor{mygray1}{\bullet}$&&&& Blind deconvolution of image from sparse Bayesian framework
\\ \hlinewd{.75pt}
\rowcolor{color_STAT}
Levin \cite{levin2011efficient, levin2011understanding} & $2011$ & B &$\textcolor{mygray1}{\bullet}$&&&$\textcolor{mygray1}{\bullet}$&&&& A simplified version of MAP is proposed to recover motion blur and latent image
\\ \hlinewd{.75pt}
\rowcolor{color_STAT}
Krishnan \cite{krishnan2009fast} & $2009$ & N &$\textcolor{mygray1}{\bullet}$&&$\textcolor{mygray1}{\bullet}$&&&$\textcolor{mygray1}{\bullet}$&& Implement fast alternating minimization to solve hyper-Laplacian prior for MAP estimation
\\ \hlinewd{.75pt}
\rowcolor{color_STAT}
Bertero \cite{bertero2009image} & $2009$ & N &$\textcolor{mygray1}{\bullet}$&&&&&&& Estimate MAP in Bayesian paradigm for Poisson imaging solved by EM algorithm
\\ \hlinewd{.75pt}
\rowcolor{color_STAT}
Bonettini \cite{bonettini2008scaled} & $2008$ & N &$\textcolor{mygray1}{\bullet}$&&&&&&& Apply scale gradient projection for iterative solution of maximum likelihood of Poisson imaging
\\ \hlinewd{.75pt}
\rowcolor{color_STAT}
Shan \cite{shan2008high} & $2008$ & B/N &$\textcolor{mygray1}{\bullet}$&&&&&$\textcolor{mygray1}{\bullet}$&& Blur and image residuals are regularized in MAP framework solved by split variable technique
\\ \hlinewd{.75pt}
%
%
\rowcolor{color_STAT}
Yuan \cite{yuan2007image} & $2007$ & B &$\textcolor{mygray1}{\bullet}$&$\textcolor{mygray1}{\bullet}$&&&&&& Create blur kernel from residual observation and deconvolve via Richardson-Lucy algorithm
\\ \hlinewd{.75pt}
%
%
\rowcolor{color_STAT}
Fergus \cite{fergus2006removing} & $2006$ & B &$\textcolor{mygray1}{\bullet}$&&&&&&& Estimate blur from gradient domain regularized in MAP framework
\\ \hlinewd{.75pt}
\rowcolor{color_STAT}
Biggs \cite{biggs1997acceleration} & $1997$ & N &$\textcolor{mygray1}{\bullet}$&&&&&&& Accelerate iterative restoration based on RL and EM algorithms
\\ \hlinewd{.75pt}
\rowcolor{color_STAT}
RL \cite{richardson1972bayesian, lucy1974iterative} & $1974$ & N &$\textcolor{mygray1}{\bullet}$&&&&&&& Richardson-Lucy (RL): Regulate image frequency in MAP framework and solve iteratively
\\ \hlinewd{1.5pt}
\end{tabular}
\end{table*}

\subsubsection{Statistical priors}\label{stat_prior}
The idea is to formulate the occurrence of the underlying image as a conditional probability of a given blurry observation by maximum-a-posterior (MAP) estimation. The early development of this method was proposed by Richardson-Lucy (RL) \cite{richardson1972bayesian, lucy1974iterative} by recasting the solution in an iterative algorithm starting from an initial guess. The accelerated RL algorithm was proposed later by Biggs \cite{biggs1997acceleration} using an adaptive line searching technique. We refer the reader to the comprehensive surveys in \cite{geman1987stochastic, kundur1996blind} for the early development of these methods. With the emergence of digital consumer electronic cameras in the early 2000s, more practical deconvolution methods were released using the blind approach \cite{fergus2006removing, shan2008high, levin2011efficient, levin2011understanding, yuan2008progressive, bonettini2008scaled, amizic2012sparse, whyte2012non, bishop2012light, dong2013nonlocally, schmidt2013discriminative}. With growing numbers of numerical solvers for alternating direction methods of multipliers (ADMM) (a variant of the splitting variable technique), the regulatory formulations were updated accordingly using different prior models \cite{shan2008high,krishnan2009fast,sun2013edge,whyte2014deblurring}. 

\subsubsection{Tikhonov regularization}\label{tikhonov_reg}
When the data fidelity is regularized in $\ell_2$-norm space to minimize a cost function, it becomes a variant of the Tikhonov regularization problem. The solution to this problem is given by quadratic minimization that can be accelerated by fast Fourier transform (FFT) and so reduce the computational complexity by an order of $\mathcal{O}(n\log{n})$. An early application of this regularization was deployed in the classical Wiener deconvolution to regulate the image spectrum in the Fourier domain using a non-linearly weighted inverse blur response  \cite{wiener1949extrapolation, gonzalez2012digital}. More recent methods employ this framework for fast reconstruction \cite{cho2009fast, bertero2010discrepancy, li2018pure}. Despite their efficiency, the reconstructed image edges are hampered by ringing artifacts, also known as the Gibbs phenomenon. 

\subsubsection{Iterative shrinkage}\label{iter_shrink}
This is a variant of sparse reconstruction which recasts the regularization problem as an iterative procedure where dominant feature coefficients are preserved during each iteration. Different regularizers can be found for image deconvolution in \cite{neelamani2004forward, dabov2008image, krishnan2011blind, zuo2013generalized, xiao2016learning}.

\subsubsection{Variational regularization}\label{var_reg}
Known as the total variation (TV) method, in which the priors for either the blur kernel or latent image are regulated by the TV-norm \cite{afonso2010fast, chan2011augmented,xu2012depth,zhang2014image, lou2015weighted}, this norm preserves sharp edges while preventing Gibbs oscillations for recovery. As a common disadvantage, these methods suffer from visual blocking or ``staircasing'' artifacts.


\subsubsection{Combined regularization}\label{comb_reg}
Combined approaches refer to the use of more than one regularization prior for recovery. This becomes more useful when both blur and image priors (in the blind case) could be fit into one regularization framework to address more complex formulations. The common practice is to use split variable techniques to recast the algorithms in parallel and independently update each sub-modular task \cite{zhang2011sparse, dong2011image, shen2012spatially, kim2013dynamic, sroubek2012robust, pan2013fast, mosleh2014image, liu2014blind, kim2015generalized, li2018learning}.

\subsubsection{Deep Convolutional Neural Network (CNN)}\label{deep_cnn}
This is a variant of deep learning methods in which a convolutional neural network is trained to encode image features in multiple layers of decomposition. Each layer contains a set of convolutional filters and activation, e.g. ReLU, to produce a feature map that is passed onto the next layer. The cascaded layers of the network provide an efficient way to decompose complex image structures for encoding/decoding \cite{xu2014deep, hradivs2015convolutional, sun2015learning, schuler2016learning, tao2018scale, wang2018training}. The common practice for developing these networks is to guide the training process by feeding in a pre-deconvolved image using Wiener/Tikhonov regularization to boost the performance results. A common disadvantage is the requirement of a training image set for training these networks. Since the latent image is not available, the train set is synthetically generated by a pre-defined blur kernel to obtain their blurry observation for training. Such an assumption however does not necessarily conform with the reality of blur observed in natural imaging applications. 

\subsubsection{Decoupled Methods}\label{decop_method}
In the literature, the problem of image deconvolution is usually coupled with denoising and deblurring, where prior assumptions are considered to regulate both inverse problems in one recovery framework. Recent developments decouple (separate) these into two sub-modular tasks, where the solution is usually cast as split variable minimization techniques for reconstruction. In fact, one can separately integrate a denoiser as a plug-in solution to address the denoising step \cite{zoran2011learning, danielyan2012bm3d, dabov2007image, romano2017little, chan2017plug, zhang2017learning}.

\subsection{Remaining Challenges and Contributions}
Despite vigorous research efforts, maintaining both precision and speed are still the main drawbacks of existing algorithms. High speed recovery simply means a ``non-iterative'' approach (or at least very few procedural algorithms) for practical implementation. Few such solutions exist, and tend to be accelerated by fast Fourier transform (FFT), such as Wiener \cite{wiener1949extrapolation, gonzalez2012digital}, Tikhonov \cite{li2018pure, cho2009fast}, Richardson-Lucy (RL) \cite{biggs1997acceleration}, and diagonalizing \cite{simoes2016framework} based algorithms. Despite their speed, they are prone to ringing artifacts and/or losing fine image details. By contrast, existing approaches with sophisticated deblurring usually recast the problem into an iterative minimization framework and are computationally expensive. Recent techniques adopt CNN models to formulate the problem in a feed-forward fashion and accelerate the recovery process using GPUs. However, such algorithms are still limited by the blur modeling of natural images. In addition, the majority of deconvolution methods involve complicated parameter tuning procedures which limit their generalization.

The contributions of this paper in addressing the above challenges are as follows:
\begin{itemize}
\item We observe that the problems of image deblurring and denoising should be decoupled for reconstruction. {This is motivated by the Wiener deconvolution method where the recovery image is regulated by the inverse PSF response in the Fourier domain. However, unlike Wiener's approach where both frequency regulation and correction are done in the Fourier domain, we define a dual spatial domain for image correction. In particular,} we design a closed-form solution of the deblurring kernel as a linear combination of high-order FIR even-derivative filters. For the numerical implementation, we employ the MaxPol library \cite{HosseiniPltaniotis_MaxPol_TIP_2017, HosseiniPltaniotis_MaxPol_SIAM_2017} and call our deblurring method ``1Shot-MaxPol'', which is available for download at\footnote{\url{https://github.com/mahdihosseini/1Shot-MaxPol}}. We consider a generalized Gaussian filter for smooth denoising before estimation of sharp deblurring edges.
\item We adopt the generalized Gaussian distribution to model PSF blur and analyze its feasibility range for recovery using the proposed deblurring method. {The rational behind such consideration is Many PSF (static blur) applications are symmetric and can be modeled by such distribution.}
\item {We formulate a new blind PSF estimation method using scale-space analysis in the Fourier domain.} We consider two variants of Gaussian and Laplacian models for blind estimation of blur statistics. {The main motivation behind such blind estimation is in many applications such as satellite imaging there is no practical means of PSF calibration and hence a blind estimation is required.}
\item An adaptive tuning parameter is introduced based on the relative image entropy calculation to control the strength of deblurring
\item Thorough experiments are conducted on $2054$ natural images across diverse wavelength imaging bands. We blindly estimate the parameters for two model PSFs and feed them into seven state-of-the-art non-blind deblurring methods and compare their performances with our 1Shot-MaxPol deblurring. Empirical results indicate the superiority of the proposed method against three performance indexes of no-reference focus quality assessment, visual perception error, and computational complexity.
\end{itemize}

The remainder of the paper is as follows. We introduce the 1Shot-MaxPol deblurring method in Section \ref{section_inverse_deconvolution_design}. The generalization of symmetric PSF model and blind estimation of its statistics are given in Section  \ref{sec_blur_model}. The experiments are provided in Section \ref{sec_experiments} and the paper is concluded in Section \ref{sec_conclusion}.

\section{Proposed Decoupled Approach}\label{section_inverse_deconvolution_design}
In this section, we propose a new approach for symmetric PSF deblurring by correcting the fall-off of the high frequency band by means of frequency polynomial approximation. We construct the dual representation in the spatial domain for inverse PSF deblurring in the form of one-shot convolution filters. The proposed method follows a similar approach to Wiener deconvolution \cite{wiener1949extrapolation, gonzalez2012digital} by regulating the effect of blurring operation in the frequency domain. However, unlike Wiener's approach, we define the dual filter representation in the spatial domain and thus avoid directly manipulating the blur image in the frequency domain. Recall that the Wiener deconvolution provides a filter $h_{\textit{W}}(x)$ such that direct convolution of the filter with the observed image yields a closed approximation of the latent image  $\bar{f}_{\textit{L}}(x)= h_{\textit{W}}(x)\ast f_{\textit{B}}(x)$. This filter is obtained by minimizing the mean square error between the latent image $f_{L}$ and the recovered (approximated) image $\bar{f}_{\textit{L}}(x)$ in the Fourier domain i.e.
\begin{align}
\hat{\epsilon}(\omega)=\mathbf{E}|\hat{f}_{\textit{L}}(\omega)-\hat{\bar{f}}_{\textit{L}}(\omega)|.
\label{eq_wiener_squre_error}
\end{align}
By substituting $\hat{\bar{f}}_{\textit{L}}(\omega)$ in (\ref{eq_wiener_squre_error}) and minimizing the error with respect to latent image, the associated filter will be expressed by
\begin{align}
\hat{h}_{\textit{W}}(\omega)=\frac{1}{\hat{h}_{\textit{PSF}}(\omega)}\hat{h}_{\textit{LPF}}(\omega)
\label{eq_wiener_filter}
\end{align}
where the low-pass filter is defined by
\begin{align}
\hat{h}_{\textit{LPF}}(\omega) = \frac{|\hat{h}_{\textit{PSF}}(\omega)|^2}{|\hat{h}_{\textit{PSF}}(\omega)|^2 + 1/\textit{SNR}(\omega)}.\notag
\label{eq_wiener_filter_LPF}
\end{align}
Here, $\textit{SNR}(\omega)$ is the signal-to-noise-ratio (obtained by the ratio of the mean-power spectral densities of the latent image to the noise). The low-pass filter attenuates high frequency information in order to combat noise artifacts, which are known to decrease the SNR at high frequencies. The numerical implementation of the Wiener filter is usually carried out by transferring the blur image observation into the Fourier domain and manipulating the frequency responses according to the filter definitions in (\ref{eq_wiener_filter}) and recovering the image using the inverse Fourier transform. However, the inverse transform is subject to the Gibbs phenomenon (also known as ringing artifacts).

Our main idea here is to decouple denoising and blur correction in (\ref{eq_wiener_filter}) by defining two separate convolutional filters in the dual spatial domain
\begin{equation}
h_{\textit{D}}(x)=h^{-1}_{\textit{PSF}}(x)\ast h_{\textit{Denoise}}(x)
\label{eq_decouple_filtering}
\end{equation}
such that the convolution of the decoupled filter $h_{\textit{D}}(x)$ in (\ref{eq_decouple_filtering}) with the blurry observed image yields the latent approximation $\bar{f}_{\textit{L}}(x)= h_{\textit{D}}(x)\ast f_{\textit{B}}(x)$. The merit of our design in (\ref{eq_decouple_filtering}) is the decoupling of the denoising and deblurring problems, enabling them to be individually addressed for recovery. One can separately apply a denoiser as a plug-in tool if the input image is perturbed with noise. 

{
\subsection{Inverse Deconvolution Kernel Design}
The inverse filter in (\ref{eq_decouple_filtering}) is defined as the inverse Fourier transform of the inverse PSF response $h^{-1}_{\textit{PSF}}(x)=\mathcal{F}^{-1}\{1/\hat{h}_{\textit{PSF}}(\omega)\}$. However, as mentioned earlier, directly calculating the inverse Fourier will introduce Gibbs artifacts. To avoid this, we define a dual representation in both Fourier and spatial domains by approximating the inverse PSF response in the Fourier domain by a series of frequency polynomials
\begin{align}
\frac{1}{\hat{h}_{\textit{PSF}}(\omega)} \approx \sum^{N}_{n=0}\alpha_{n}\omega^{2n}.
\label{eq_inverse_deconv_kernel_3}
\end{align}
Note that we considered only even polynomials for approximation as the symmetry of the PSF ensures that the inverse response will be an even function. The unknown coefficients $\{\alpha_{n}\}^N_{n=1}$ are determined by fitting the inverse PSF response to the polynomial series up to certain frequency range
\begin{align}
\arg\min_{\alpha_n} \|\frac{1}{\hat{h}_{\textit{PSF}}(\omega)} - \sum^{N}_{n=0}\alpha_{n}\omega^{2n}\|_2,~\omega\in[0,\omega_T],
\label{eq_inverse_deconv_kernel_minimization}
\end{align}
where the range $\omega_T$ is tweaked to avoid fitting instabilities. The numerical solution to the fitting problem in (\ref{eq_inverse_deconv_kernel_minimization}) is provided by solving a linear least square problem in \cite{kelley1999iterative}.
}

\begin{figure}[htp]
\centerline{
\subfigure[GG-Blurr kernel $h$]{\includegraphics[width=0.2\textwidth]{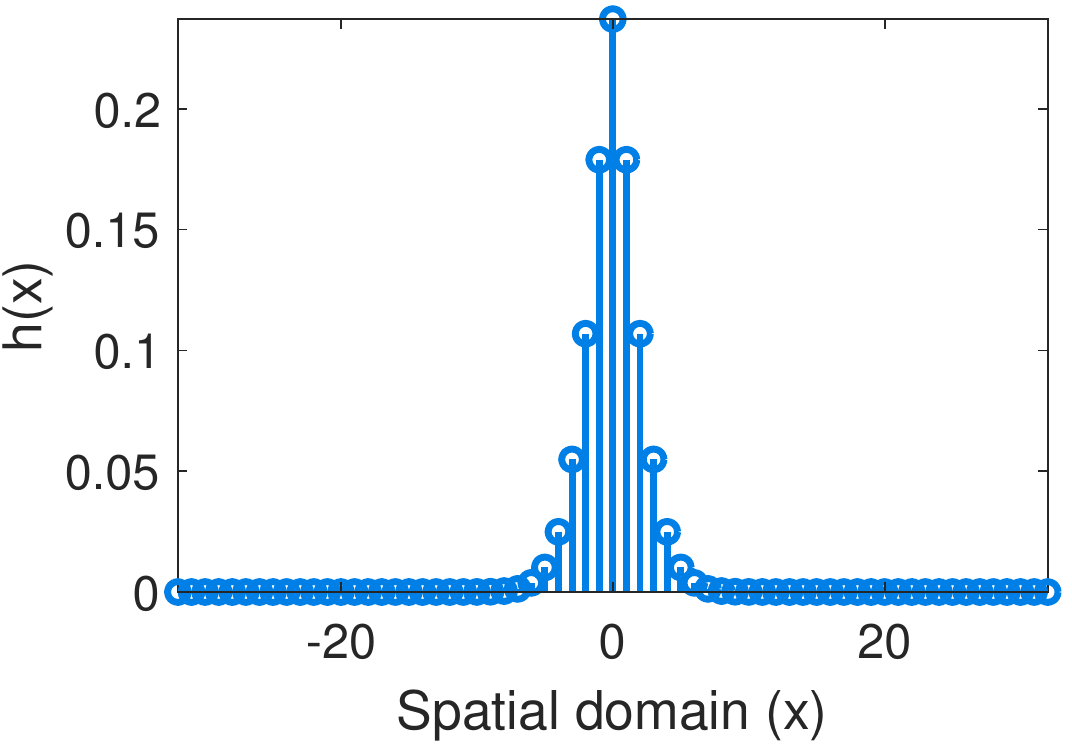}}
\subfigure[Inverse deblurring kernel $D$]{\includegraphics[width=0.2\textwidth]{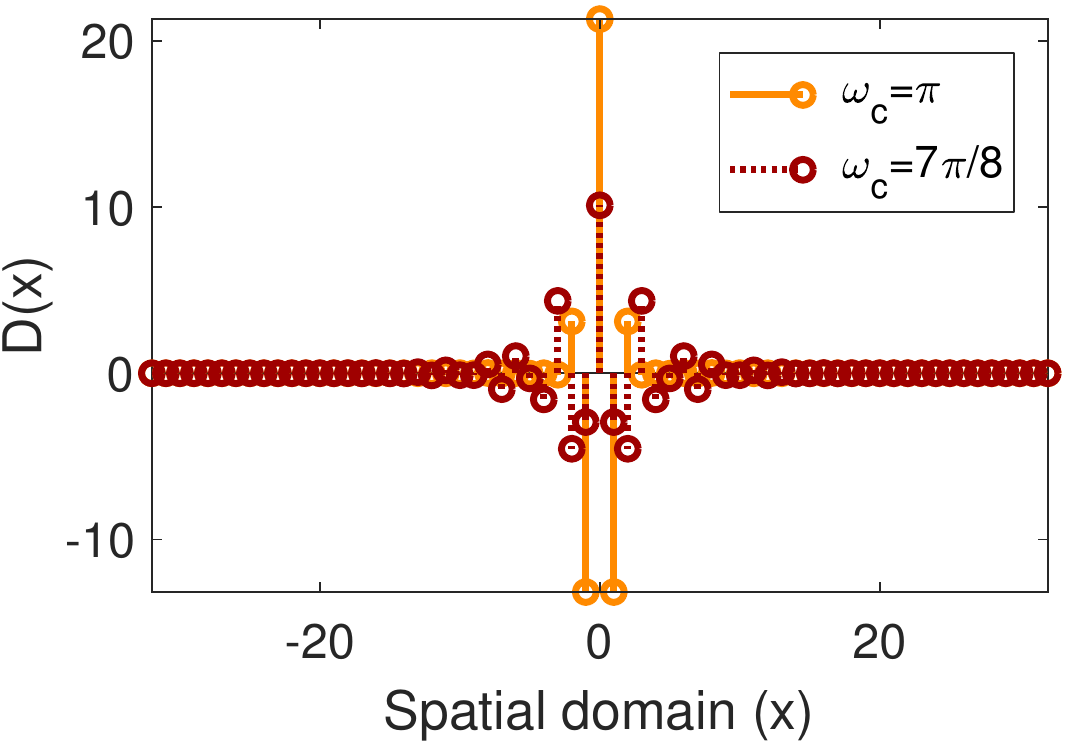}}
}\vspace{-.05in}
\centerline{
\subfigure[Spectrum of $|\mathcal{F}\{h\ast D\}|$]{\includegraphics[width=0.2\textwidth]{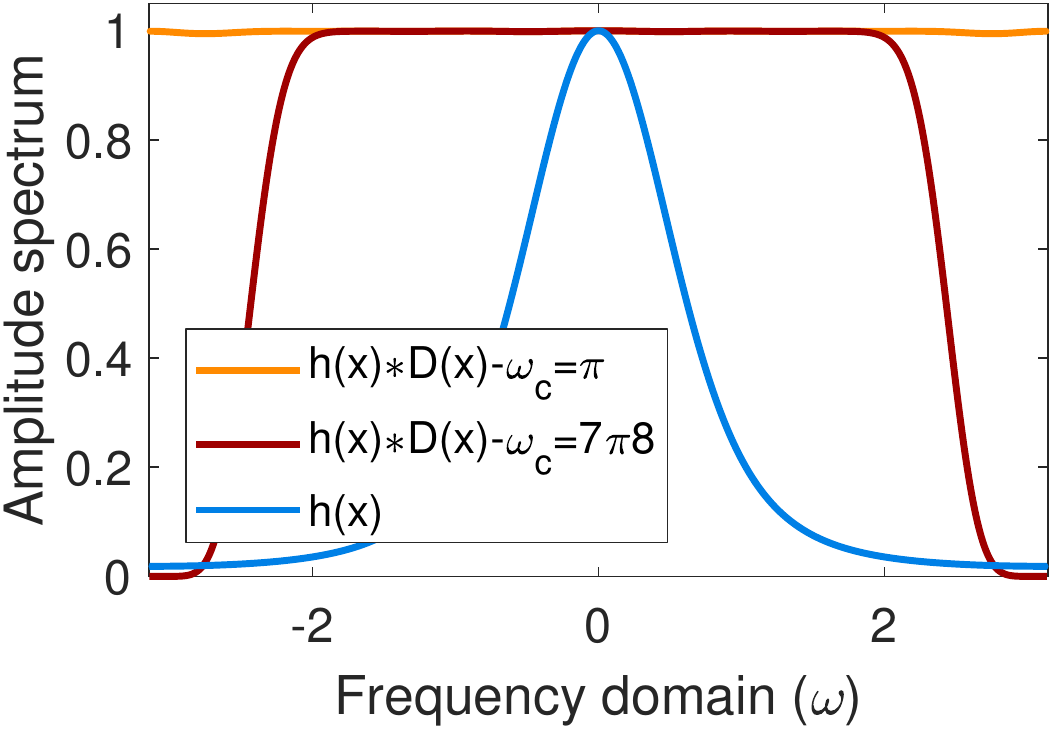}}
\subfigure[Spectrum of inverse kernel]{\includegraphics[width=0.2\textwidth]{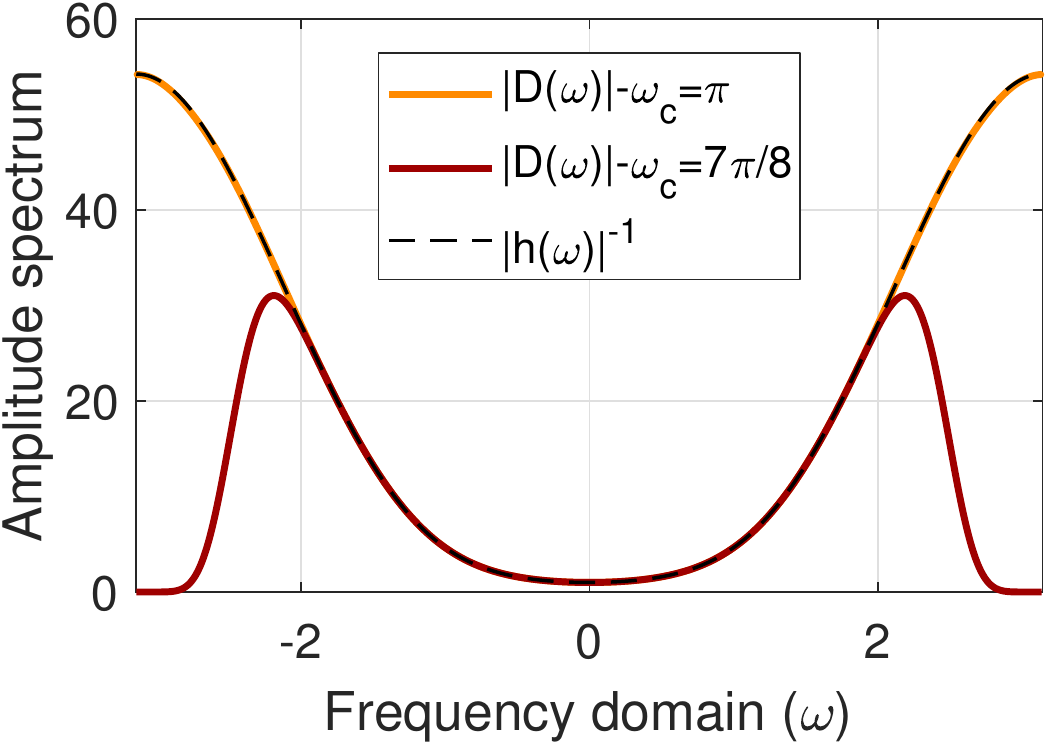}}
}\vspace{-.05in}
\caption{Inverse deconvolution $D(x)$ kernel design. The blurring kernel here is a generalized Gaussian form with parameters $\alpha=2$ and $\beta=1.5$. This type of kernel is usually common in many real imaging applications such as optical aberration and turbulent medium blur. Derivatives up to the $14$th order are used, i.e. $N=7$, to design $D(x)$ with two different cut-offs $\omega_c=\{7\pi/8,\pi$\}.}
\label{fig_inverse_deblurring_kernel_design}
\end{figure}

{
The dual representation of the frequency polynomials in the spatial domain is equivalent to a linear combination of derivative operators. Therefore, the dual representation of the inverse filter defined in (\ref{eq_inverse_deconv_kernel_3}) can be equally represented by
\begin{align}
h^{-1}_{\textit{PSF}}(x) \approx \delta(x) + \sum^{N}_{n=1}\alpha_{n}(-1)^n\frac{\partial^{2n}}{\partial x^{2n}}.
\label{eq_inverse_deconv_kernel_dual_spatial}
\end{align}
Note that the energy of PSF is assumed to be normalized $\int h_{\textit{PSF}}(x)dx = 1$ in (\ref{eq_inverse_deconv_kernel_dual_spatial}), which is separated from the deconvolution process. The continuous derivative operators in (\ref{eq_inverse_deconv_kernel_dual_spatial}) can be numerically approximated using Finite Impulse Response (FIR) convolution filters
\begin{align}
h^{-1}_{\textit{PSF}}[k] = \delta[k] + D[k],
\label{eq_inverse_deconv_kernel_dual_spatial_FIR}
\end{align}
where $D[k] = \sum^{N}_{n=1}\alpha_{n}(-1)^n d^{2n}[k]$ is the associated FIR deblurring kernel. Here, $d^{2n}[k]$ is the discrete approximation to the $2n$-th order derivative operator applied in the bounded continuous domain $x\in[-T,T]$ i.e. $d_{2n}(x)\equiv{\partial^{2n}}/{\partial x^{2n}}$.
For a numerical solution of the derivative filters $d^{2n}[k]$ in (\ref{eq_inverse_deconv_kernel_dual_spatial_FIR}), we used MaxPol\footnote{Available online at \url{https://github.com/mahdihosseini/MaxPol}}, a package to solve numerical differentiation \cite{HosseiniPltaniotis_MaxPol_TIP_2017, HosseiniPltaniotis_MaxPol_SIAM_2017}.} In particular, MaxPol provides a closed-form solution to the FIR derivative kernels that can be regulated in terms of different parameter designs such as arbitrary order of differentiation $n$, different cut-off frequency, and polynomial accuracy $\ell$ for high frequency resolution. For more information, we refer the reader to \cite{HosseiniPltaniotis_MaxPol_TIP_2017, HosseiniPltaniotis_MaxPol_SIAM_2017} and the references therein. Figure \ref{fig_inverse_deblurring_kernel_design} demonstrates an example of approximating an inverse deblurring kernel with two different cut-off parameters.

\begin{figure}[htp]
\centerline{
{\includegraphics[width=0.25\textwidth]{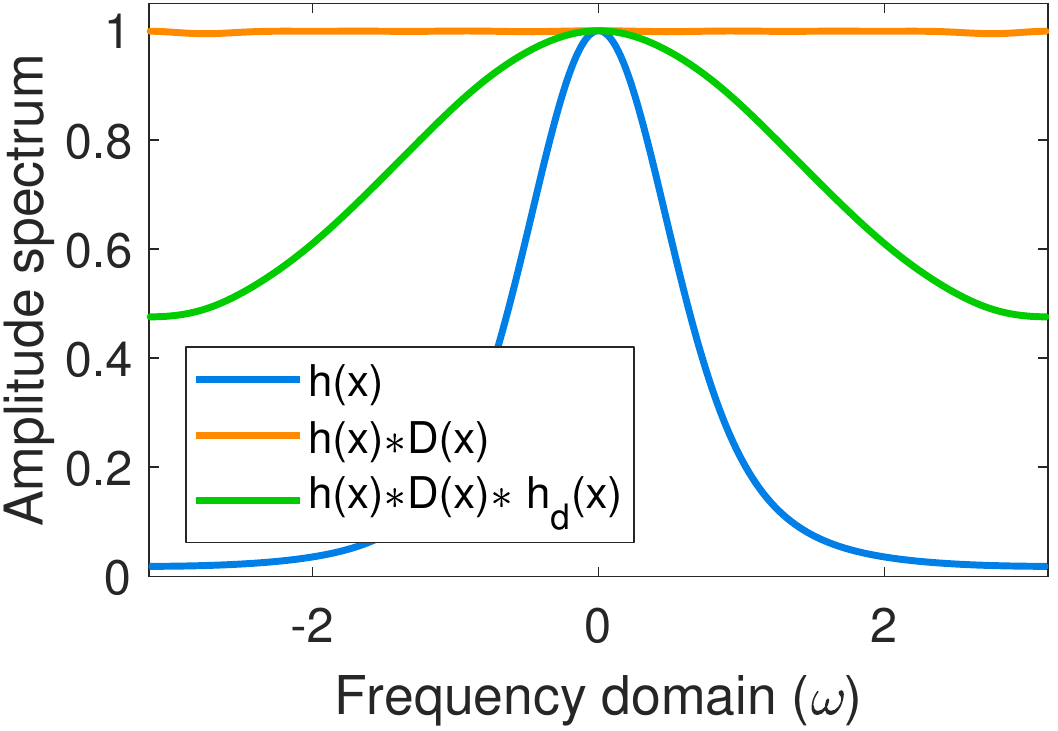}}
}\vspace{-.05in}
\caption{Adding a denoising filter $h_d$ as a decoupled module after fall-off correction of the blurring kernel.}
\label{fig_denoising_module}
\end{figure}

{
\subsection{Decoupled Smoothing Filter}}
{Here we introduce an efficient and yet simple denoising approach} to avoid computational complexity and maintain recovery accuracy. The whole idea of decoupled design in (\ref{eq_decouple_filtering}) is to balance the amplitude fall-off of high frequency components caused by the PSF kernel. Ideally speaking, if no denoising/cut-off is considered, all of the frequency domain will be deconvolved according to the inverse kernel response. However, such full correction should be avoided due to noise contamination in real applications. Once the image is deconvolved by an inverse filter,  we apply (convolve) similar symmetric blur kernels for denoising with less blur scale than that considered for deconvolution. This guarantees that the fall-off of the high frequency amplitude will be balanced between noise cancellation and amplifying meaningful edge information. See Figure \ref{fig_denoising_module} for an example. We suggest using generalized Gaussian kernel for such denoising where the associated FIR kernel has no vanishing moment and hence does not cause edge hallucinations. For more information on the generalized Gaussian, please refer to Section \ref{sec_blur_model}. Nevertheless one can deploy more sophisticated denoising methods such as a pre-trained CNN model in \cite{zhang2017learning} or the overcomplete dictionary design in \cite{danielyan2012bm3d} as a decoupled module.

\subsection{Two dimensional deblurring framework}
While our design in the previous section is applied in one dimension, for imaging applications this should be extended to two dimensions (2D). Let $f(x,y)\in\mathbb{R}^{N_1\times N_2}$ represent the image in the 2D domain, with $N_1$ and $N_2$ being the number of discrete pixels along the vertical and horizontal axes, respectively. {
The PSF blur in many optical imaging systems is considered to be rotationally symmetric - the associated blurring operator is identical in any arbitrary rotational angle i.e. $h_{\textit{PSF}}(r,\theta)=h_{\textit{PSF}}(r)$. For instance, a Gaussian-like PSF kernel $h_{\textit{PSF}}(r,\theta)=1/\sqrt{2\pi \sigma}e^{-r^2/2\sigma^2}$ is rotationally-invariant. In fact, the Gaussian PSF can be constructed by means of two separable (independent) kernels along the horizontal and vertical Cartesian axes, e.g. $h_{\textit{PSF}}(x,y)=1/\sqrt{2\pi \sigma}e^{-(x^2+y^2)/2\sigma^2}=h_{\textit{PSF}}(x)h_{\textit{PSF}}(y)$. In general, we assume the 
} the blur operator is independently applied in both dimensions (separable mode). So, the linear model in (\ref{eq_psf_model_1}) is revised to
\begin{align}
f_{\textit{B}}(x,y) = f_{\textit{L}}(x,y) \ast h_{\textit{PSF}}(x) \ast h_{\textit{PSF}}(y) + \eta(x, y).
\label{eq_2D_psf_model_1}
\end{align}
The corresponding deblurring kernels in both directions are designed by means of the approximation method in the previous section and applied to the blurry image for reconstruction
\begin{align}
f_{R}(x,y) = h_{\textit{D}}(x) \ast h_{\textit{D}}(y) \ast f_{B}(x,y),
\label{eq_2D_psf_model_2}
\end{align}

The energy level of the blurring kernel is usually unknown \textit{a priori} for natural imaging problems. We define a tuning parameter $\gamma\in[0,1]$ to control the significance of the deconvolution level
\begin{align}
f_{R}(x,y) = f_{B}(x,y) + \gamma \nabla_{D}f_{B}(x,y),
\label{eq_2D_psf_model_3}
\end{align}
where $\nabla_{D}f_{B}(x,y) = f_{B}(x,y)\ast\left[D_x+D_y+D_{xy}\right]$ gives the reconstructed image edges and $D_{xy} = D_x\ast D_y$ is the crossed deconvolution operator independently applied to the horizontal and vertical axes. Therefore, all of the convolution operations in (\ref{eq_2D_psf_model_3}) are conducted in one dimension with a total computational complexity of $\mathcal{O}(4L)$ when $L$ is the tap-length of the FIR deconvolution operators. 

\begin{figure}[htp]
\centerline{
\subfigure[Original Image $f_{\textit{L}}$]{\includegraphics[width=0.15\textwidth]{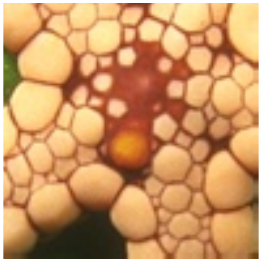}}\hspace{-.02in}
\subfigure[Blurred by (\ref{eq_2D_psf_model_1})]{\includegraphics[width=0.15\textwidth]{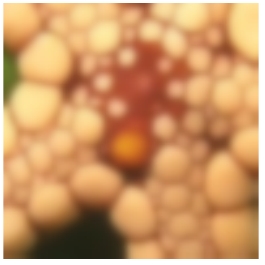}}\hspace{-.02in}
\subfigure[Deconvolved by (\ref{eq_2D_psf_model_3})]{\includegraphics[width=0.15\textwidth]{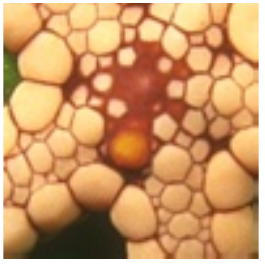}}
}\vspace{-.05in}
\centerline{
\subfigure[$f_B\ast D_x$]{\includegraphics[width=0.15\textwidth]{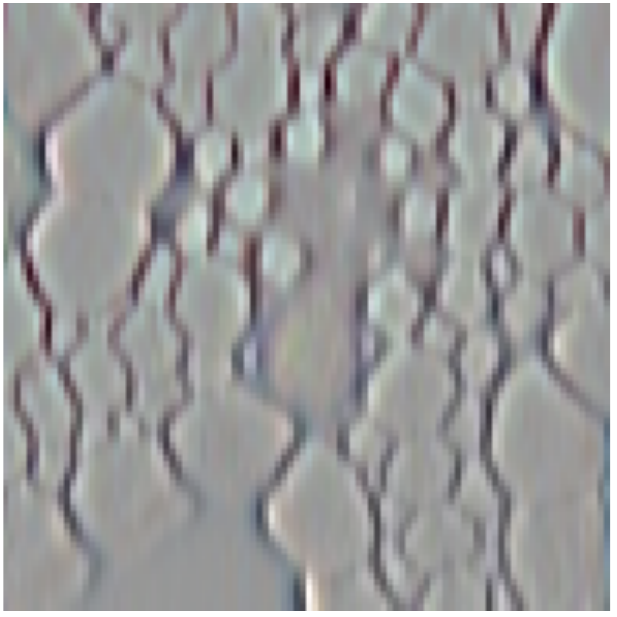}}\hspace{-.02in}
\subfigure[$f_B\ast D_y$]{\includegraphics[width=0.15\textwidth]{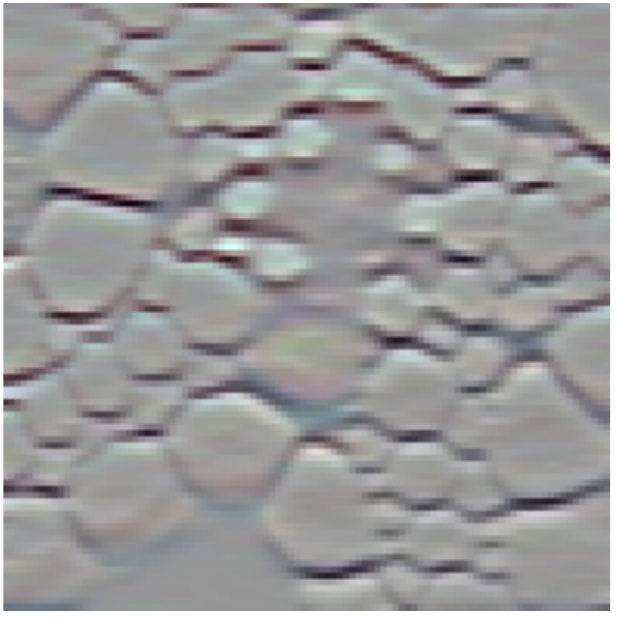}}\hspace{-.02in}
\subfigure[$f_B\ast D_{xy}$]{\includegraphics[width=0.15\textwidth]{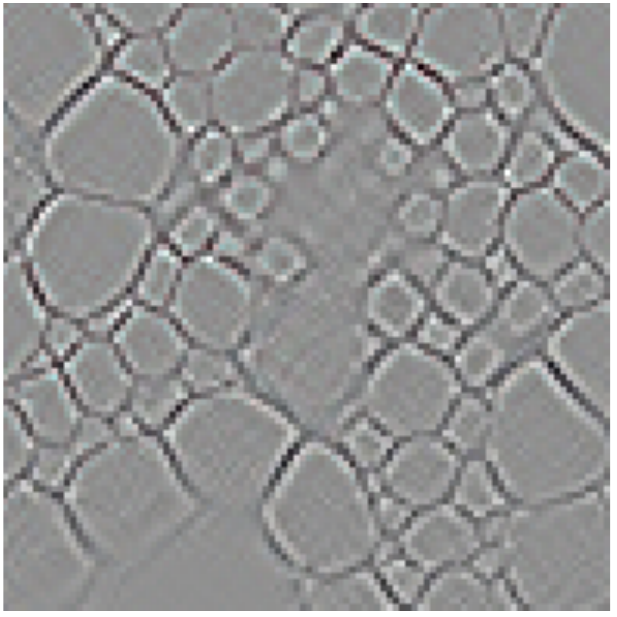}}
}\vspace{-.05in}
\caption{Deconvolving a blurred image using the inverse kernel $D$ with fullband design $\omega_c=\pi$ and significance level $\gamma=1$. The purpose of this experiment is to perform a sanity check for full image recovery assuming that no noise is added, i.e. $\eta=0$. The objective quality of blurred image is SSIM$=0.7247$, PSNR=$24.09$, where the reconstructed image is a one-to-one match i.e. SSIM$=1$, PSNR$=\infty$. The same blur model as that shown in Figure \ref{fig_inverse_deblurring_kernel_design} is chosen.}
\label{fig_Starfish_deconvolution_POC}
\end{figure}

\begin{figure}[htp]
\centerline{
\subfigure[Red channel]{\includegraphics[width=0.15\textwidth]{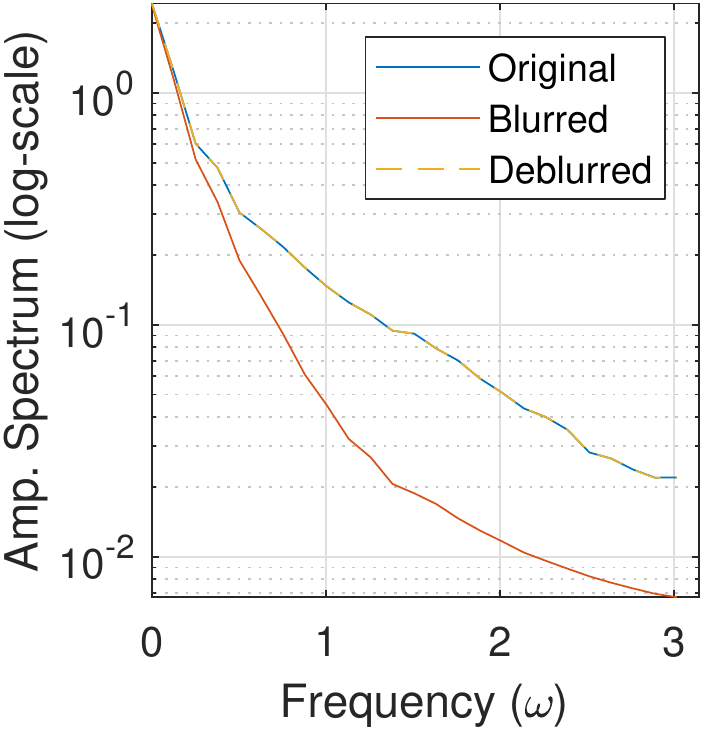}}\hspace{-.02in}
\subfigure[Green channel]{\includegraphics[width=0.15\textwidth]{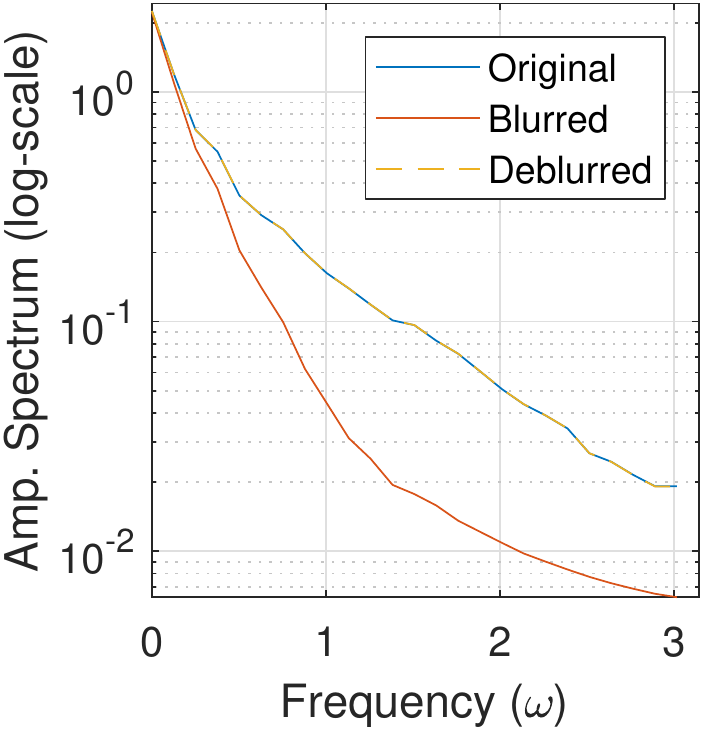}}\hspace{-.02in}
\subfigure[Blue channel]{\includegraphics[width=0.15\textwidth]{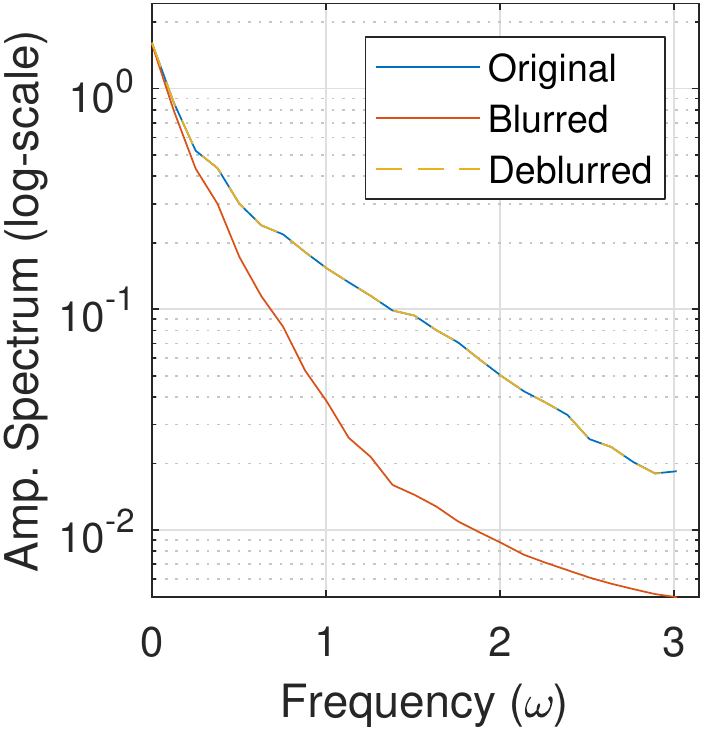}}
}\vspace{-.05in}
\caption{Amplitude spectrum of original, blurred, and deconvolved images shown for all three color channels. The spectrum of the deblurred image is very close to the original spectrum.}
\label{fig_Starfish_amplitude_spectrum_POC}
\end{figure}

We demonstrate a proof of concept in Figure \ref{fig_Starfish_deconvolution_POC} to deconvolve a starfish image under severe blurring conditions. The visual appearance of the blurred image makes it almost impossible to detect fine edges compared to its reference frame. The result of deblurring is also shown in the same figure where the recovery is a one-to-one match with its reference frame. This particular experiment validates a perfect recovery under the linear assumptions in (\ref{eq_2D_psf_model_3}) for deblurring. In Figure \ref{fig_Starfish_amplitude_spectrum_POC}, we also demonstrate the spectral responses of blurred and deconvolved images compared with their references in all three color channels. As shown, the deblurring model is capable of recovering a perfect spectral range of different frequency bands. To calculate the spectral response of each image channel, we integrate the amplitude spectrum of Fourier transform of the image in a subband frequency along a circular ring, shown in Figure \ref{fig_circular_ring}.

\subsection{Adaptive level tuning}
Here we define an adaptive measure to tune the deblurring significance level $\gamma$ by calculating the relative ratio of two image entropy
\begin{align}
\gamma \triangleq \frac{E\left(f_{B}\right)}{E\left(\nabla_{D}f_{B}\right)+T},
\label{eq_2D_psf_model_4}
\end{align}
where $E(I) = -\sum\limits_{k\in\Omega} p(k)\log{p(k)}$ is the entropy of the input image and $p(k)$ is the histogram count for gray level $k$. The threshold level `$T$' is defined here to avoid the singularity that could be caused by sparse deblurring edges. The entropy calculates the histogram dispersion (a.k.a average rate) of the image. The dispersion ratio in (\ref{eq_2D_psf_model_4}) defines the relative measure for proper adjustment of the blur image with respect to its deblurring edges.

\begin{figure}[htp]
\centerline{
\subfigure[Raw Image (unprocessed)]{\includegraphics[width=0.25\textwidth]{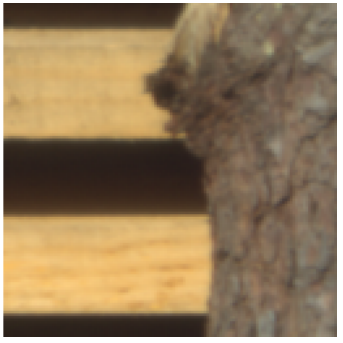}}\hspace{-.04in}
\subfigure[Deblurred by (\ref{eq_2D_psf_model_3})]{\includegraphics[width=0.25\textwidth]{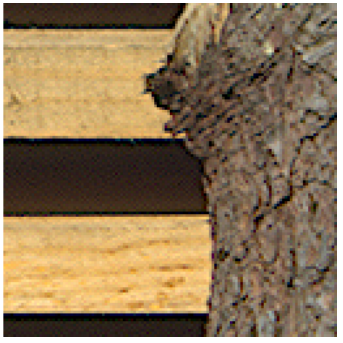}}
}\vspace{-.05in}
\centerline{
\subfigure[$f_B\ast D_x$]{\includegraphics[width=0.165\textwidth]{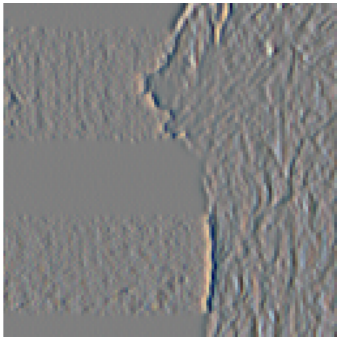}}\hspace{-.02in}
\subfigure[$f_B\ast D_y$]{\includegraphics[width=0.165\textwidth]{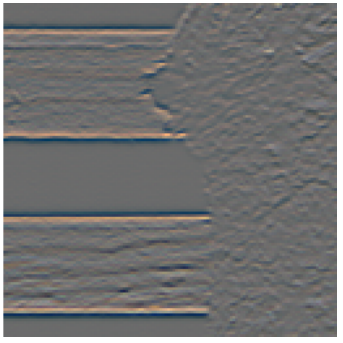}}\hspace{-.02in}
\subfigure[$f_B\ast D_{xy}$]{\includegraphics[width=0.165\textwidth]{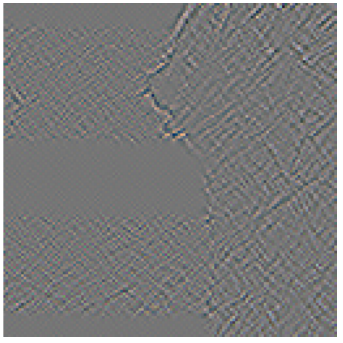}}
}\vspace{-.05in}
\caption{Deblurred image of naturally blurred hyper-spectral image (rendered RGB). The PSF is modeled by a generalized Gaussian distribution with shape $\beta=1.8$ and scale $2.64$. For better visual comparison, please turn off image-smoothing option from Adobe Acrobat view software.}
\label{fig_Spectral_Image_Delurring_POC}
\end{figure}

Figure \ref{fig_Spectral_Image_Delurring_POC} shows an example of deblurred image of naturally blurred hyper-spectral image (renfered RGB) using the proposed correction in (\ref{eq_2D_psf_model_3}). The PSF here is modeled by a generalized Gaussian blur (shape $\beta=1.8$ and scale $2.64$) introduced in next section. 

\section{Blur Modeling and Estimation}\label{sec_blur_model}
In this section, we model the PSF blur kernel in natural imaging applications by the generalized Gaussian (GG) distribution. We study two model shapes of Gaussian and Laplacian distributions as particular cases of GG for blind estimation.

\subsection{Modeling blur by Generalized Gaussian (GG)}
\begin{figure}[htp]
\centerline{
\subfigure[$h_{GG}(x), \beta=1$]{\includegraphics[width=0.15\textwidth]{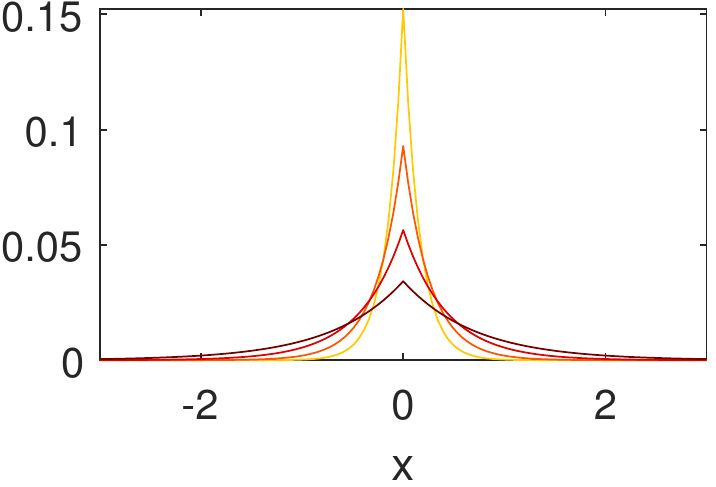}}
\subfigure[$h_{GG}(x), \beta=1.5$]{\includegraphics[width=0.15\textwidth]{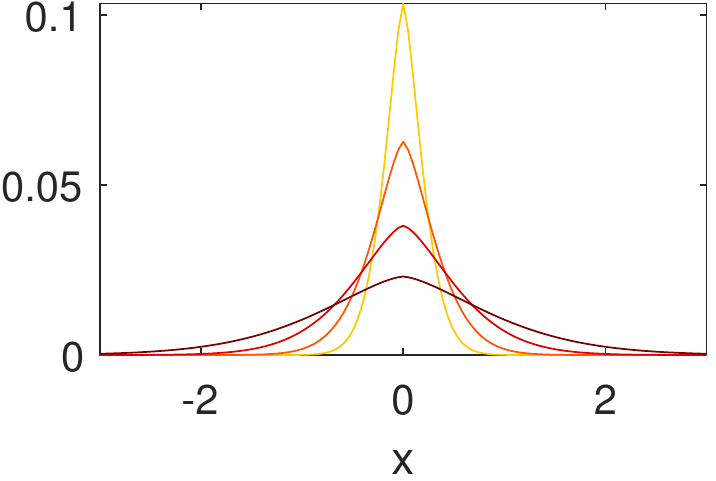}}
\subfigure[$h_{GG}(x), \beta=2$]{\includegraphics[width=0.15\textwidth]{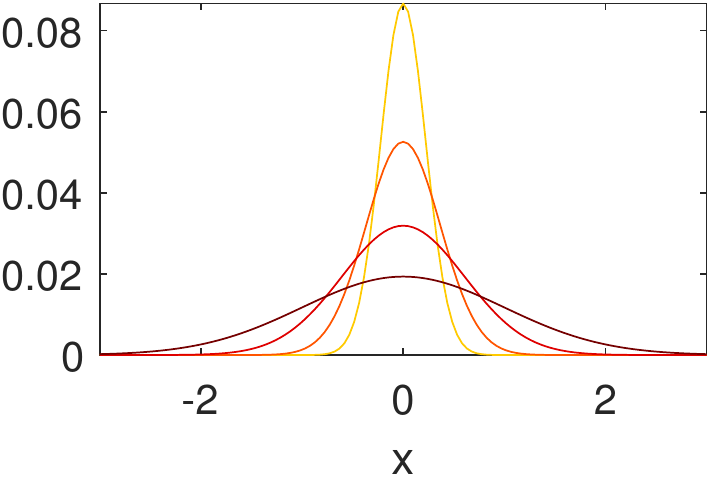}}
}\vspace{-.05in}
\centerline{
\subfigure[$|\hat{h}_{GG}(\omega)|$]{\includegraphics[width=0.145\textwidth]{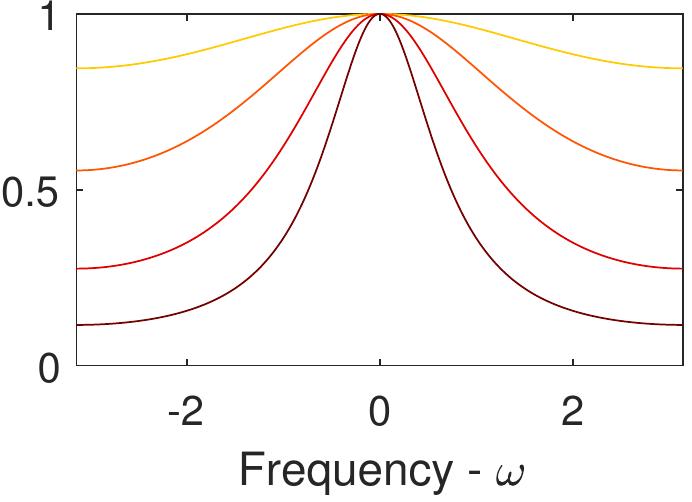}}
\subfigure[$|\hat{h}_{GG}(\omega)|$]{\includegraphics[width=0.145\textwidth]{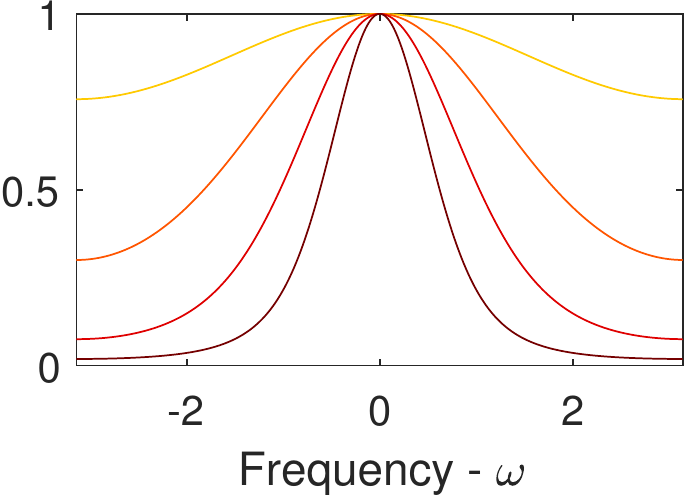}}
\subfigure[$|\hat{h}_{GG}(\omega)|$]{\includegraphics[width=0.145\textwidth]{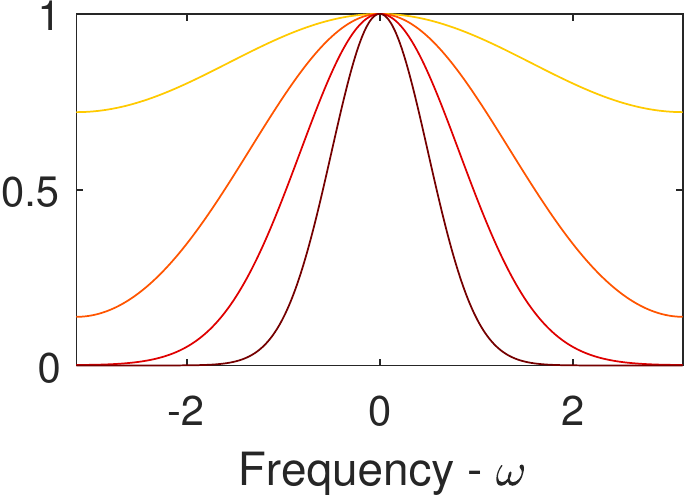}}
}\vspace{-.05in}
\centerline{
\subfigure[$|\hat{h}_{GG}(\omega)|^{-1}$]{\includegraphics[width=0.14\textwidth]{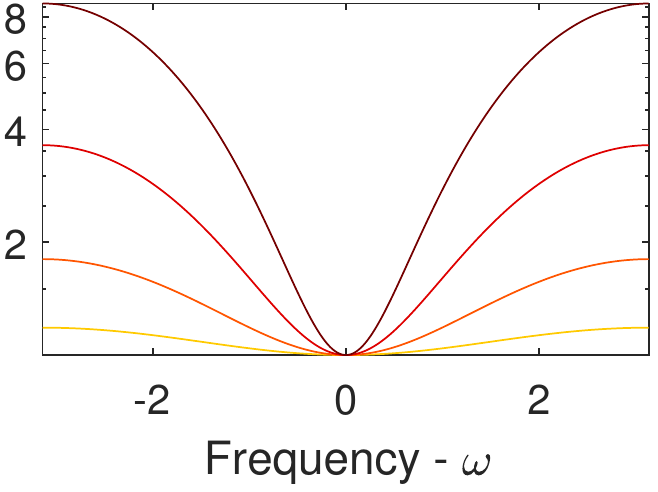}}
\subfigure[$|\hat{h}_{GG}(\omega)|^{-1}$]{\includegraphics[width=0.14\textwidth]{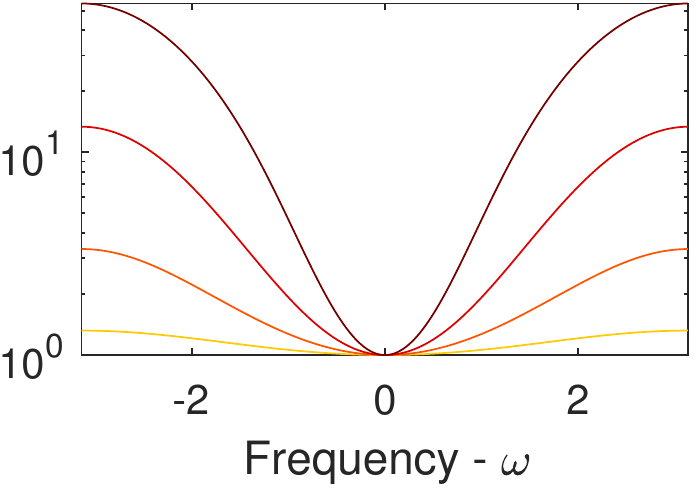}}
\subfigure[$|\hat{h}_{GG}(\omega)|^{-1}$]{\includegraphics[width=0.14\textwidth]{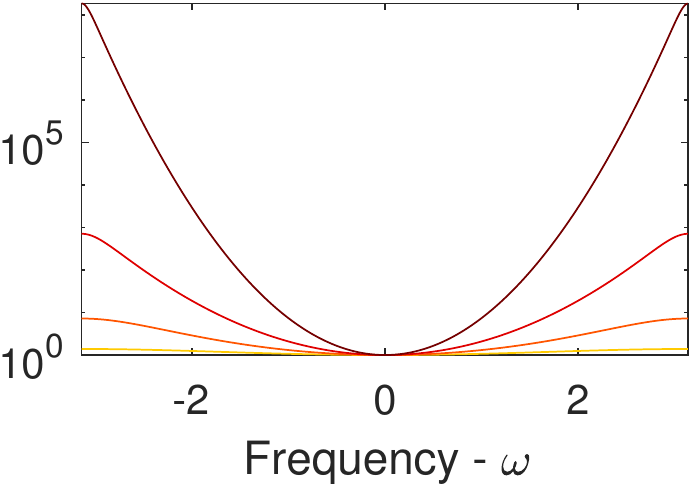}}
}\vspace{-.05in}
\caption{Generalized Gaussian kernel considered for blurring model. The kernels are generated with different shapes $\beta$ and scales $\alpha$. The range of selected scales here is $\alpha=\exp(-0.75, -0.5, \hdots,:0.75)$ and correspond to the transition of color shades from dark to yellow, respectively. The impulse responses are shown in the first row, amplitude spectrum in second row, and the inverse amplitude spectrum in the third row.}
\label{fig_generalized_Gaussian_blur}
\end{figure}

The generalized Gaussian (GG) distribution was introduced by Subbotin \cite{subbotin1923law} to revise the power law of Gauss's distribution into the more generalized sense
\begin{align}\label{blur_model_eq_1}
h_{GG}(x) = \frac{1}{2\Gamma(1+1/\beta)A(\beta,\sigma)}\exp{-\Big{\vert}\frac{x}{A(\beta,\sigma)}\Big{\vert}^\beta},
\end{align}
where $\beta$ defines the \textit{shape} of the distribution function, $A(\beta,\sigma) = \left(\sigma^2\Gamma(1/\beta)/\Gamma(3/\beta)\right)^{1/2}$ is the scaling parameter, and $\Gamma(\cdot)$ is the Gamma function $\Gamma(z) = \int^{\infty}_{0}e^{-t}t^{z-1}dt, \forall z>0$. For instance, the standard Gaussian distribution, i.e. second order model, is determined by $\beta=2$ and $A(2, \sigma)$. For more information on the distribution and how it is used in different engineering applications, please refer to \cite{kotz2012laplace} and the references therein. Figure \ref{fig_generalized_Gaussian_blur} demonstrates examples of GG blur for a variety of selected shapes and scales. The \textit{shape} and the \textit{scale} of the distribution control the decay rate and energy concentration of the distribution, respectively. The amplitude spectra of the kernels are also shown in the same figure (second row). The spectral responses of the blur kernels are inversely related to their scales, where low scales maintain wider frequencies for transformation.

\begin{figure}[htp]
\centerline{
\subfigure[Feasibility Range]{\includegraphics[width=0.225\textwidth]{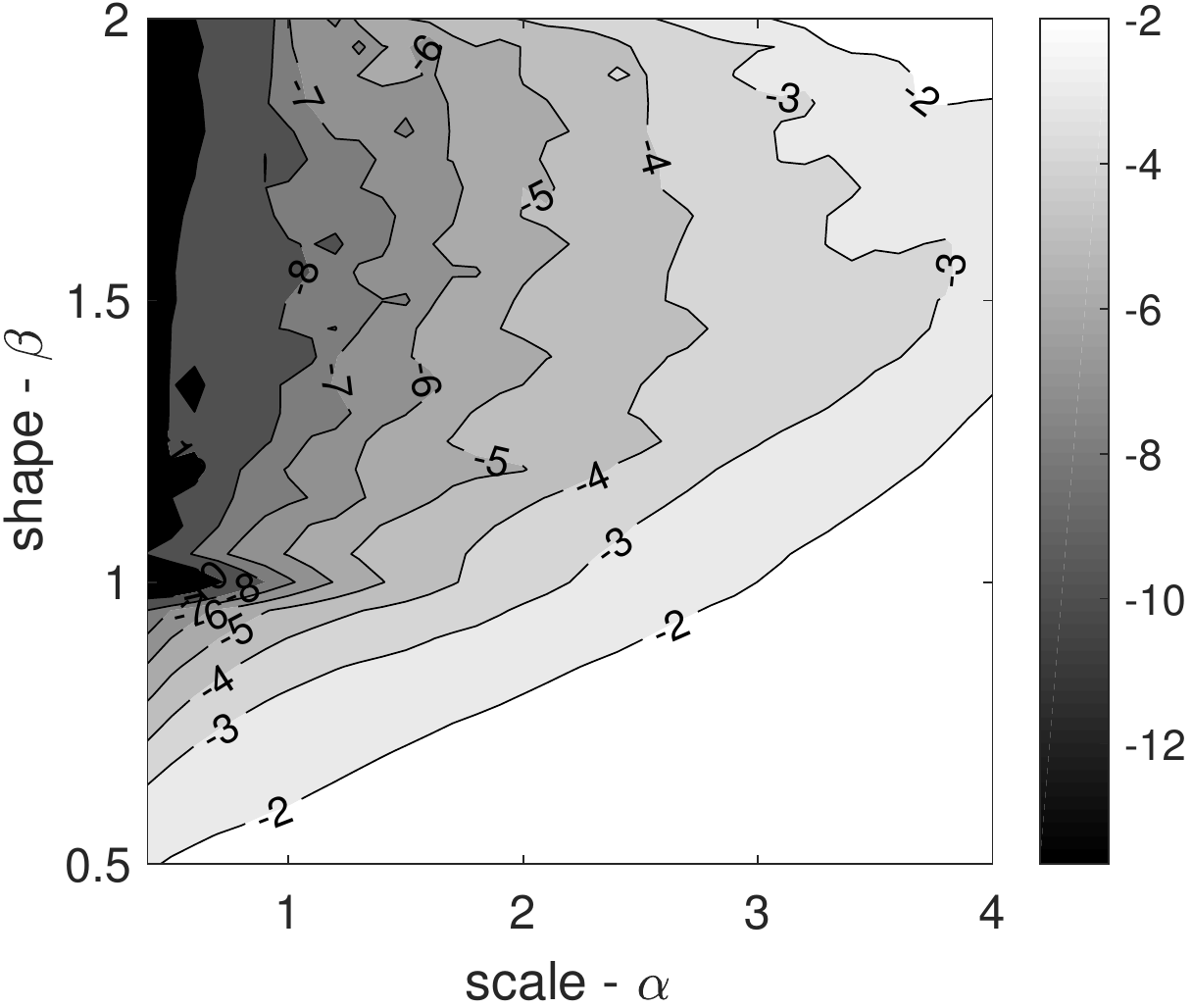}\label{fig_GG_feasibility_range}}
\subfigure[Radial Spectrum]{\includegraphics[width=0.225\textwidth]{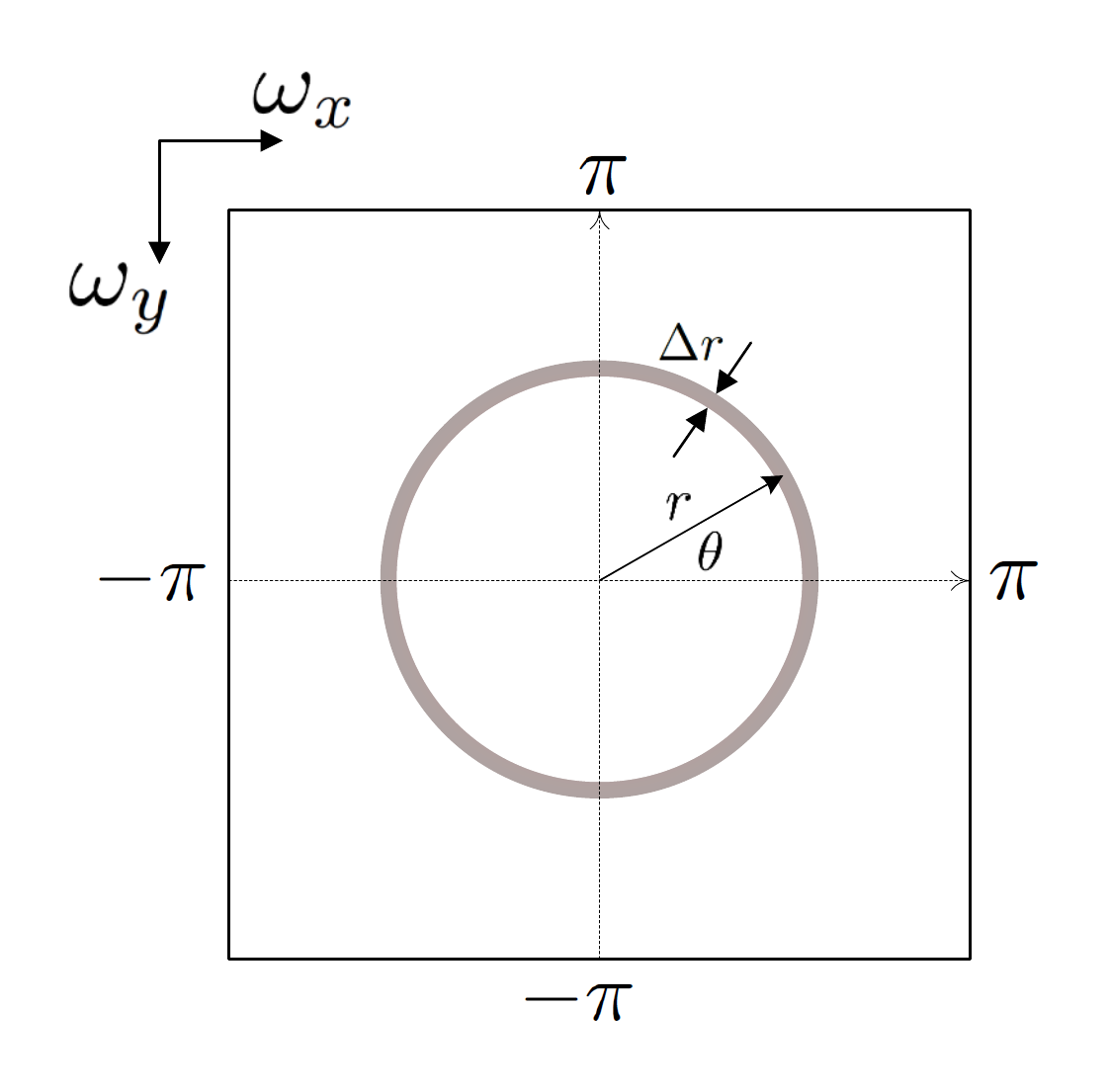}\label{fig_circular_ring}}
}\vspace{-.05in}
\caption{(a) Error plot between ideal inverse blur response $|\hat{h}(\omega)|^{-1}$ and approximated inverse deconvolution kernel response $\hat{D}(\omega)$ using different shapes and scale parameters of the GG blur model. The error map is shown in $\log_{10}$ scale where lower values indicate high accuracy approximation. (b) Two dimensional (2D) Fourier transform domain quantized into radial bins to obtain one dimensional (1D) radial spectrum}
\end{figure}

The GG model is used in several imaging applications to model static blur in natural imaging, such as atmospheric turbulence and optical aberrations \cite{hufnagel1964modulation, savakis1993blur, elder1998local, metari2007new, stallinga2010accuracy, levin2011understanding}.  A common approach is to employ such kernels in a non-blind fashion for image deconvolution. The shape and the scale are the two different characteristics that fit different blur applications. For instance, in weather-conditioned environments, the shape of atmospheric turbulence in haze imaging is close to $\beta\approx 1.5$ \cite{metari2007new}. One of our goals in this section is to study the range of feasibility for designing deconvolution kernels, introduced in Section \ref{section_inverse_deconvolution_design}, using different GG types. Figure \ref{fig_GG_feasibility_range} demonstrates the error of fitting GG blur with different contour levels identifying the error between the approximated inverse kernel and the ideal inverse response i.e. $\||\hat{h}(\omega)|^{-1} - \hat{D}(\omega)\|/\||\hat{h}(\omega)|^{-1}\|$. Examples of inverse responses are shown in Figure \ref{fig_generalized_Gaussian_blur} (third row). The shades of gray in Figure \ref{fig_GG_feasibility_range} show that a wide range of selected kernels with different shapes and scales can be associated to approximate their inverse response using MaxPol kernels.

\subsection{Blind PSF estimation}
In this section we introduce our blind approach to estimate the blur level of the PSF kernel from naturally blurred images. Our approach relies on image scale-space analysis using two different scales that are the originally sampled image $f_B(x, y)$ and its down-sampled version $f_B(s x, s y)$ for $s>1$. This scale is reversed in the Fourier domain, i.e. $\hat{f}_{B}(\omega_x/s, \omega_y/s)$, where we transfer the coordinates from Cartesian to polar $(\omega_x, \omega_y)\mapsto(r,\theta)$ to obtain $\hat{f}_{B}(r/s, \theta)$. We integrate the blur image along a closed circle to calculate its radial spectrum and expand the terms using the linear convolution model in (\ref{eq_psf_model_1}) which gives
\begin{align}\label{blur_model_eq_2}
\int^{2\pi}_0{\hat{f}_{B}(r/s,\theta)d\theta}=\int^{2\pi}_0{\hat{f}_{L}(r/s, \theta)\hat{h}(r/s, \theta)+\hat{\eta}d\theta}
\end{align}
The integral in (\ref{blur_model_eq_2}) calculates the radial spectrum of the input image along the radial ring. Natural images (without blur) usually follow a decay spectrum of $\hat{f}_{L}\approx 1/r$ \cite{field1987relations, brady1995s, field1997visual}. Furthermore, we consider the noise contamination to be additive white Gaussian noise (AWGN). Substituting these assumptions into the expansion in (\ref{blur_model_eq_2}) gives
\begin{align}\label{blur_model_eq_3}
\int^{2\pi}_0{\hat{f}_{B}(r/s,\theta)d\theta} \approx 
\int^{2\pi}_0{\frac{s}{r}\hat{h}(r/s, \theta)d\theta} + c,
\end{align}
where $c\propto SNR^{-1}$ is proportional to the inverse signal-to-noise-ratio (SNR) level of the sample measurements. For good quality images, this coefficient is negligible ($c\rightarrow 0$). Next, we define a ratio spectrum of two different scales of original and subsampled domains:
\begin{align}\label{blur_model_eq_4}
R(r) \equiv \frac{\int^{2\pi}_0{\hat{f}_{B}(r,\theta)d\theta}}{\int^{2\pi}_0{\hat{f}_{B}(r/s,\theta)d\theta}} \approx
\frac{\int^{2\pi}_0{\hat{h}(r, \theta)d\theta} + cr}{s\int^{2\pi}_0{\hat{h}(r/s, \theta)d\theta} + cr}
\end{align}
The right-hand-side (RHS) of equation (\ref{blur_model_eq_4}) is used to fit a certain blur model as a prior knowledge of the equation, where the data fidelity term is provided by approximating the ratio $R(r)$ using two radial spectra of different image scales. To calculate the radial spectrum, we define a radial ring with fixed bin size $\Delta r$ shown in Figure \ref{fig_circular_ring} and integrate the spectrum along the selected ring. The bin size determines the quantized level of the radial spectrum.

The next two subsections study two different cases of generalized Gaussian for blur scale estimation: Gaussian ($\beta=2$) and Laplacian ($\beta=1$).

\subsubsection{Gaussian blur kernel}
The 2D Gaussian blur kernel is defined by $h(x,y) = \frac{1}{2\pi \alpha^2}\exp{-(x^2+y^2})/2\alpha^2$ and its 2D Fourier transform is $\hat{h}(\omega_x,\omega_y)=\exp{-\alpha^2/2(\omega^2_x+\omega^2_y)}$. Converting the domain from Cartesian to polar, we obtain a rotation-invariant kernel $\hat{h}(r)=\exp{-\alpha^2 r^2/2}$ and substituting this into ratio spectrum (\ref{blur_model_eq_4}) simplifies to
\begin{align}\label{blur_model_eq_5}
R(r) \approx \frac{\exp{-\alpha^2 r^2/2} + c^{\prime}r}{s\exp{-\alpha^2 r^2/2s^2} + c^{\prime}r}.
\end{align}
The discrete measurements of the ratio $R(r)$ in (\ref{blur_model_eq_5}) are also calculated by the ratio between the radial spectrum of the original image and its downsampled image with scale $s>1$. 

\subsubsection{Laplacian blur kernel}
The 2D Laplacian blur kernel in the separable mode is defined by $h(x,y) = 1/{2\alpha^2}\exp{-\sqrt{2}/\alpha(|x|+|y|)}$ and its 2D Fourier transform is $\hat{h}(\omega_x,\omega_y)=4/(2+\alpha^2\omega^2_x)(2+\alpha^2\omega^2_y)$ \cite{dubeau2011fourier}. Converting from Cartesian to polar coordinates yields $\hat{h}(r,\theta)=4/(4+2\alpha^2 r^2 + \alpha^4 r^4 \cos^2(\theta)\sin^2(\theta))$, which has a rotationally-dependent spectrum, unlike the Gaussian in the previous section. Therefore, we need to calculate the radial spectrum of the blur image defined in (\ref{blur_model_eq_3}) for this particular case. First, we revise the integral in (\ref{blur_model_eq_3}) by
\begin{align}\label{blur_model_eq_6}
\int^{2\pi}_0{\hat{f}_{B}(r/s,\theta)d\theta} \approx 
\int^{2\pi}_0{\frac{A}{B+\sin^2(\theta)}d\theta} + c
\end{align}
where
\begin{align}\label{blur_model_eq_7}
A = 16s^5/\alpha^4 r^5~~\text{and}~~B=(16s^4+8\alpha^2 r^2s^2)/\alpha^4 r^4.
\end{align}
The term in (\ref{blur_model_eq_6}) is a definite integral and it can be identified as a line integral by change of variable $z=\exp(i\theta)$ and substituting this into the Euler formula, we have $\sin(\theta)=(z-z^{-1})/2i$. By plugging this into the integral in (\ref{blur_model_eq_6}), we have
\begin{align}\label{blur_model_eq_8}
\int^{2\pi}_0{\frac{A}{B+\sin^2(\theta)}d\theta} = 
\int_{C}{\frac{-4Azdz}{i[z^4-(4B+2)z^2+1]}},
\end{align}
where the integral is applied around a closed unit circle. By taking another change of variable $u=z^2$, the integral in (\ref{blur_model_eq_8}) simplifies to
\begin{align}\label{blur_model_eq_9}
\int^{2\pi}_0{\frac{A}{B+\sin^2(\theta)}d\theta} = 
4iA\int_{C}{\frac{du}{u^2-(4B+2)u+1}}.
\end{align}
We find the poles inside the unit circle from the denominator and apply the Residue theorem to calculate the integral value. The roots of the denominator are
\begin{align}\label{blur_model_eq_10}
\left\{
\begin{array}{l}
r_1 = 2B+1+2\sqrt{B^2+1} \\
r_1 = 2B+1-2\sqrt{B^2+1}
\end{array}
\right.,
\end{align}
where the root $r_1$ is outside the unit circle and does not apply. The second root $r_2<1$ for any $B$ and hence the residue of the integral (\ref{blur_model_eq_9}) at $r_2$ can be computed by
\begin{align}\label{blur_model_eq_11}
\Res\limits_{r_2} f(u) = \lim\limits_{u\rightarrow r_2} \frac{u-r_2}{(u-r_2)(u-r_1)}=\frac{-1}{4\sqrt{B^2+1}}.
\end{align}
Substituting (\ref{blur_model_eq_11}) into the integral in (\ref{blur_model_eq_9}) gives 
\begin{align}\label{blur_model_eq_12}
4iA\int_{C} = 4iA\left[2\pi i\sum\Res\right] = \frac{2\pi A}{\sqrt{B^2+1}}.
\end{align}
Finally, the radial spectrum in (\ref{blur_model_eq_6}) is obtained by 
\begin{align}\label{blur_model_eq_13}
\int^{2\pi}_0{\hat{f}_{B}(r/s,\theta)d\theta} \approx 
\frac{2\pi A}{\sqrt{B^2+1}} + c
\end{align}

The radial spectrum in (\ref{blur_model_eq_13}) can now be used to approximate the ratio spectrum $R(r)$ defined in (\ref{blur_model_eq_4}).

\begin{figure}[htp]
\centerline{
\setbox1=\hbox{\includegraphics[height=2.7cm]{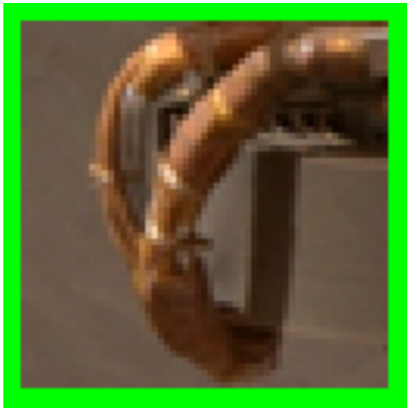}}
\subfigure[Original]{\includegraphics[height=2.7cm]{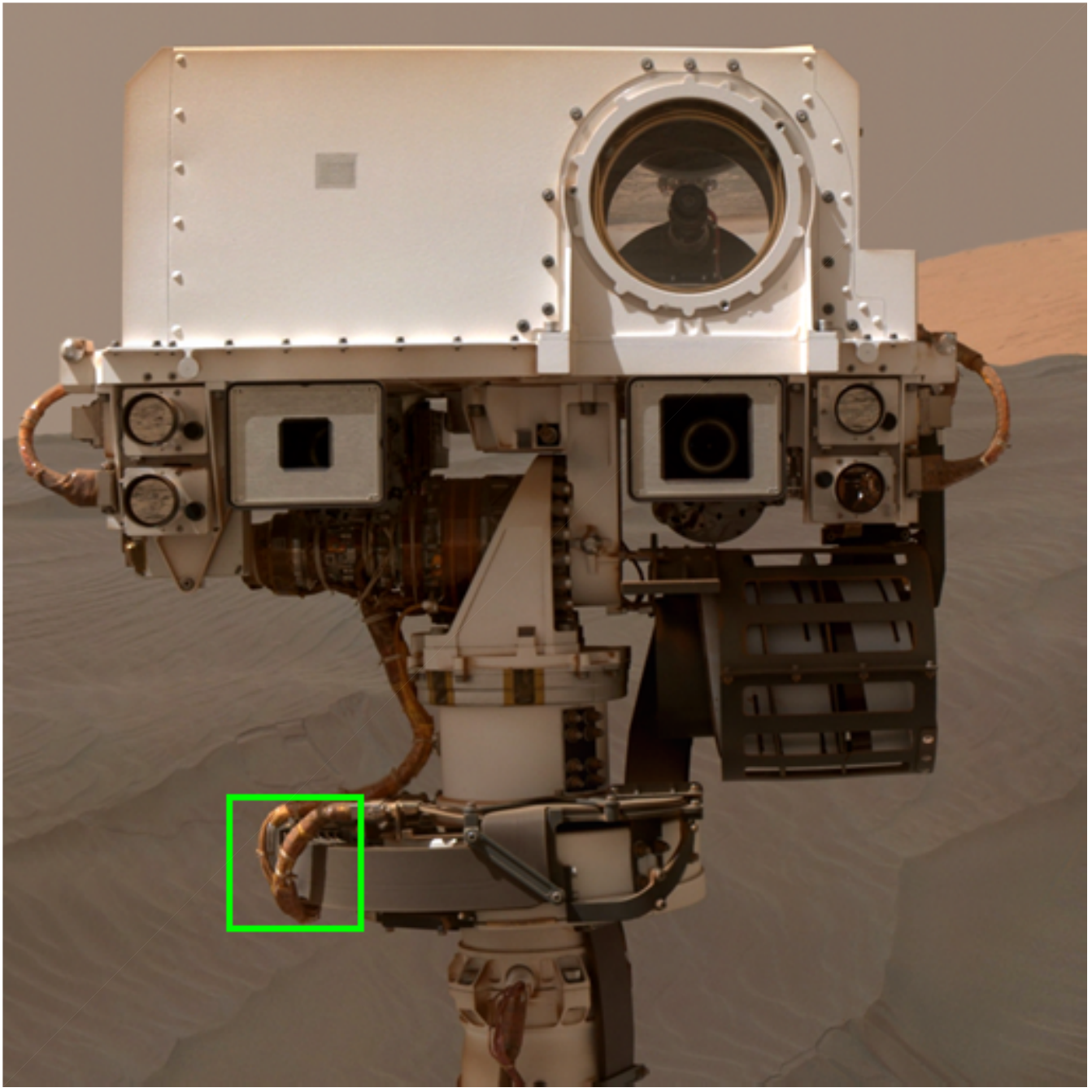}\llap{\makebox[\wd1][r]{\raisebox{0cm}{\includegraphics[height=1.25cm]{image_scan_patch_POC_Gaussian_blur_estimation}}}}\label{fig_image_scan_POC_Gaussian_blur_estimation}}
\subfigure[Blurred]{\includegraphics[height=2.7cm]{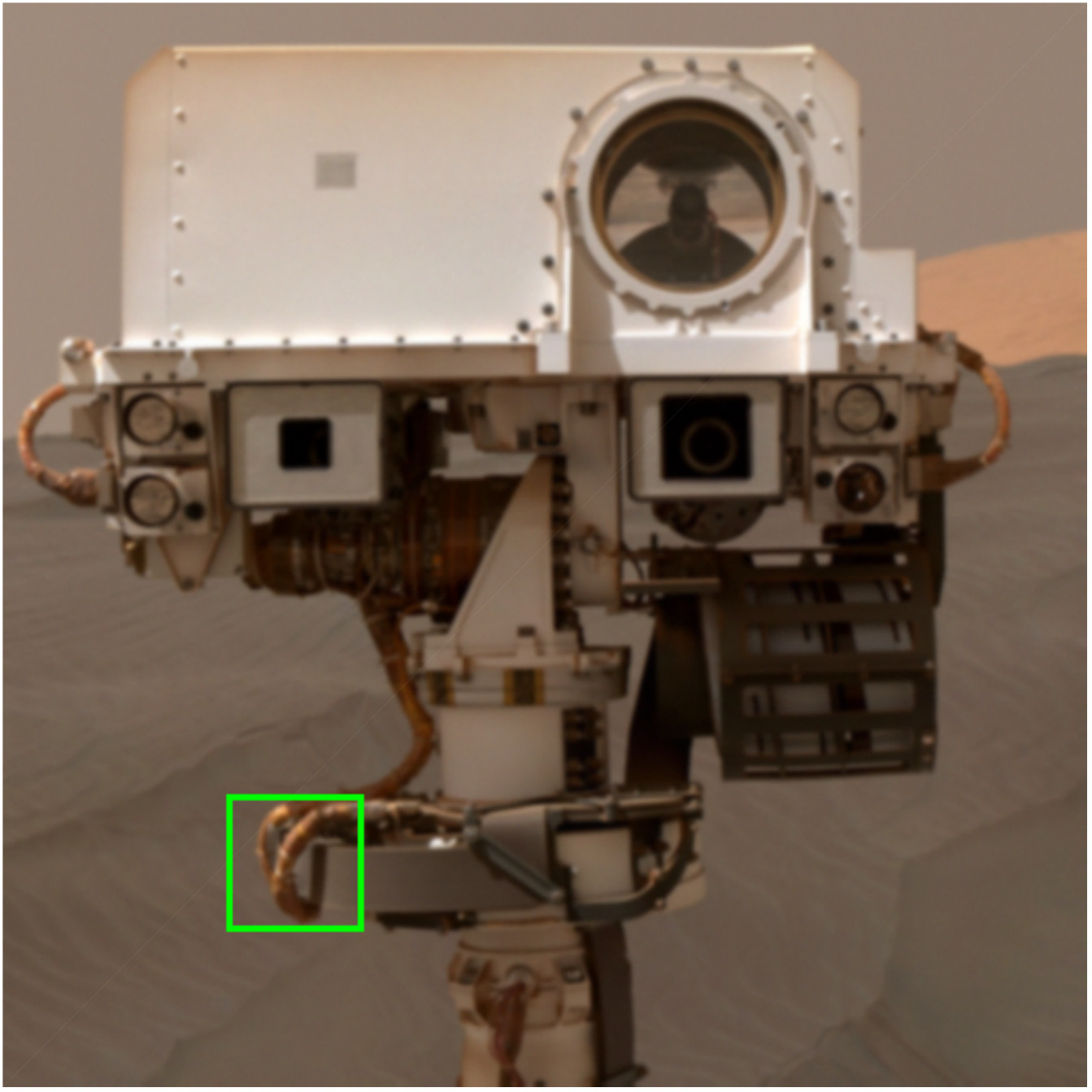}\llap{\makebox[\wd1][r]{\raisebox{0cm}{\includegraphics[height=1.25cm]{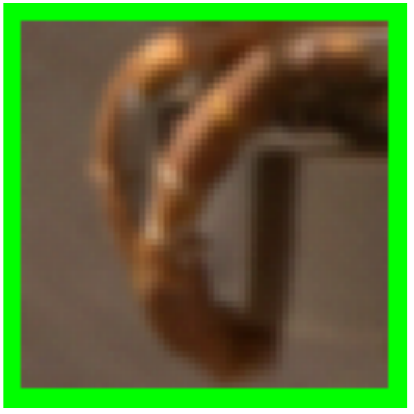}}}}\label{fig_image_blurry_POC_Gaussian_blur_estimation}}
\subfigure[Scaled]{\includegraphics[height=2.7cm]{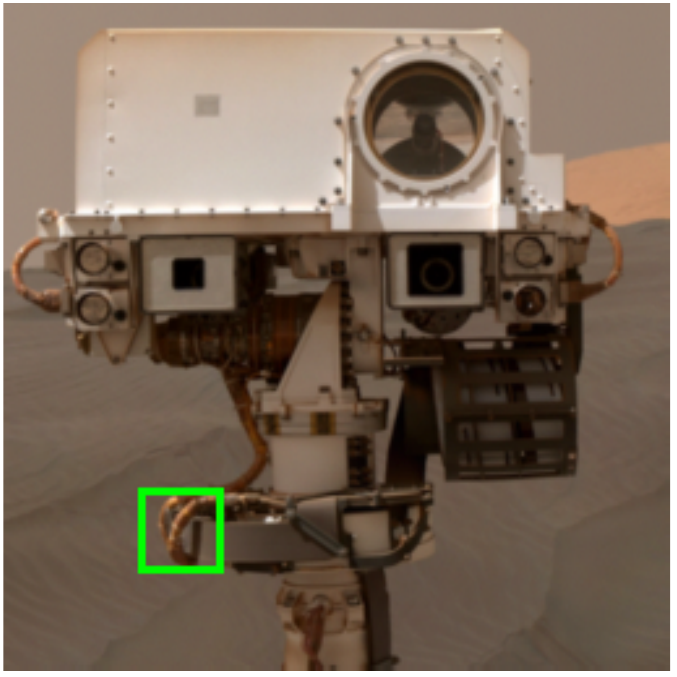}\llap{\makebox[\wd1][r]{\raisebox{0cm}{\includegraphics[height=1.25cm]{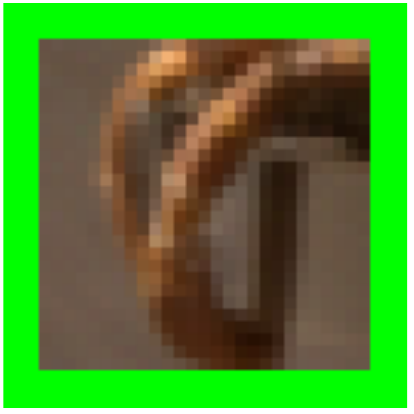}}}}\label{fig_image_scaled_POC_Gaussian_blur_estimation}}
}\vspace{-.05in}
\centerline{
\subfigure[Red channel]{\includegraphics[width=0.15\textwidth]{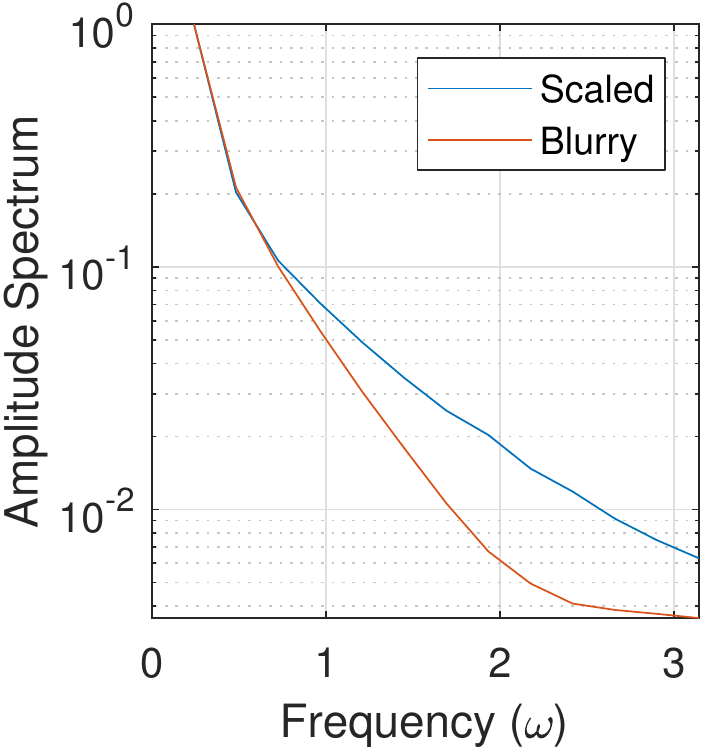}\label{fig_Radial_Spectrum_Red_POC_Gaussian}}\hspace{-.02in}
\subfigure[Green channel]{\includegraphics[width=0.15\textwidth]{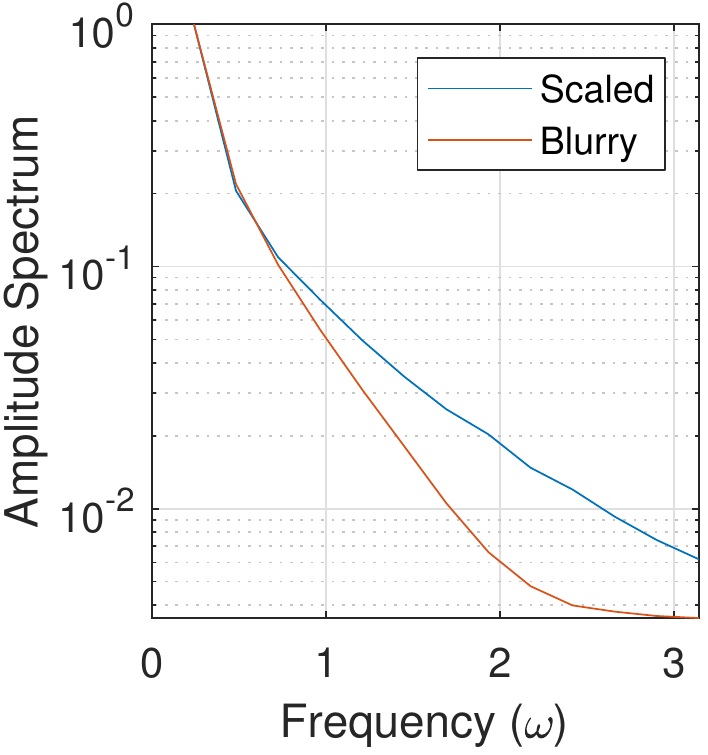}\label{fig_Radial_Spectrum_Green_POC_Gaussian}}\hspace{-.02in}
\subfigure[Blue channel]{\includegraphics[width=0.15\textwidth]{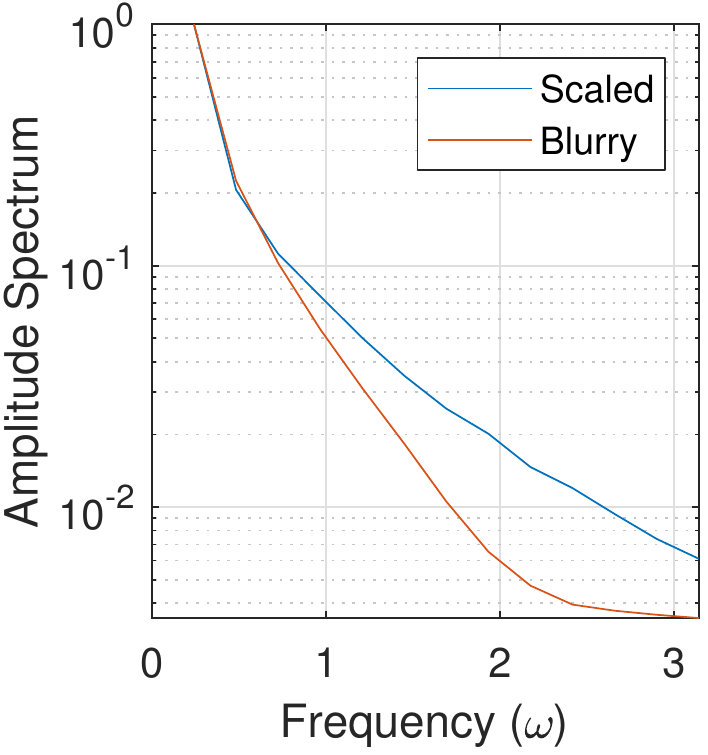}\label{fig_Radial_Spectrum_Blue_POC_Gaussian}}
}\vspace{-.05in}
\centerline{
\subfigure[Red:   $\bar{\alpha}=1.04$]{\includegraphics[width=0.15\textwidth]{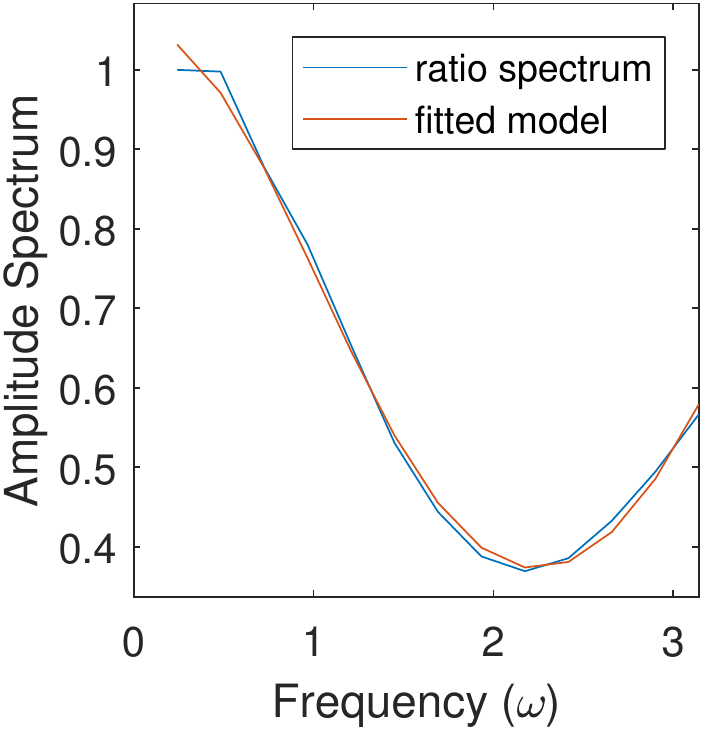}}\hspace{-.02in}
\subfigure[Green: $\bar{\alpha}=1.05$]{\includegraphics[width=0.15\textwidth]{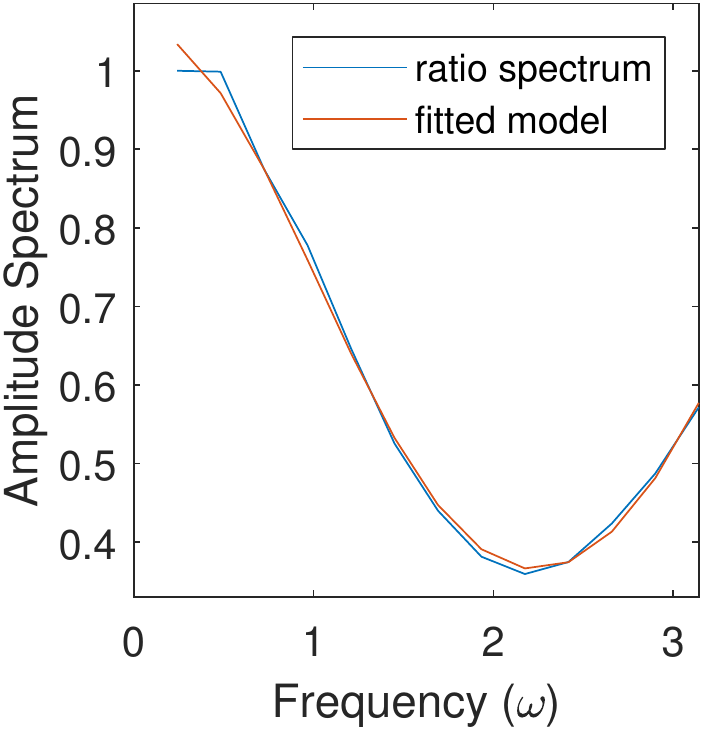}}\hspace{-.02in}
\subfigure[Blue:  $\bar{\alpha}=1.05$]{\includegraphics[width=0.15\textwidth]{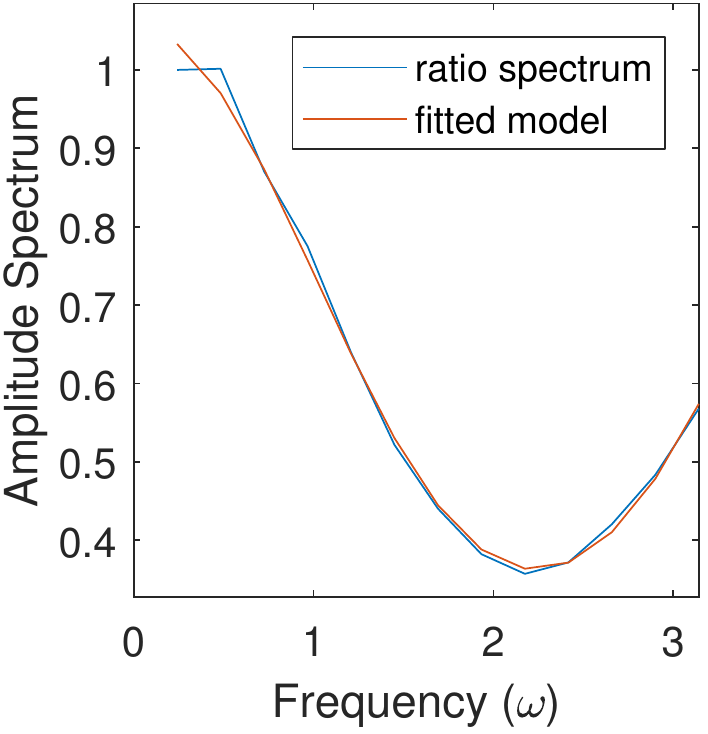}}
}\vspace{-.05in}
\caption{Demo example of blind Gaussian blur estimation. The original image is blurred with $\alpha=1$ and white Gaussian noise with $\sigma=1/255$ is added. The radial spectra of both blurred and scaled images are shown in (e)-(f) for all color channels. The ratio spectrum $R(r)$ is also estimated and fitted with the model in (\ref{blur_model_eq_5}) for all color channels as shown in (g)-(i)}
\label{fig_demo_Gaussian_blur_estimation}
\end{figure}

\subsection{Synthetic validation}
In both blur models (Gaussian/Laplacian), the unknown parameters scale $\alpha$ and noise level $c^{\prime}$ are obtained by solving a linear least square problem from \cite{kelley1999iterative}. Figure \ref{fig_demo_Gaussian_blur_estimation} demonstrates an example for blur parameter estimation. It is worth noting that we employ a good in-focus image for experimenting and synthetically blur it to estimate its scale for validation. If the original image is naturally blurred, the final blur will be the combination of natural and synthetic blur. The original image is shown in Figure \ref{fig_image_scan_POC_Gaussian_blur_estimation} and is blurred with a Gaussian kernel of scale $\alpha=1$ and perturbed by AWGN noise with standard deviation $\sigma=1/255$ in unit scale shown in Figure \ref{fig_image_blurry_POC_Gaussian_blur_estimation}. The blurred image is downsampled by a factor of $s=2$, as shown in Figure \ref{fig_image_scaled_POC_Gaussian_blur_estimation}. We employ MaxPol kernel of $0$th-order derivative with tap-length $l=16$ and cut-off parameter $P=24$ for downsampling to preserve most of the frequency spectrum. The radial spectra of both images are calculated and shown in Figure \ref{fig_Radial_Spectrum_Red_POC_Gaussian}-\ref{fig_Radial_Spectrum_Blue_POC_Gaussian}. The radial spectrum is fitted to the model in (\ref{blur_model_eq_5}) and both unknown parameters $\bar{\alpha}$ and $\bar{c^{\prime}}$ are approximated. It worth noting the radial spectrum defined for both the Gaussian and Laplacian yield valid estimations through blur assessment. These models will be used in the experiment section to truly validate the blur levels of natural images and design their corresponding inverse kernels for deblurring.

\section{Experiments}\label{sec_experiments}
We evaluate the proposed 1shot-MaxPol\footnote{Source code available from \url{https://github.com/mahdihosseini/1Shot-MaxPol}} deblurring by conducting experiments in terms of reconstruction accuracy, computational complexity, and scalability for two different Gaussian and Laplacian blur models. We evaluate the reconstruction accuracy by adopting a no-reference sharpness quality assessment (NR-FQA) metric using the maximum local variation (MLV) method introduced in \cite{bahrami2014fast}, where high values indicate better focus resolution and low values indicate the opposite. The processing speed of this metric is quite fast and can be used to evaluate large image databases such as the one introduced in Section \ref{sec_natural_databases}. For comparison, we select eight non-blind image deconvolution methods, including Krishnan \cite{krishnan2009fast, krishnan2011blind}, EPLL \cite{zoran2011learning}, Chan-DeconvTV \cite{chan2011augmented}, IDD-BM3D \cite{danielyan2012bm3d}, MLP \cite{schuler2013machine}, Simoes \cite{simoes2016framework}, Chan-PlugPlay \cite{chan2017plug}, and IRCNN \cite{zhang2017learning}. For more information on the procedural steps and functionality of these methods, please refer to the Section ``Comparison Methods'' in the supplementary document of this paper. The associated PSF used in all non-blind deconvolution methods is provided by the blind estimation approach proposed in Section \ref{sec_blur_model} for each test image. Using the same PSF for all methods provides a benchmark for a fair comparison of reconstruction quality. We estimate both Gaussian and Laplacian blur models with two different scales of $s=\{2,4\}$.

\begin{figure*}[htp]
\centerline{
\subfigure[LROC]{\includegraphics[width=0.19\textwidth]{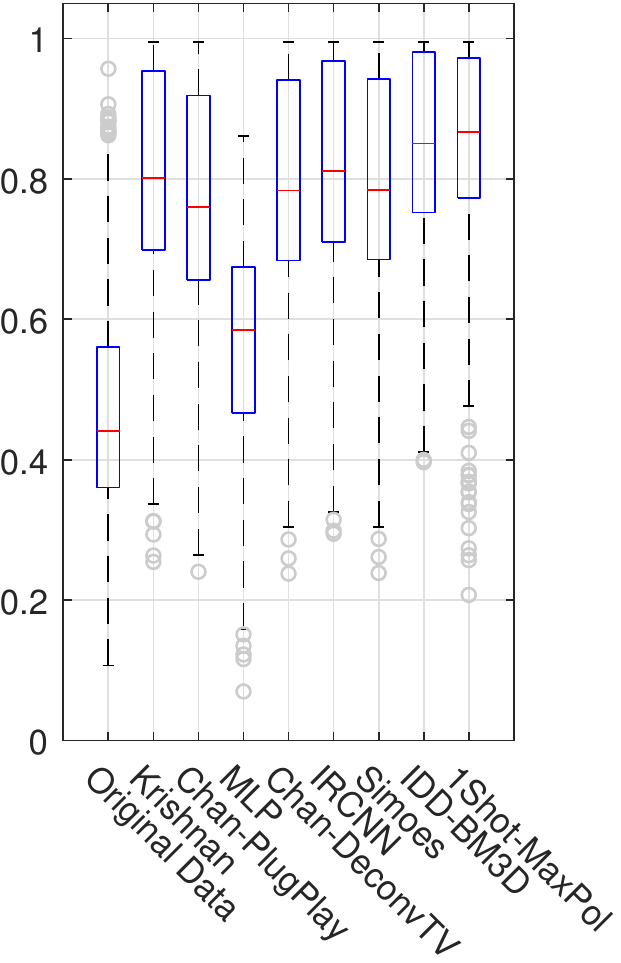}}\hspace{-.25in}
\subfigure[McMaster]{\includegraphics[width=0.19\textwidth]{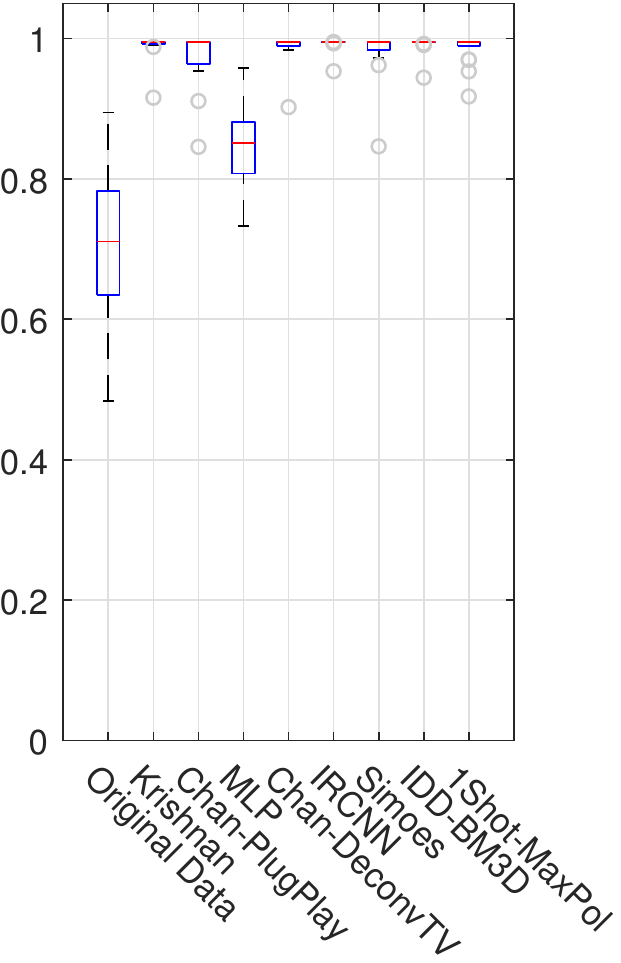}}\hspace{-.25in}
\subfigure[NIR-Scene]{\includegraphics[width=0.19\textwidth]{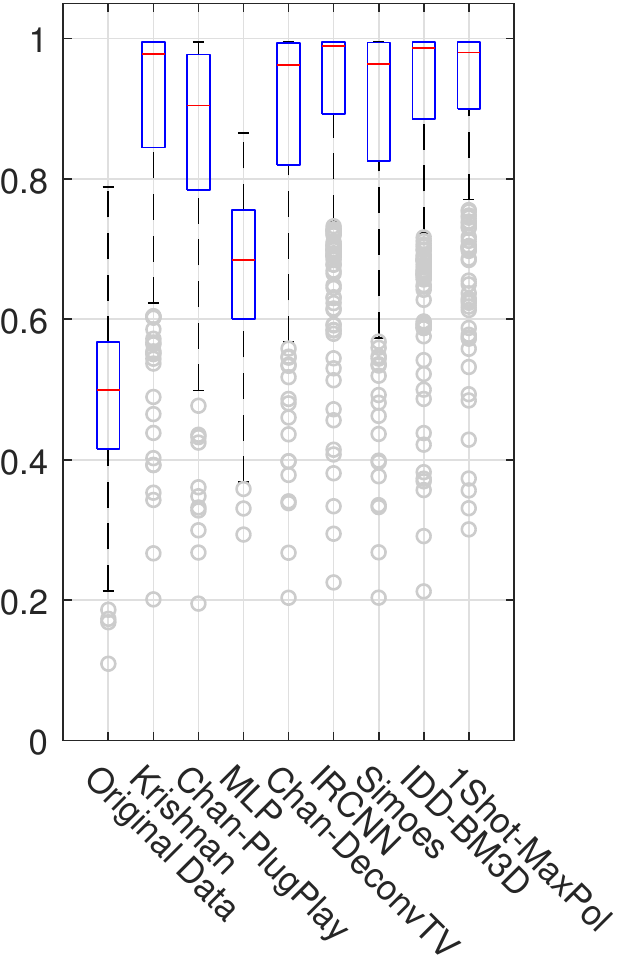}}\hspace{-.25in}
\subfigure[RGB-Scene]{\includegraphics[width=0.19\textwidth]{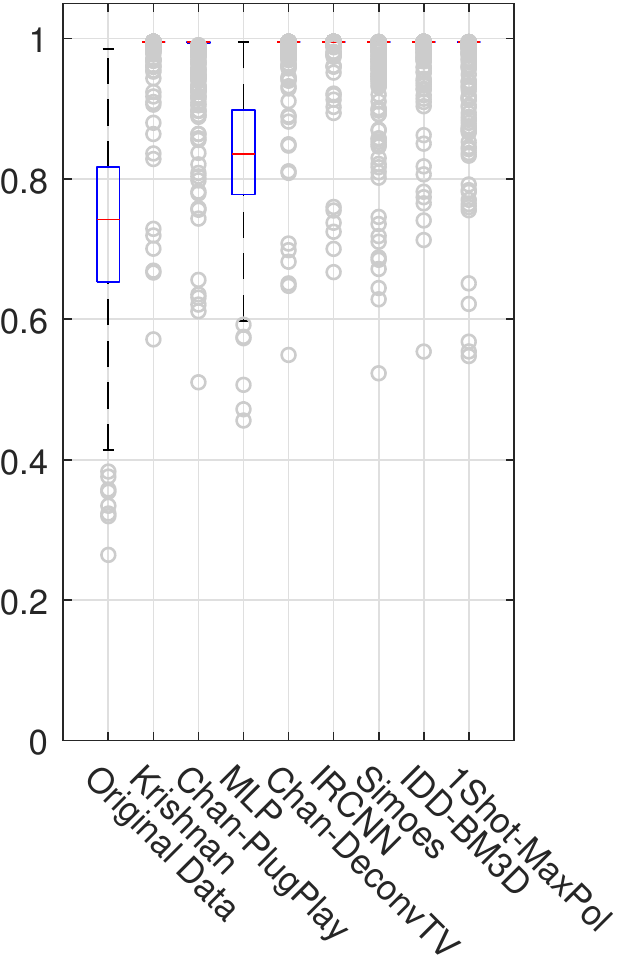}}\hspace{-.25in}
\subfigure[Hyperspectral]{\includegraphics[width=0.19\textwidth]{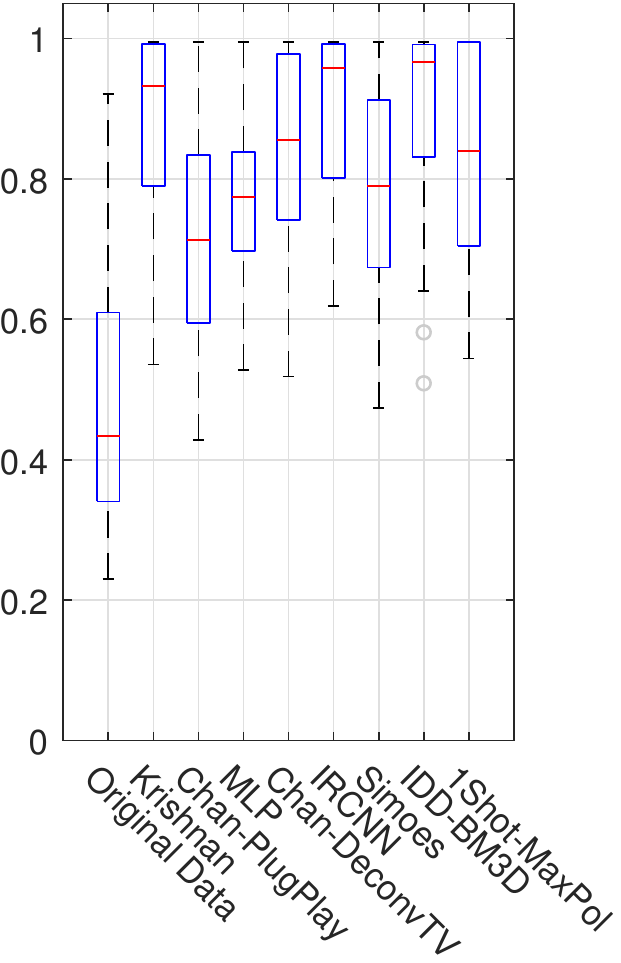}}\hspace{-.25in}
\subfigure[Haze]{\includegraphics[width=0.19\textwidth]{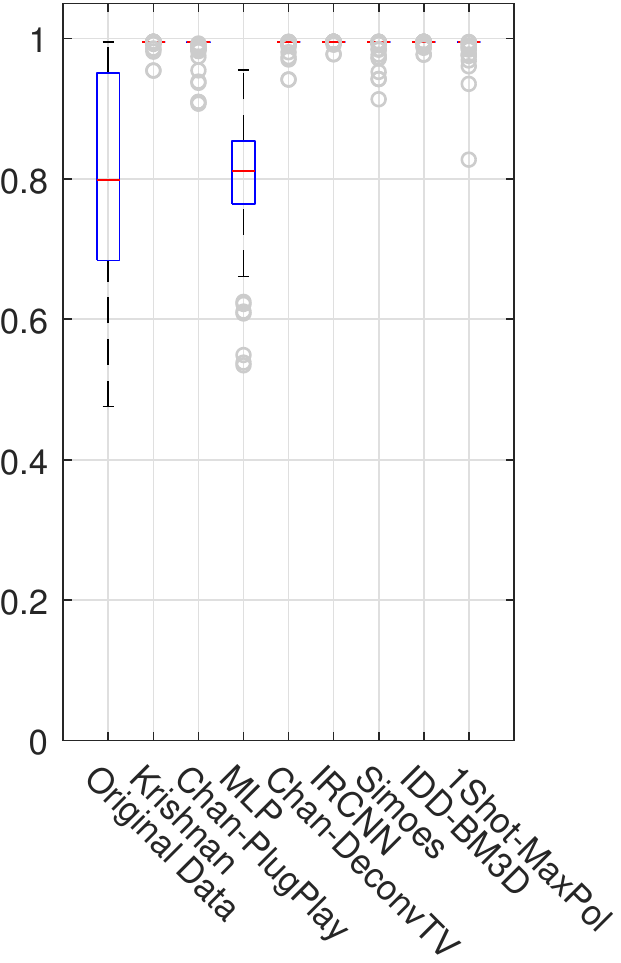}}
}\vspace{-.05in}
\centerline{
\subfigure[LROC]{\includegraphics[width=0.19\textwidth]{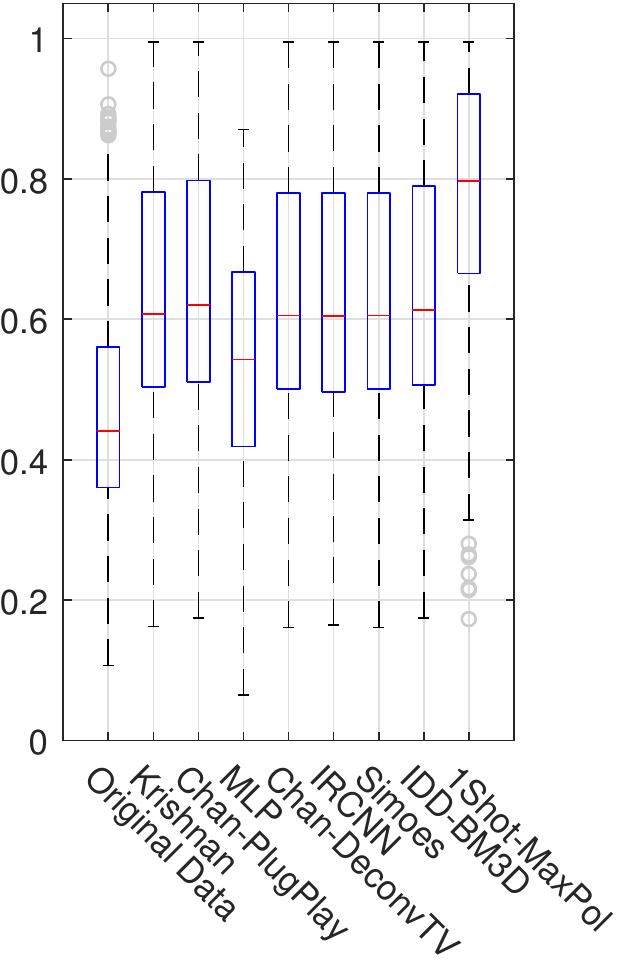}}\hspace{-.25in}
\subfigure[McMaster]{\includegraphics[width=0.19\textwidth]{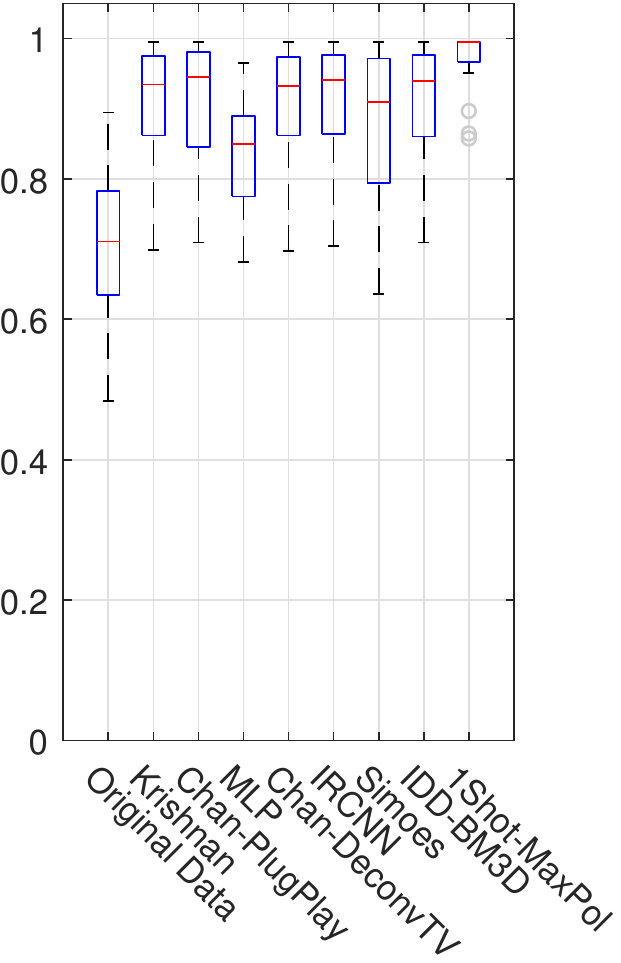}}\hspace{-.25in}
\subfigure[NIR-Scene]{\includegraphics[width=0.19\textwidth]{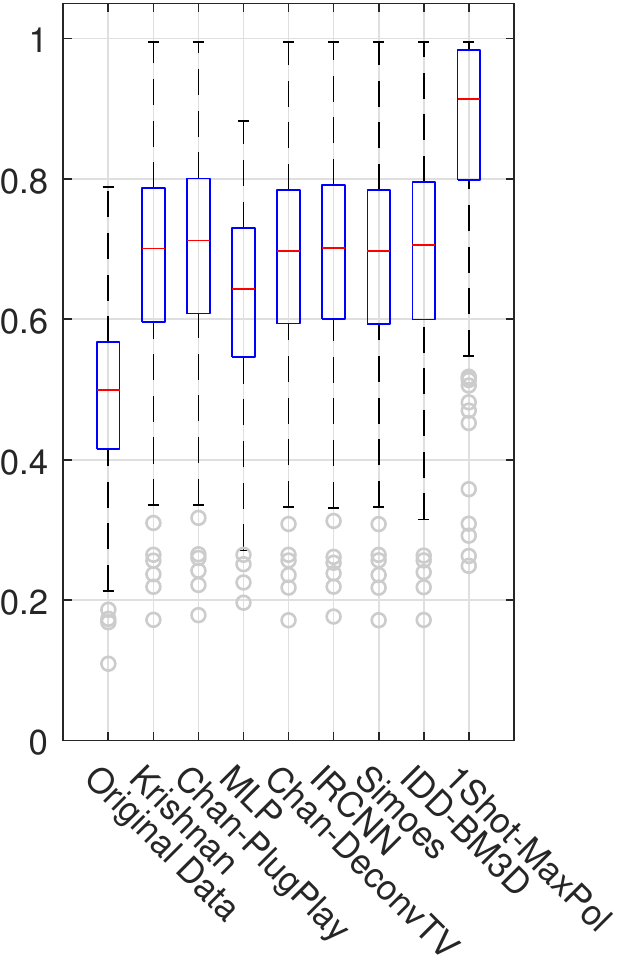}}\hspace{-.25in}
\subfigure[RGB-Scene]{\includegraphics[width=0.19\textwidth]{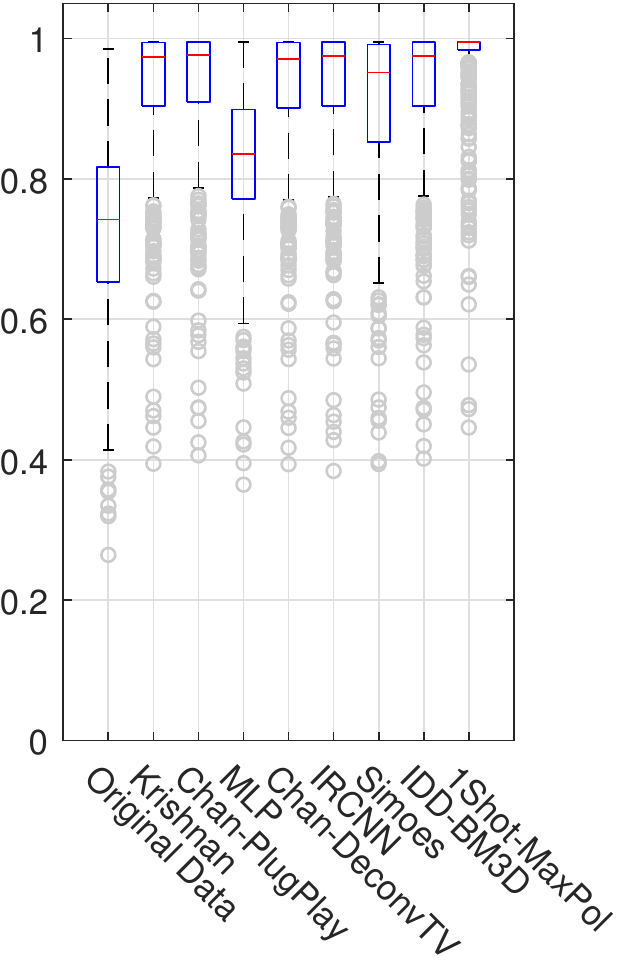}}\hspace{-.25in}
\subfigure[Hyperspectral]{\includegraphics[width=0.19\textwidth]{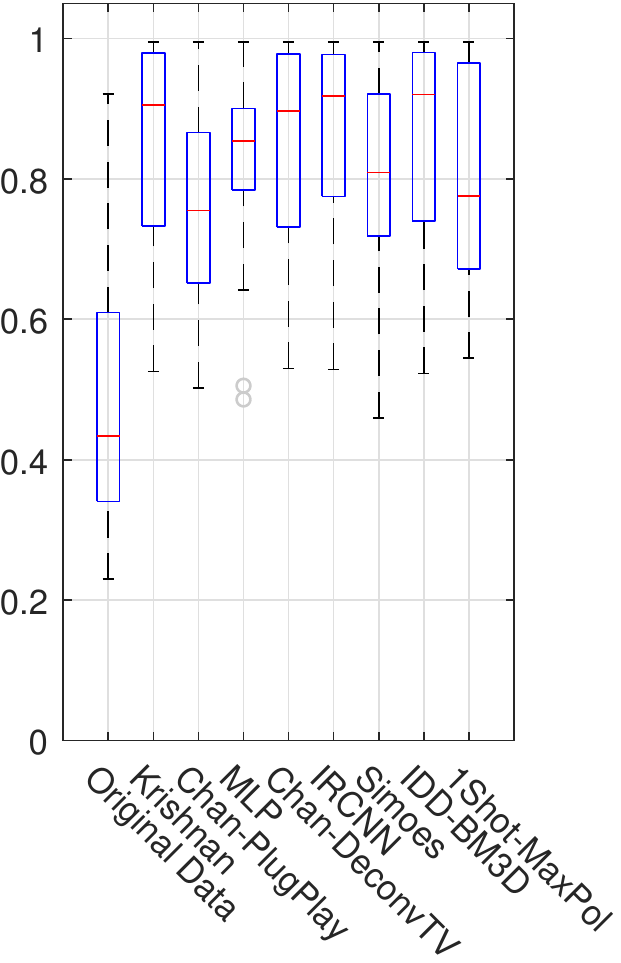}}\hspace{-.25in}
\subfigure[Haze]{\includegraphics[width=0.19\textwidth]{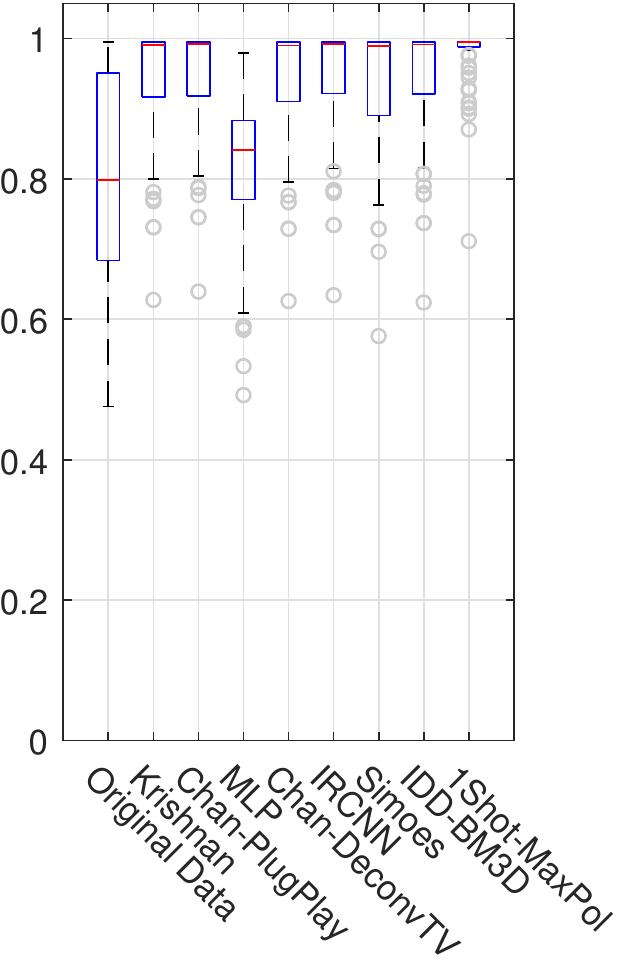}}
}\vspace{-.05in}
\caption{Performance analysis on each category of image database listed in Table \ref{table_natural_database} by means of eight deblurring methods using blind Gaussian (top row) and Laplacian (bottom row) PSF estimation. The box-plots demonstrate the statistical distribution of recovery scores for each category using MLV NR-FQA metric. The analyzed scale for blind PSF calculation here is $s=2$.}
\label{fig_NR_FQA_boxplot}
\end{figure*}

\subsection{Selected Natural Image Database}\label{sec_natural_databases}
Here we describe the selected natural image databases for deblurring. We have collected $2054$ images across different imaging modalities with spectral wavelengths between $350$nm and $850$nm, such as visible light (RGB), hyperspectral (multi-channel), and near-infrared (NIR) (single channel) images. Blurring is caused in these modalities by lens imperfections and turbid medium. An overview of the selected natural databases is listed in Table \ref{table_natural_database}. 

\begin{table}[htp]
\renewcommand{\arraystretch}{1.1}
\caption{$2054$ selected natural images for deblurring across different modalities containing optical blur}
\label{table_natural_database}
\centering
\scriptsize
\begin{tabular}{lccccc}
\hlinewd{1.5pt}
Database                        & Camera       & Format & \#Bit & Patch Size           & $\#$ \\ \hlinewd{1.5pt}
LROC \cite{robinson2010lunar}      & NarrowAngle & TIF        &     8-Gray &$512\times{512}$ & $937$      \\ \hlinewd{.75pt}
McMaster \cite{zhang2011color}  & Kodak Film & TIF    &     8-RGB     &$500\times{500}$ & $18$       \\ \hlinewd{.75pt}
Hyperspec. \cite{nascimento2016spatial}   & Hamamatsu & TIF  &     12-RGB     &$1024\times{1344}$ & $63$ \\ \hlinewd{.75pt}
RGB-Scene \cite{BS11}       & Nikon D90 & TIF        &     8-RGB     &$682\times{1024}$ & $477$      \\ \hlinewd{.75pt}
NIR-Scene \cite{BS11}       & Canon T1i & TIF        &     8-NIR     &$682\times{1024}$ & $477$      \\ \hlinewd{.75pt}
Haze \cite{zhu2015fast, ren2016single}       & N/A& TIF        &     8-RGB     & $\sim500\times{600}$ & $82$      \\ \hlinewd{1.5pt}
\end{tabular}
\end{table}

\subsubsection{Lunar Reconnaissance Orbiter Camera (LROC) Images}
The LROC images are high resolution photos captured by two narrow angle cameras (NAC) mounted on the Lunar Reconnaissance Orbiter (LRO) satellite launched by NASA on 18 June 2009 \cite{robinson2010lunar, humm2016flight, mahanti2016inflight, speyerer2012flight}. The mission objective of LROC is to map the surface of Moon for the identification of future landing sites and the scientific exploration of key targets. The images are released by Arizona State University every three months to the NASA Planetary Data System (PDS), which is publicly available for downloading at \footnote{\url{http://lroc.sese.asu.edu/}}. The NAC camera projects a $700$-mm focal-length telescope imaging onto a linear array CCD camera with spectral response between $400-760$nm. The camera provides a near diffraction-limited performance with $0.5$m pixel resolution over a combined $5$-km swath at a nominal $50$-km Lunar altitude. The NAC images are sampled at 12 bits and converted to 8 bits, then lossless compression is applied prior to downlink. The modular transfer function (MTF) of the camera (a.k.a PSF) yields a wide span over the Nyquist band \cite{robinson2010lunar}. Such a wide span guarantees that the majority of image frequency information is preserved, and we show in our experiments that these images can be well recovered. We have cropped $937$ gray channel image patches of $512\times 512$ pixel resolution from six different Lunar craters: Hell Q, Luminous Pierazzo, Jackson, Tycho, Burg, and Rozhdestvenskiy W.

\subsubsection{Hyperspectral Imaging Camera}
Hyperspectral radiance images are captured in $33$ different wavelength filters sampled within $[400,720]$nm \cite{nascimento2016spatial}. The images are acquired by a monochrome camera with CCD arrays and are sized $1024\times 1344$ pixels. The database consists of $30$ natural scenes including rural and urban photos captured in the Minho region of Portugal. The optical lens aberration integrated in camera has a line spread function close to a Gaussian function. We have rendered an RGB image from all $33$ Hyperspectral frames. Note that we have applied image deblurring on gamma corrected RGB images as suggested by the authors in \cite{nascimento2016spatial}. 

\subsubsection{McMaster Images}
The McMaster database is constructed with true RGB color from Kodak film \footnote{\url{http://www4.comp.polyu.edu.hk/~cslzhang/CDM_Dataset.htm}}. The database contains $18$ image patches of size $500\times 500$ pixels and are cropped from eight image scenes. The images have high saturation with abrupt color transitions and sharp structures \cite{zhang2011color}.

\subsubsection{NIR-RGB Scene Dataset}
This dataset is introduced in \footnote{\url{https://ivrl.epfl.ch/supplementary_material/cvpr11/}}\cite{BS11}, where the images are captured by Nikon D90 and Canon T1i cameras using both visible and near infrared (NIR) filters with $750$nm cut-off between the two filters. Images are processed after image acquisition using white balance correction. Both NIR and RGB images are registered using SIFT features and the final images are re-sampled. Note that such re-sampling deteriorates the high frequency spectrum which can negatively impact the quality of image deblurring.

\subsubsection{Haze Images}
Haze and foggy images refer to outdoor imaging in bad weather conditions. Turbid media such as atmospheric particles, smoke, and water droplets absorb the scattered light and prevent it from reaching the camera sensor. Therefore, the acquired image is unclear and needs to be de-hazed \cite{zhu2015fast, ren2016single}. We obtain $41$ haze images used in the literature and de-haze them using two methods described in \cite{zhu2015fast, ren2016single}, thus producing $82$ de-hazed images in total. We show here that by adding the deconvolution module, the recovered images become clearer. 

\begin{table}[htp]
\renewcommand{\arraystretch}{1.1}
\centering
\caption{Average performance on all recovered images by means of different methods across different blur models. The average NR-FQA score for original databases is $0.5408$.}
\label{table_overall_performance}
\begin{tabular}{l|cc|cc}
\hlinewd{1.5pt}
Model & \multicolumn{2}{c|}{Gaussian ($\beta=2$)} & \multicolumn{2}{c}{Laplacian ($\beta=1$)} \\ \hlinewd{1.5pt}
Scale &   $s=2$   &   $s=4$   &  $s=2$   &   $s=4$    \\ \hlinewd{1.5pt}
Krishnan \cite{krishnan2009fast, krishnan2011blind}  & $0.9009$ &  $0.9026$ & $0.7415$ & $0.7346$ \\ \hlinewd{1pt}
Chan-PlugPlay \cite{chan2017plug}                    & $0.8583$ &  $0.8590$ & $\textbf{0.7464}$ & $\textbf{0.7396}$  \\ \hlinewd{1pt}
MLP \cite{schuler2013machine}                        & $0.6832$ &  $0.6862$ & $0.6582$ & $0.6539$  \\ \hlinewd{1pt}
Chan-DeconvTV \cite{chan2011augmented}               & $0.8867$ &  $0.8886$ & $0.7390$ & $0.7323$ \\ \hlinewd{1pt}
IRCNN \cite{zhang2017learning}                       & $\textbf{0.9090}$ & $\textbf{0.9114}$ & $0.7413$ & $0.7359$ \\ \hlinewd{1pt}
Simoes \cite{simoes2016framework}                    & $0.8856$ &  $0.8887$ & $0.7315$ & $0.7254$ \\ \hlinewd{1pt}
IDD-BM3D \cite{danielyan2012bm3d}                    & $\textbf{0.9264}$ & $\textbf{0.9307}$ & $\textbf{0.7461}$ & $\textbf{0.7387}$ \\ \hlinewd{1pt}
1Shot-MaxPol                                         & $\textbf{0.9283}$ &  $\textbf{0.9265}$ & $\textbf{0.8792}$ &  $\textbf{0.8561}$ \\ \hlinewd{1.5pt}
\end{tabular}
\end{table}

\begin{figure*}[htp]
\centerline{
\setbox1=\hbox{\includegraphics[height=4cm]{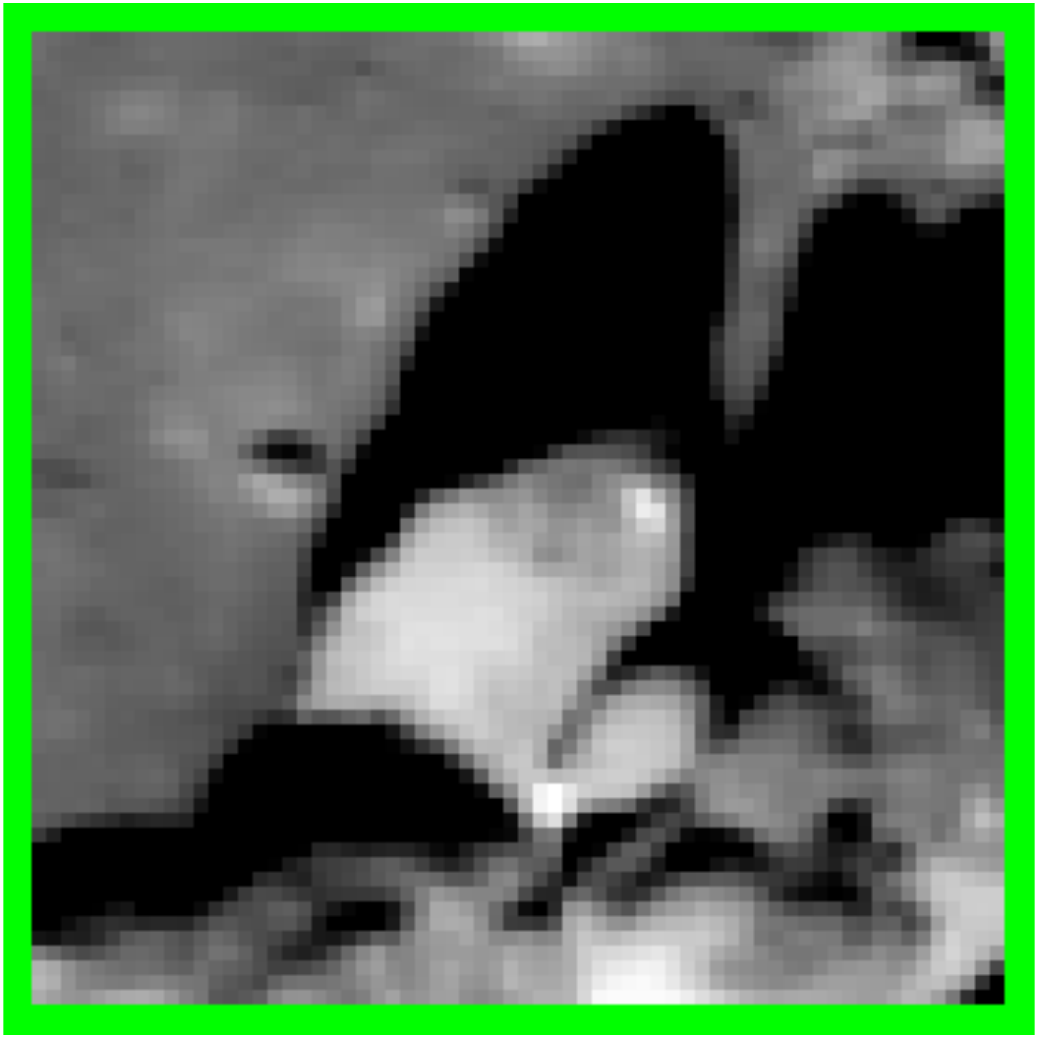}}
\subfigure[Original]{\includegraphics[height=4cm]{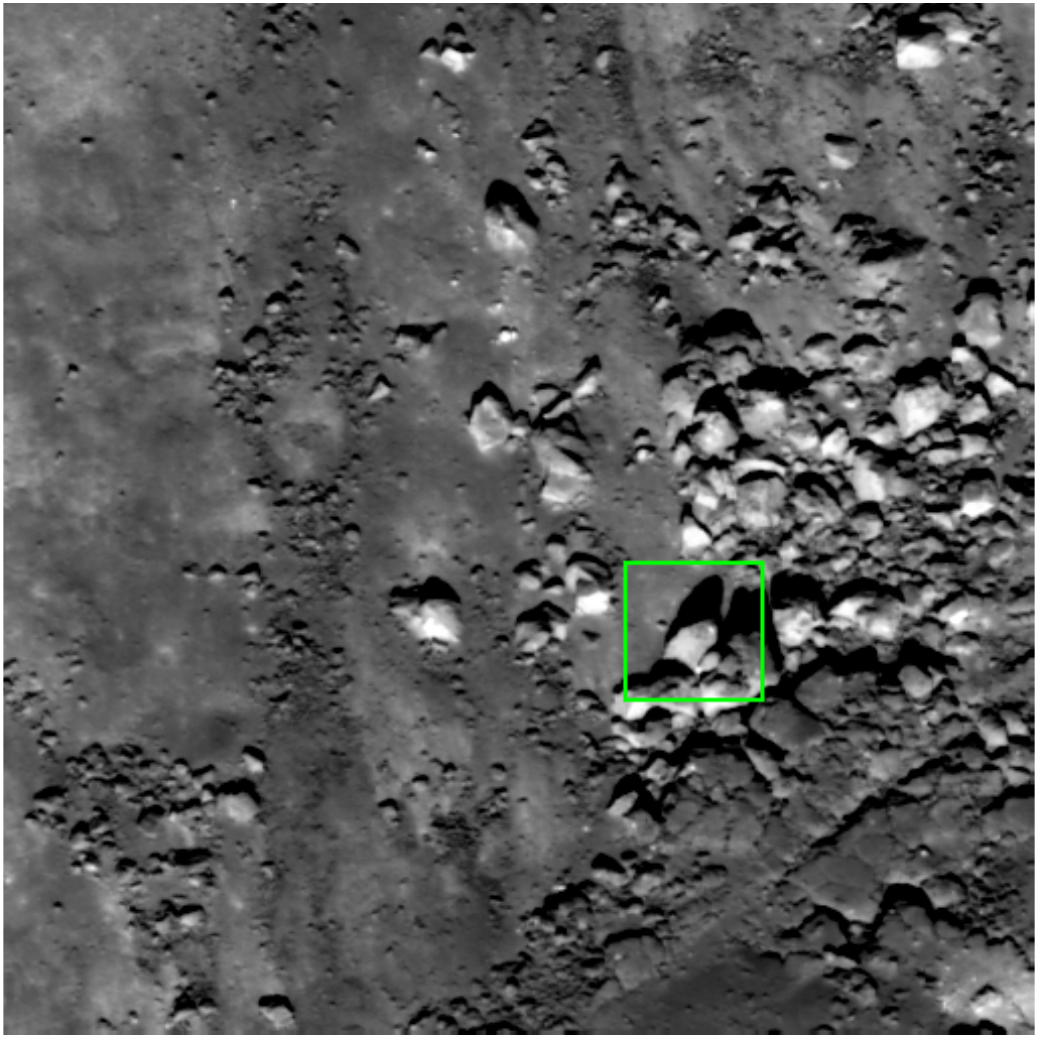}\llap{\makebox[\wd1][l]{\raisebox{2cm}{\includegraphics[height=2cm]{image_scan_blurry_category_001_image_number_148_patch}}}}\label{fig_LROC_raw_category_001_image_number_148_Scale_25_Gaussian}}
\subfigure[Krishnan \cite{krishnan2009fast, krishnan2011blind}]{\includegraphics[height=4cm]{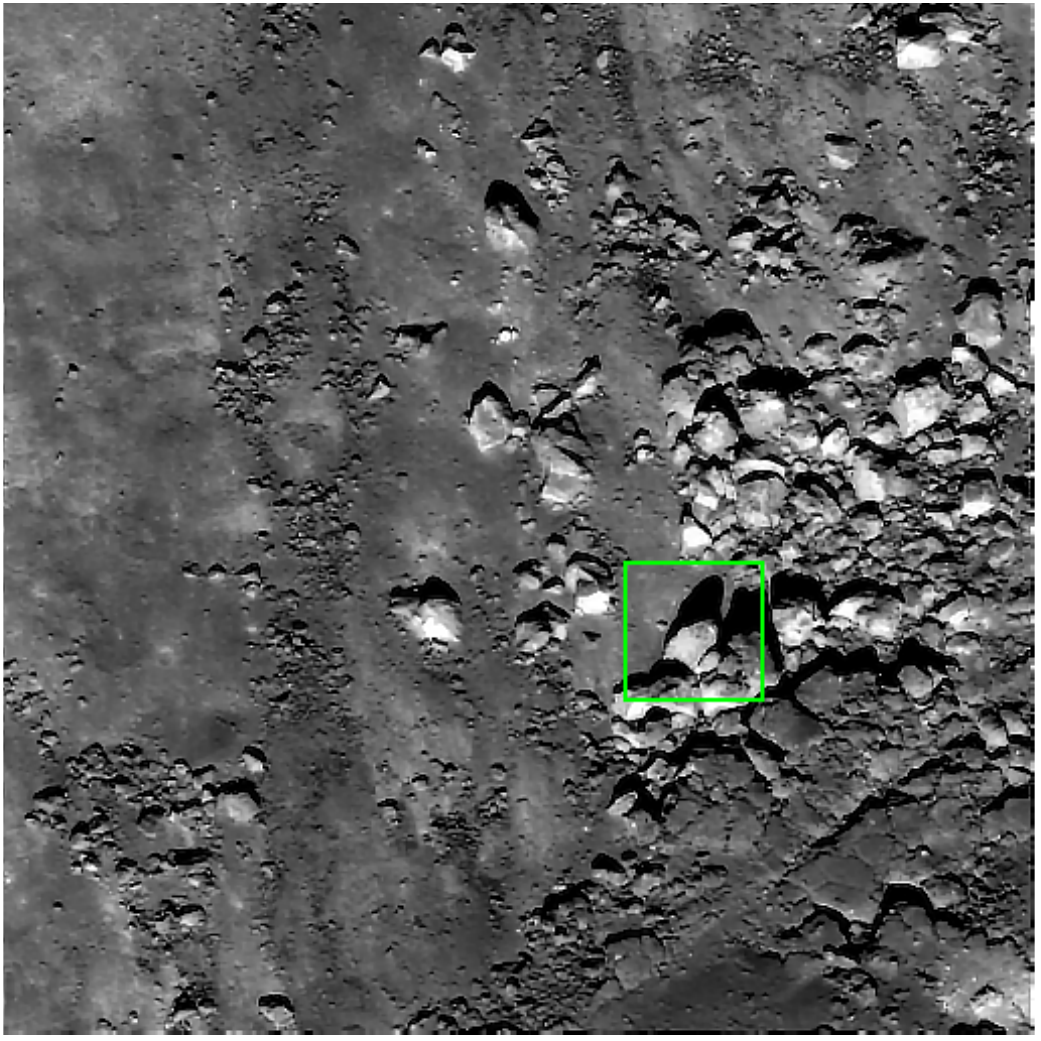}\llap{\makebox[\wd1][l]{\raisebox{2cm}{\includegraphics[height=2cm]{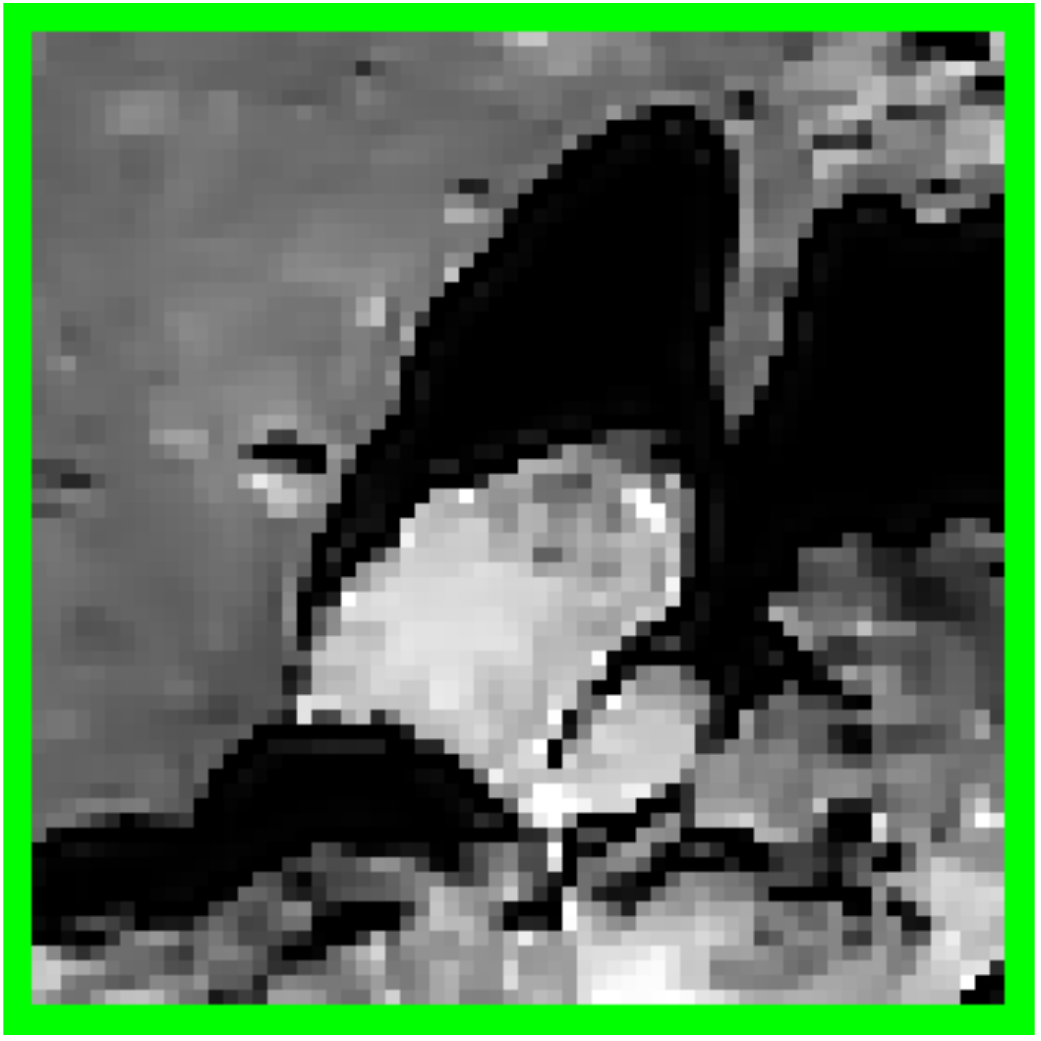}}}}\label{fig_LROC_Krishnan_category_001_image_number_148_Scale_25_Gaussian}}
\subfigure[Chan-PlugPlay \cite{chan2017plug}]{\includegraphics[height=4cm]{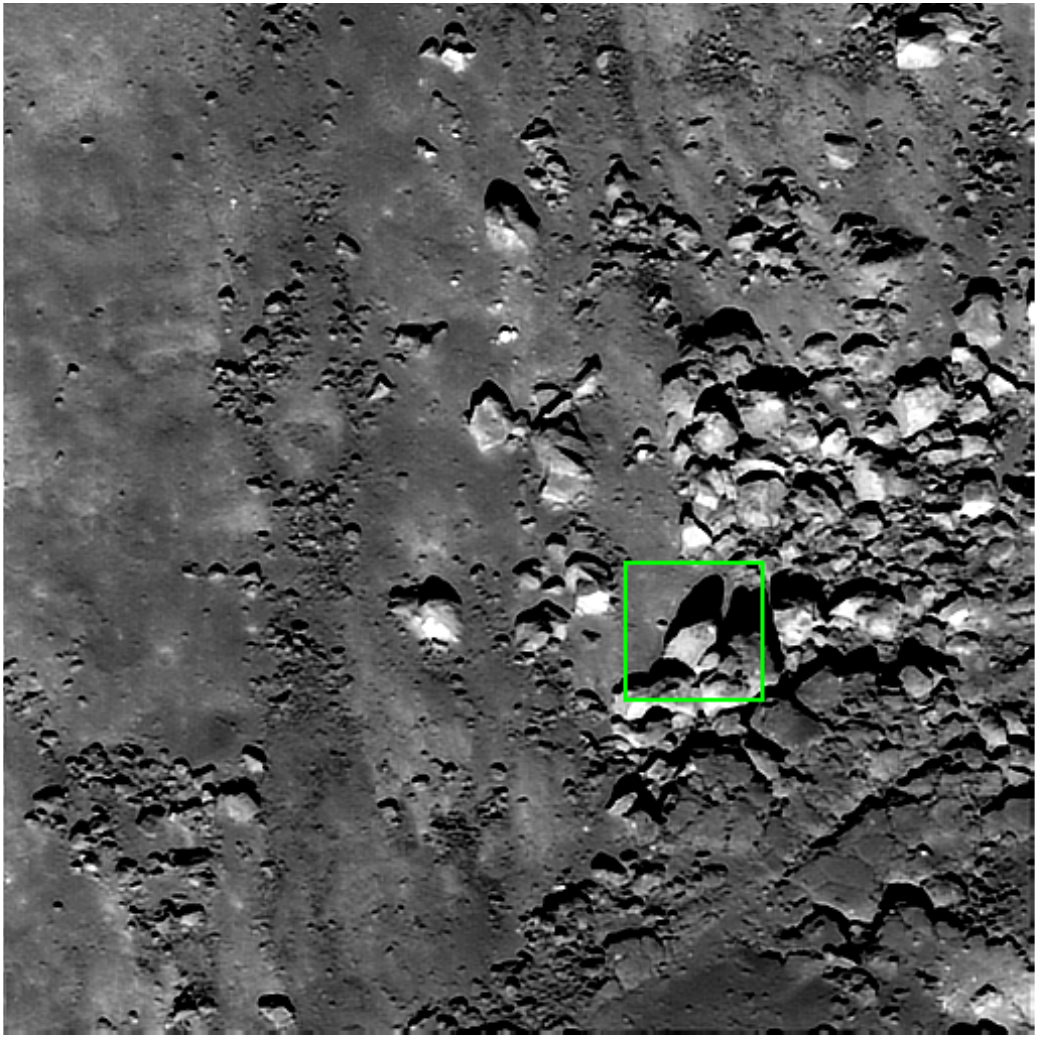}\llap{\makebox[\wd1][l]{\raisebox{2cm}{\includegraphics[height=2cm]{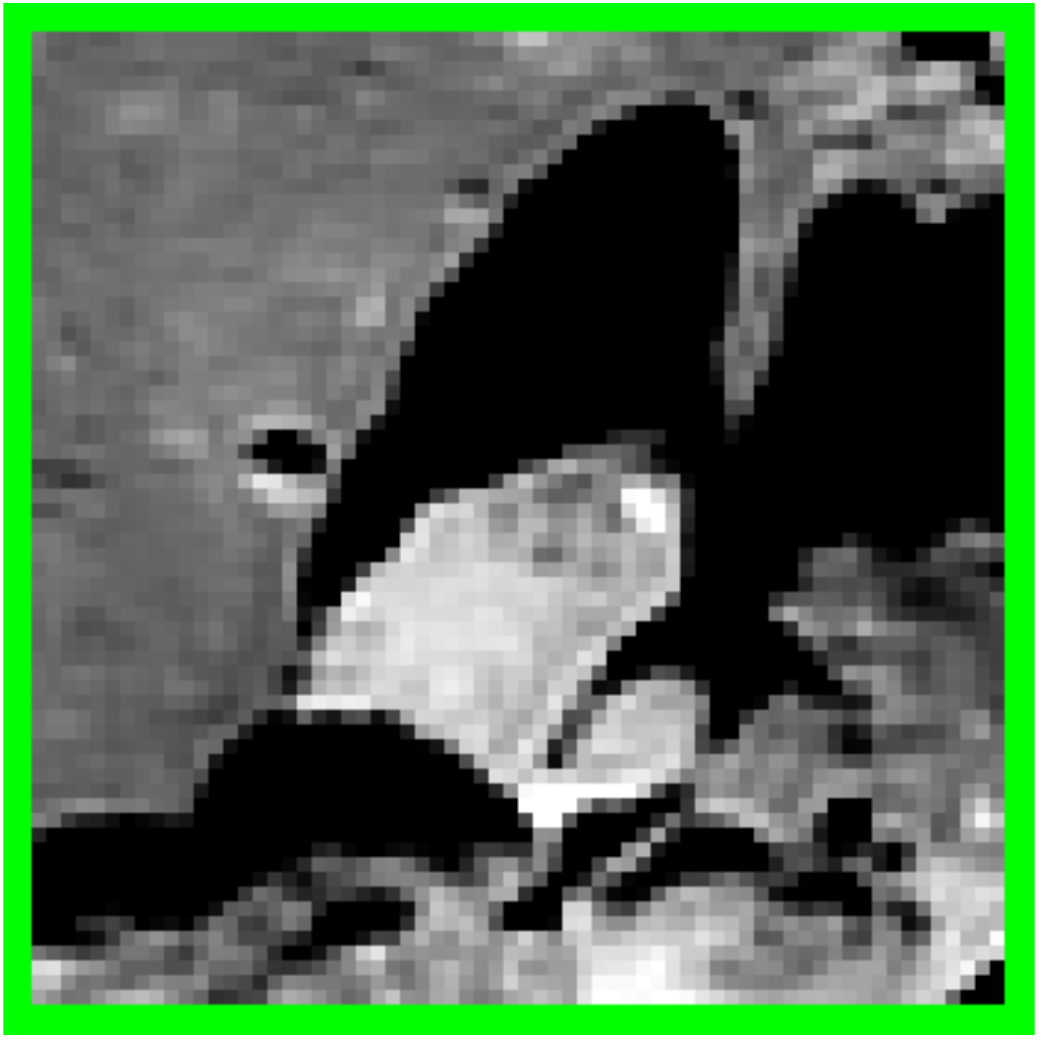}}}}\label{fig_LROC_Chan_PlugPlay_category_001_image_number_148_Scale_25_Gaussian}}
\subfigure[MLP\cite{schuler2013machine}]{\includegraphics[height=4cm]{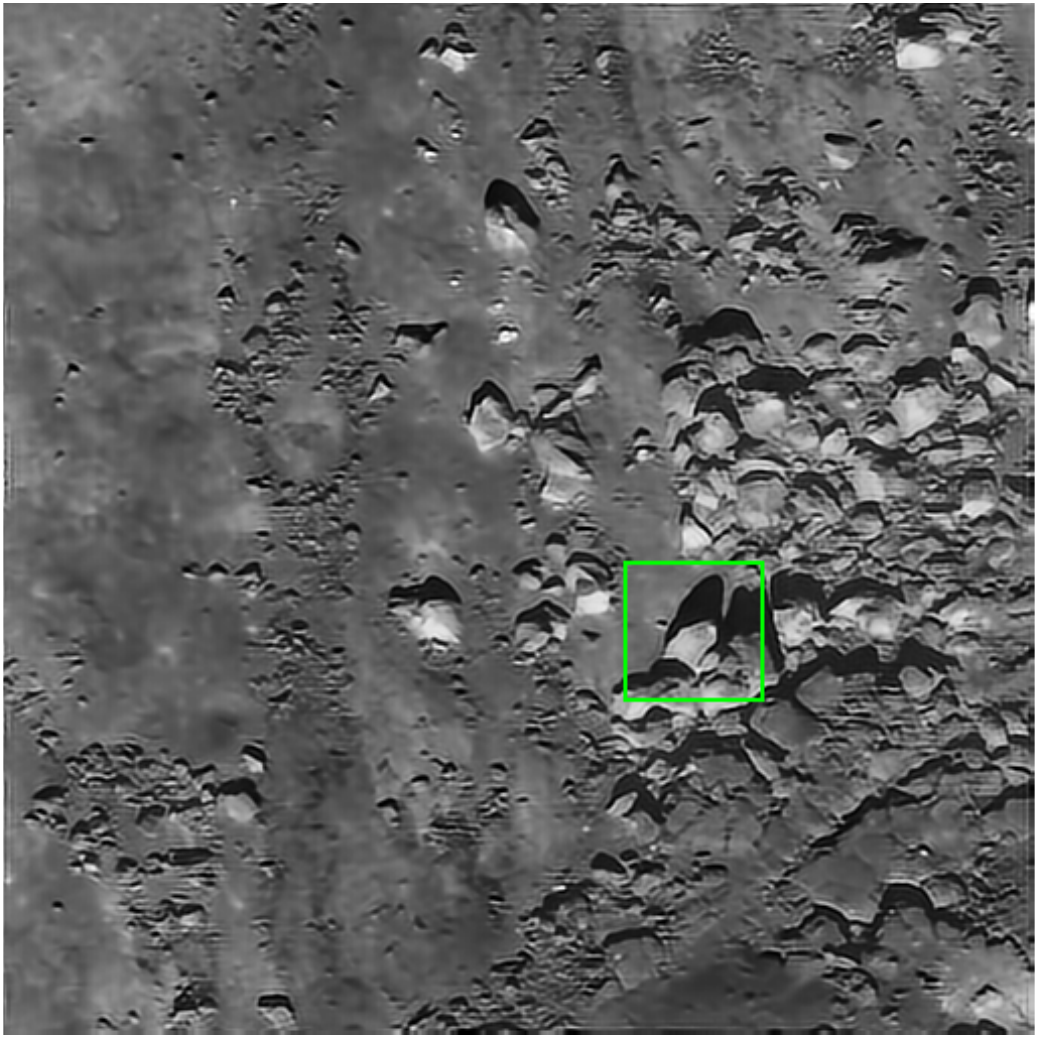}\llap{\makebox[\wd1][l]{\raisebox{2cm}{\includegraphics[height=2cm]{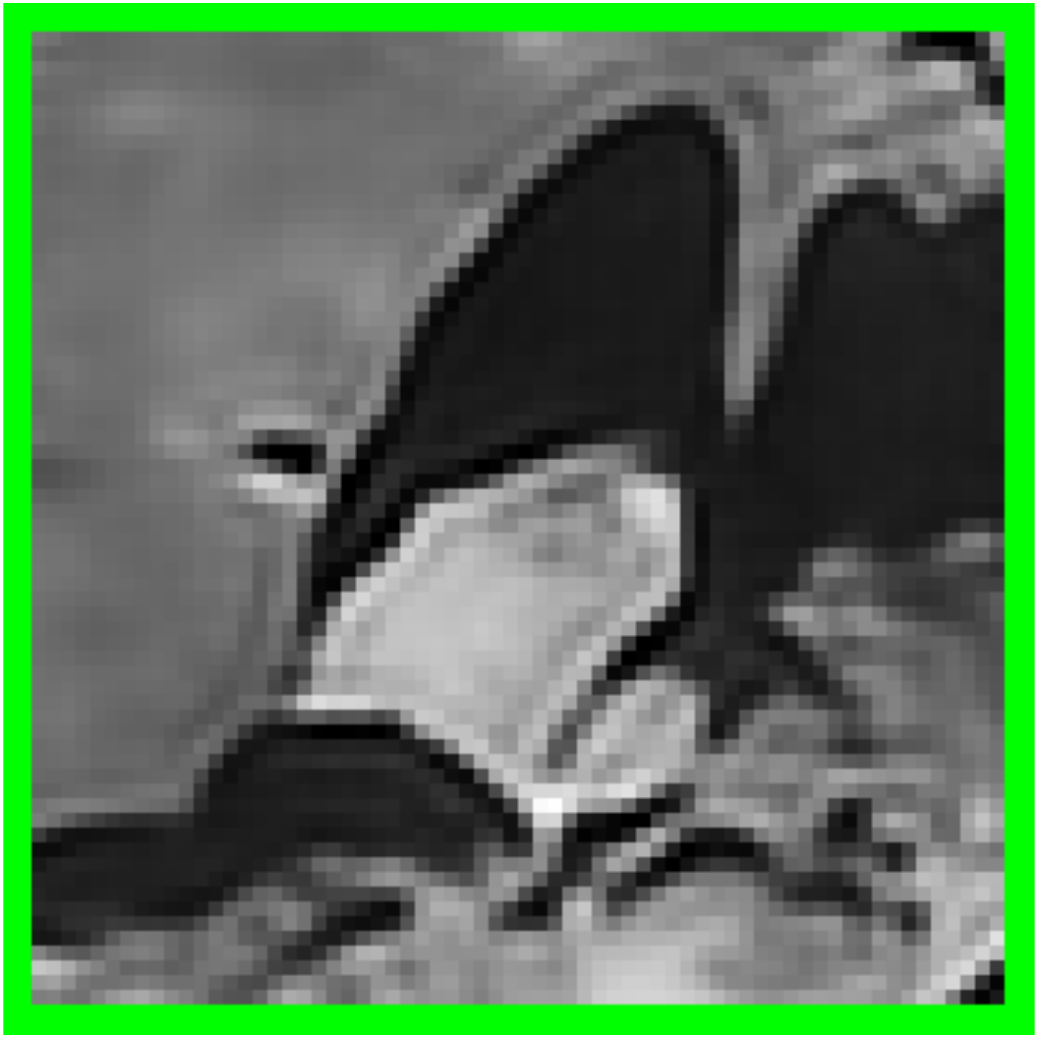}}}}\label{fig_LROC_MLP_category_001_image_number_148_Scale_25_Gaussian}}
}\vspace{-.05in}
\centerline{
\setbox1=\hbox{\includegraphics[height=4cm]{image_scan_blurry_category_001_image_number_148_patch}}
\subfigure[Chan-DeconvTV \cite{chan2011augmented}]{\includegraphics[height=4cm]{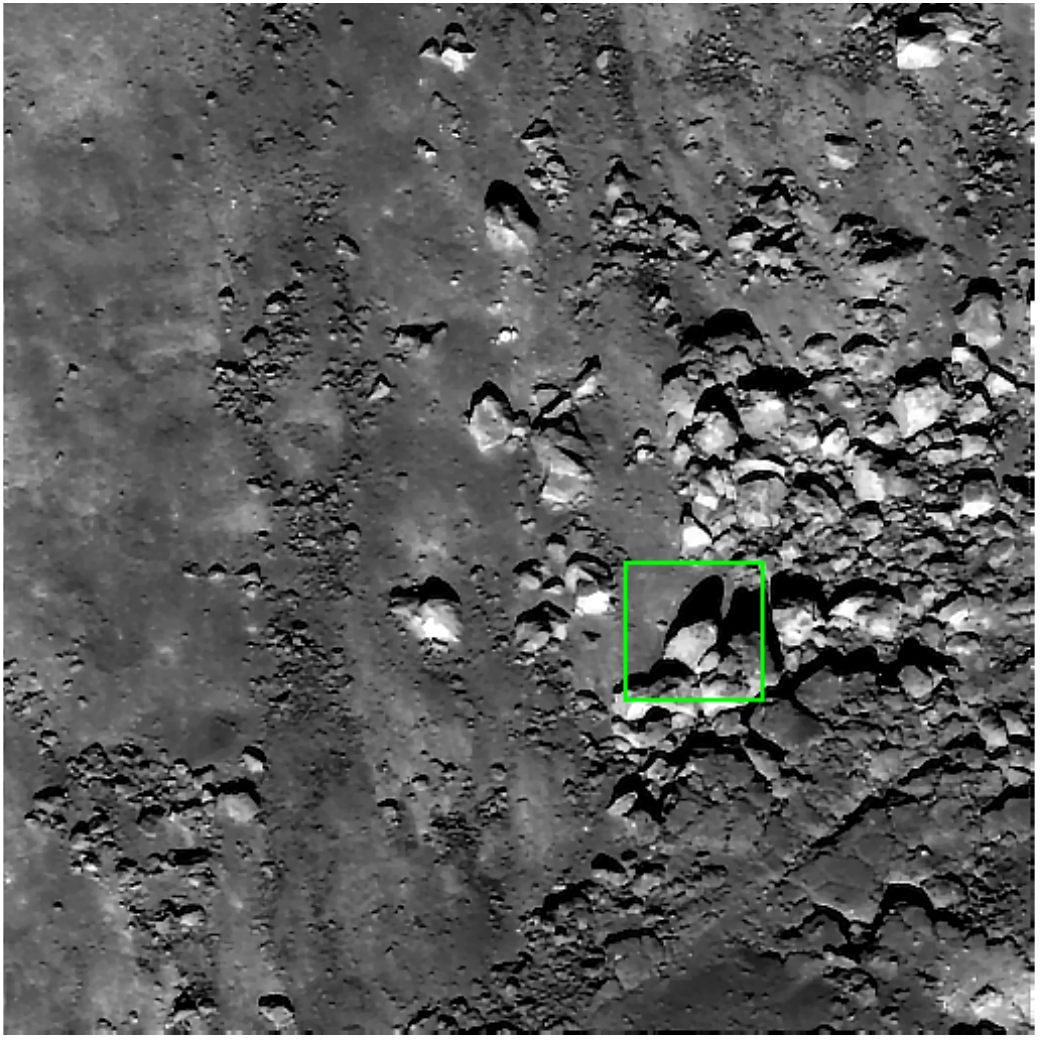}\llap{\makebox[\wd1][l]{\raisebox{2cm}{\includegraphics[height=2cm]{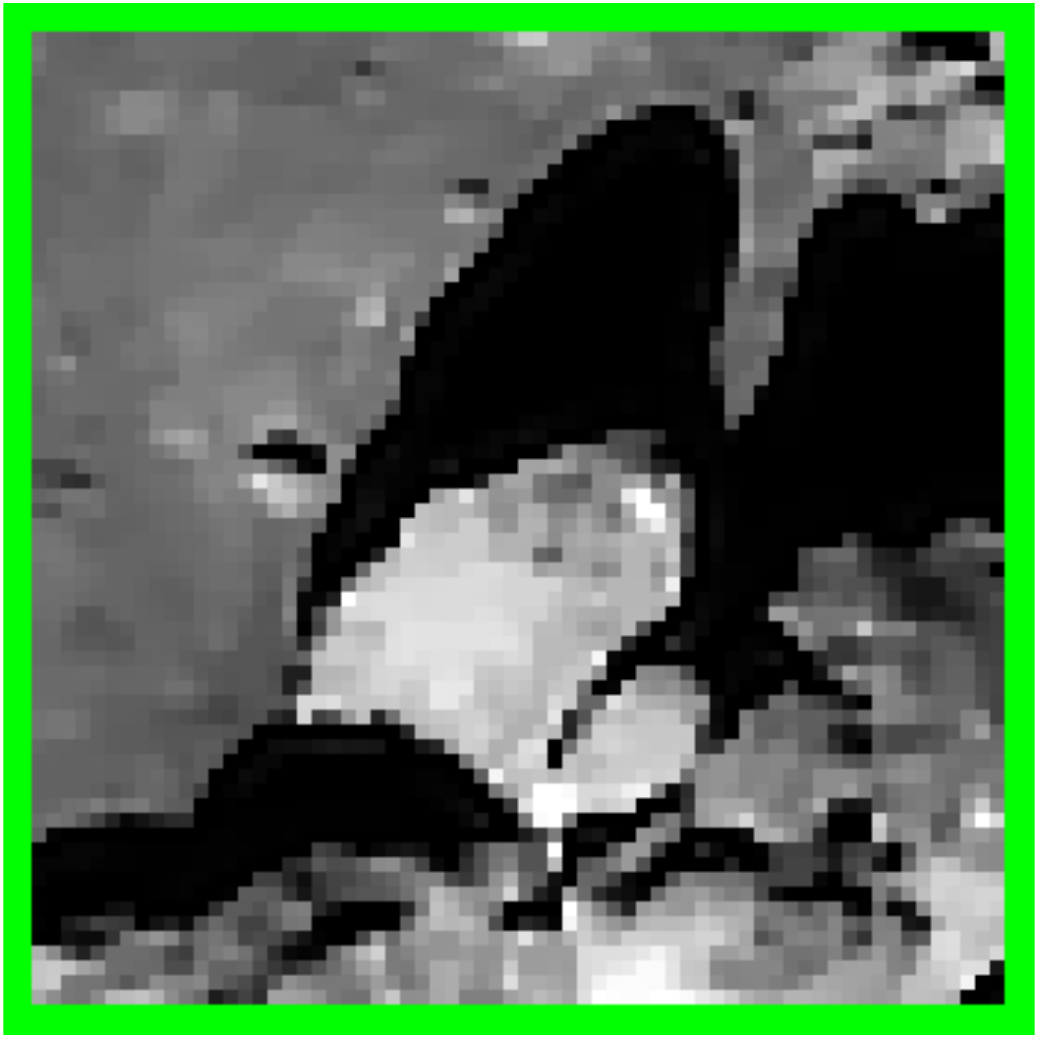}}}}\label{fig_Chan_DeconvTV_category_001_image_number_148_Scale_25_Gaussian}}
\subfigure[IRCNN \cite{zhang2017learning}]{\includegraphics[height=4cm]{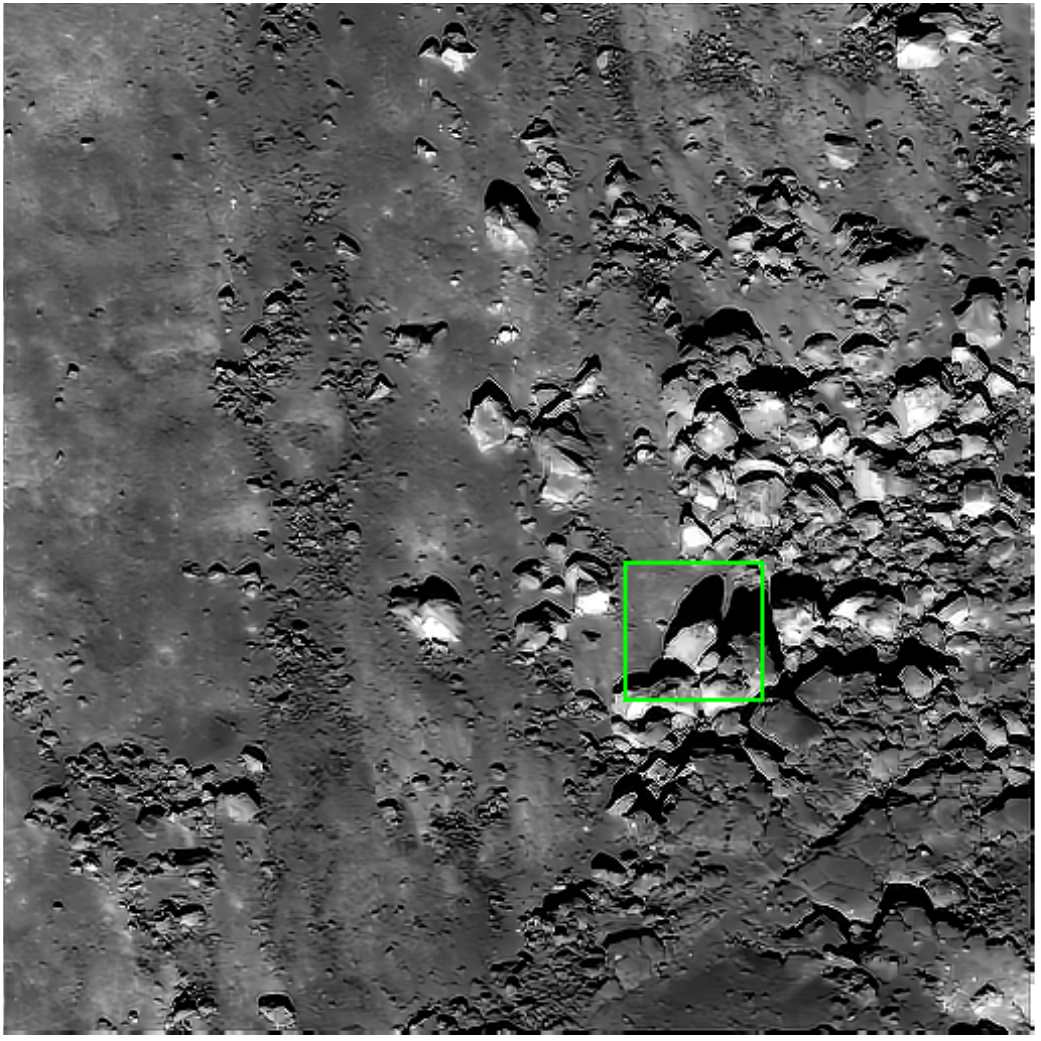}\llap{\makebox[\wd1][l]{\raisebox{2cm}{\includegraphics[height=2cm]{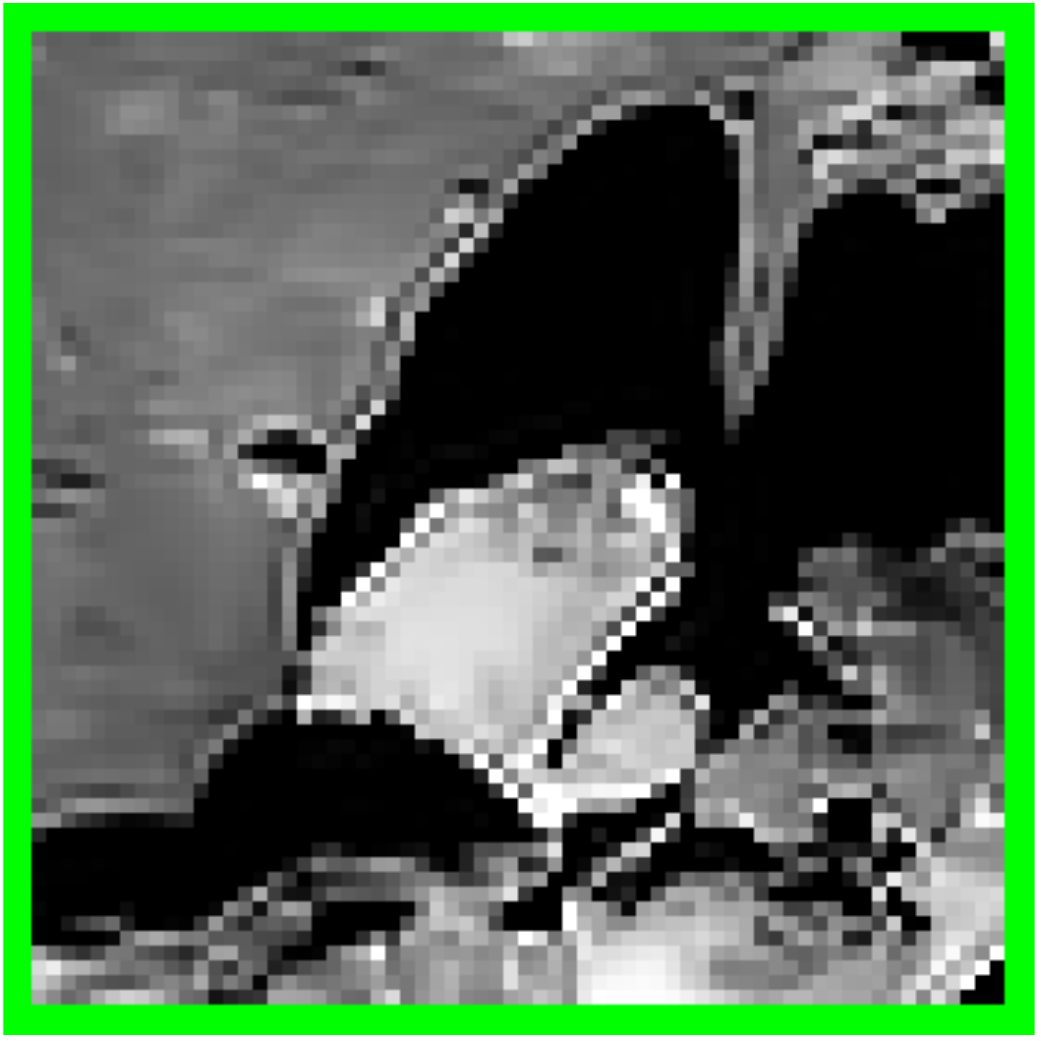}}}}\label{fig_LROC_IRCNN_category_001_image_number_148_Scale_25_Gaussian}}
\subfigure[Simoes\cite{simoes2016framework}]{\includegraphics[height=4cm]{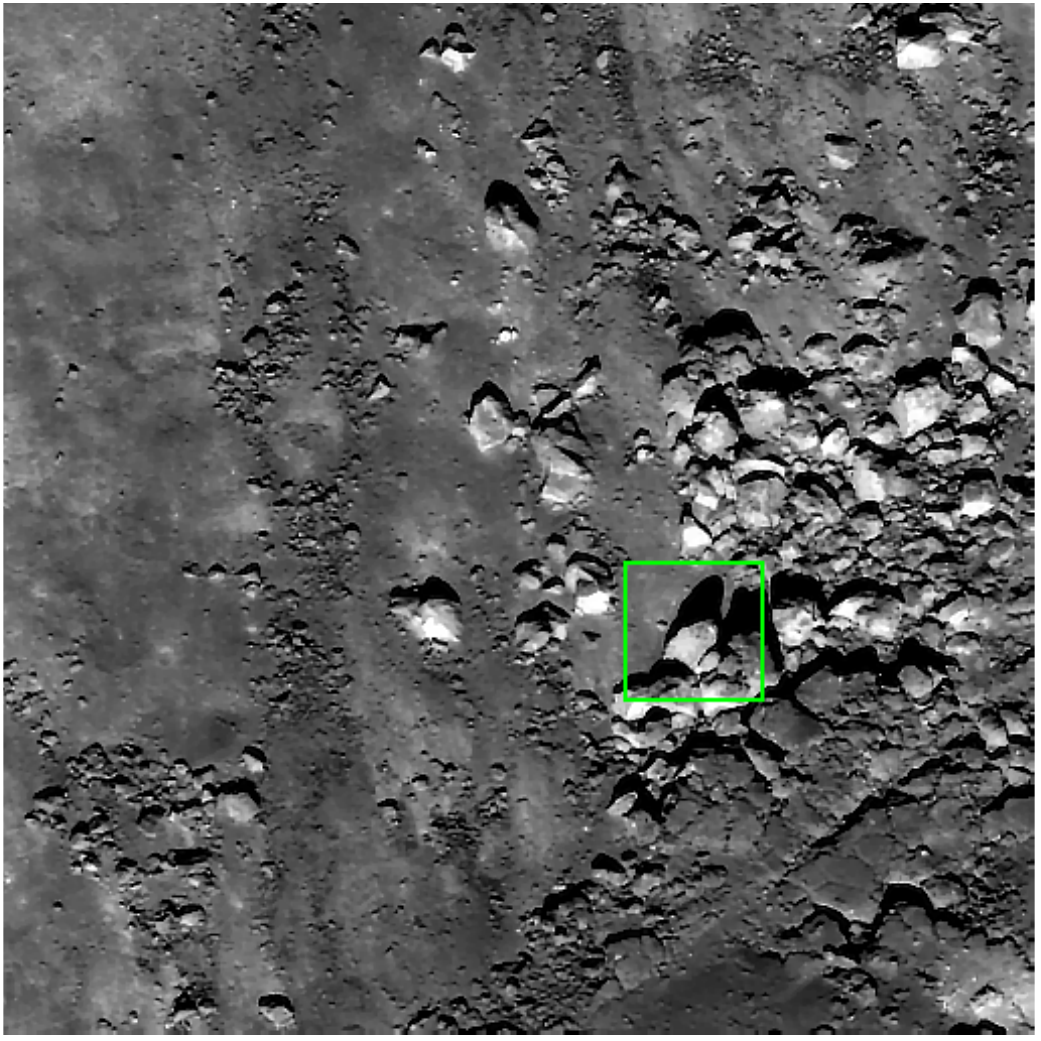}\llap{\makebox[\wd1][l]{\raisebox{2cm}{\includegraphics[height=2cm]{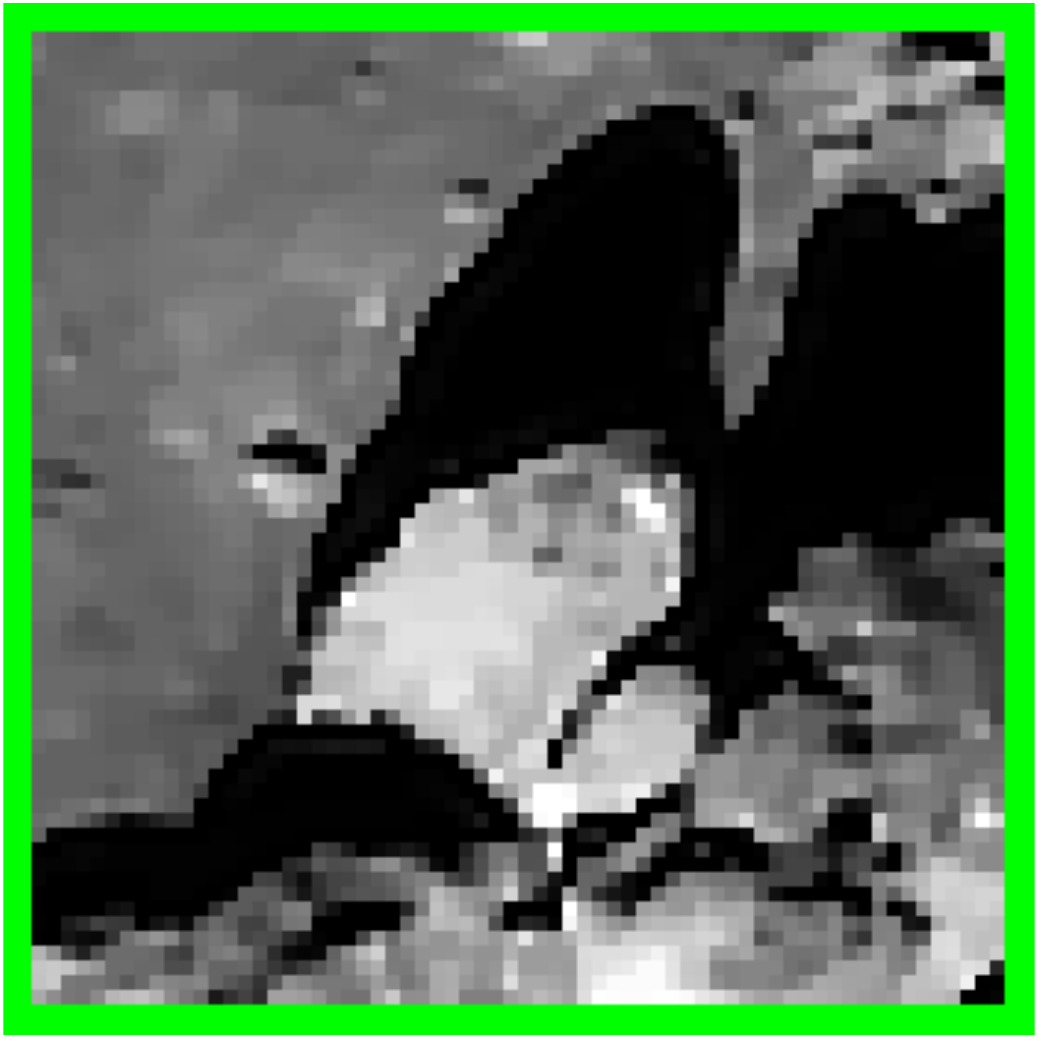}}}}\label{fig_LROC_Simoes_category_001_image_number_148_Scale_25_Gaussian}}
\subfigure[IDD-BM3D\cite{danielyan2012bm3d}]{\includegraphics[height=4cm]{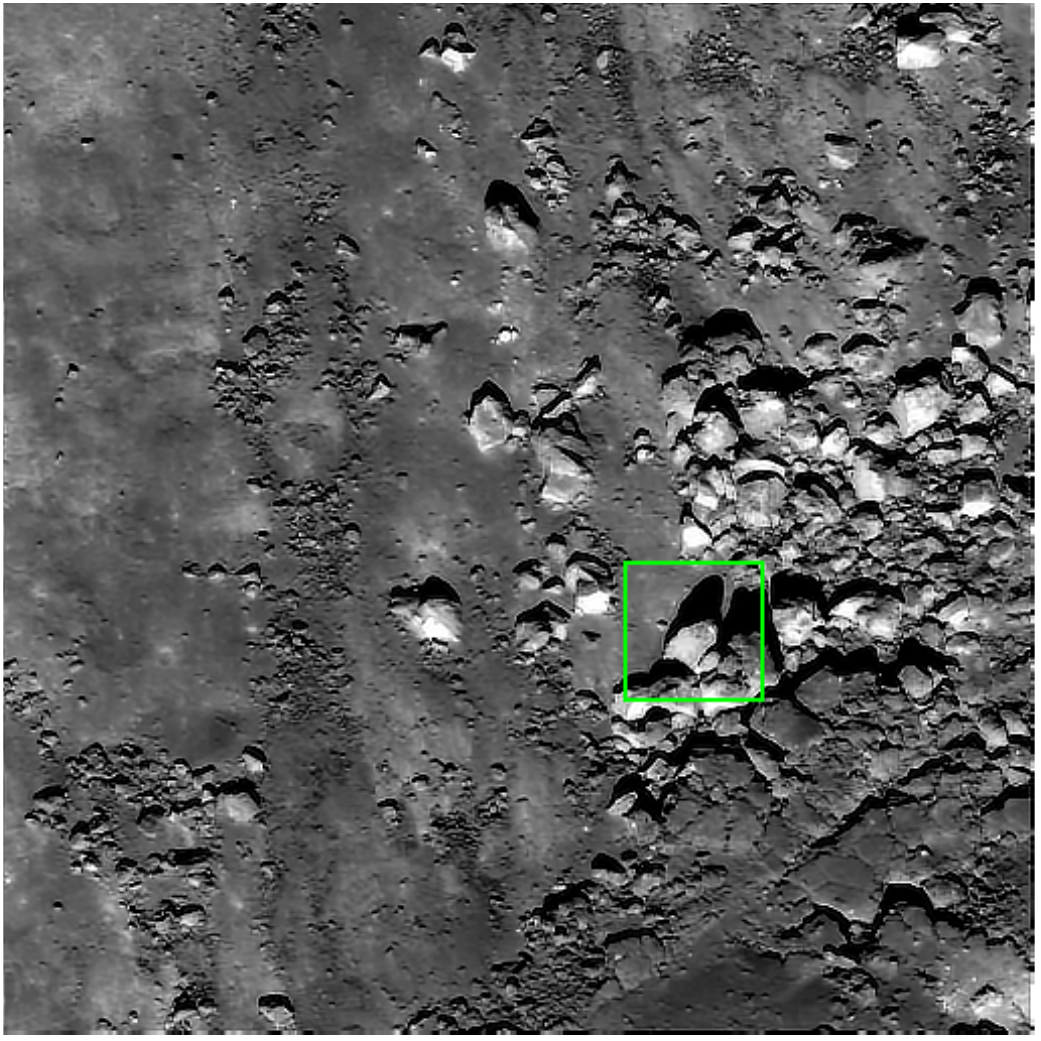}\llap{\makebox[\wd1][l]{\raisebox{2cm}{\includegraphics[height=2cm]{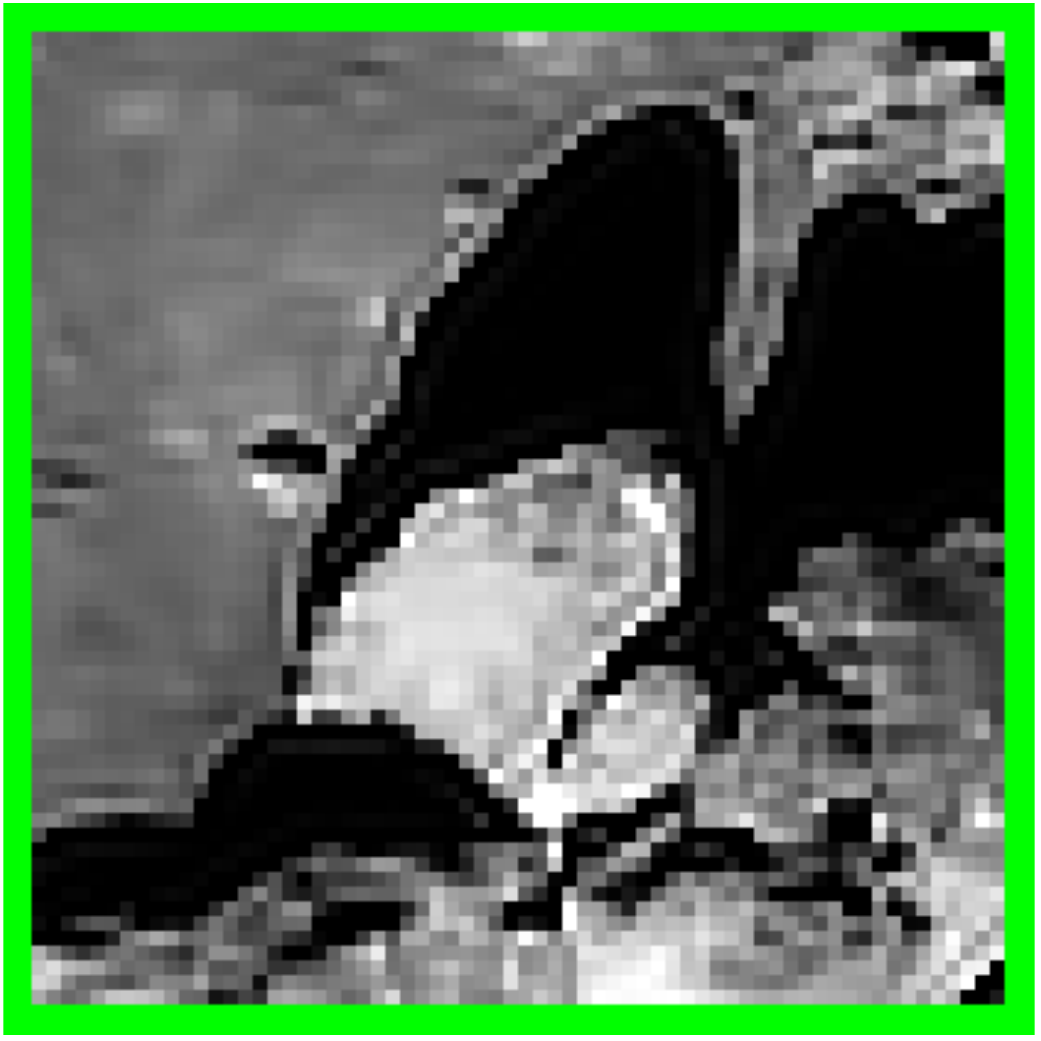}}}}\label{fig_LROC_IDD_BM3D_category_001_image_number_148_Scale_25_Gaussian}}
}\vspace{-.05in}
\centerline{
\setbox1=\hbox{\includegraphics[height=4cm]{image_scan_blurry_category_001_image_number_148_patch}}
\subfigure[1Shot-MaxPol]{\includegraphics[height=4cm]{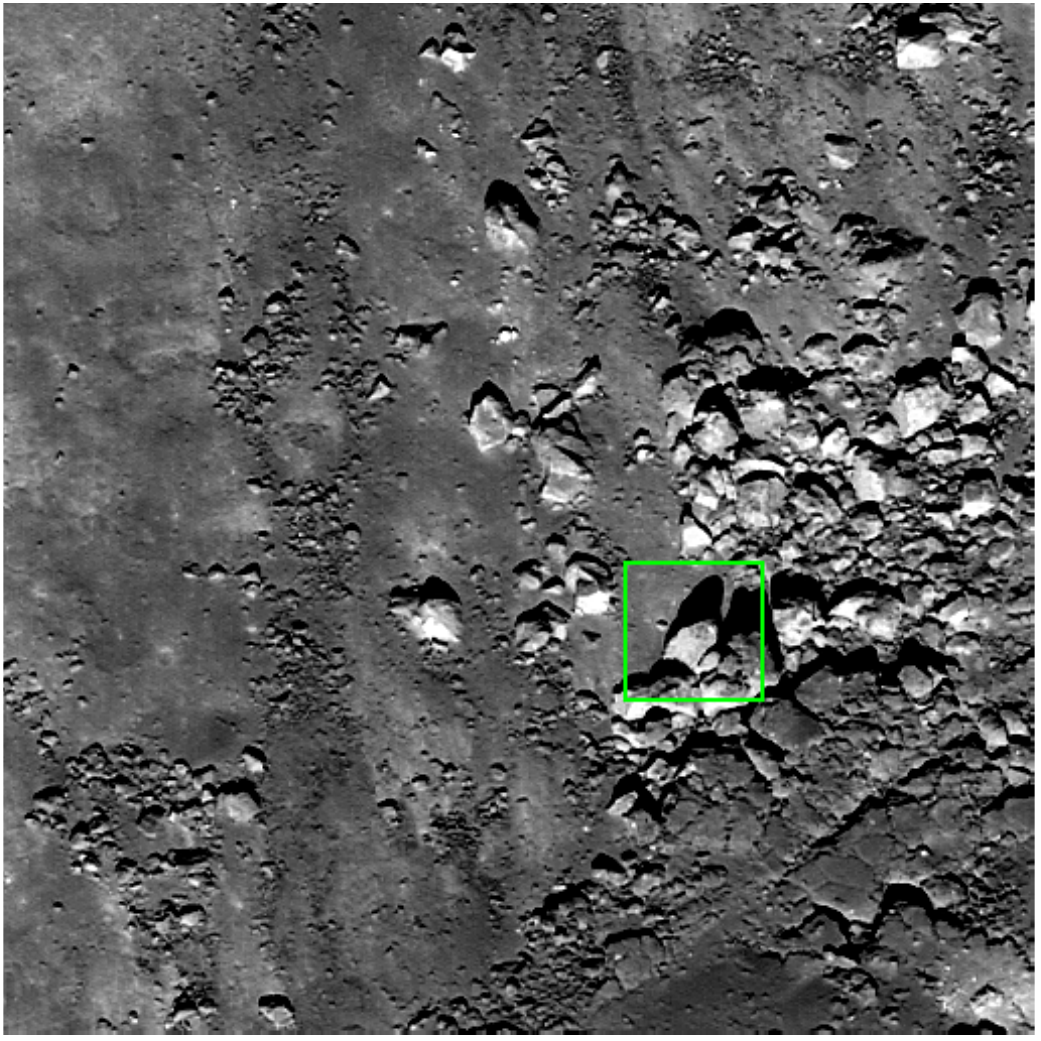}\llap{\makebox[\wd1][l]{\raisebox{2cm}{\includegraphics[height=2cm]{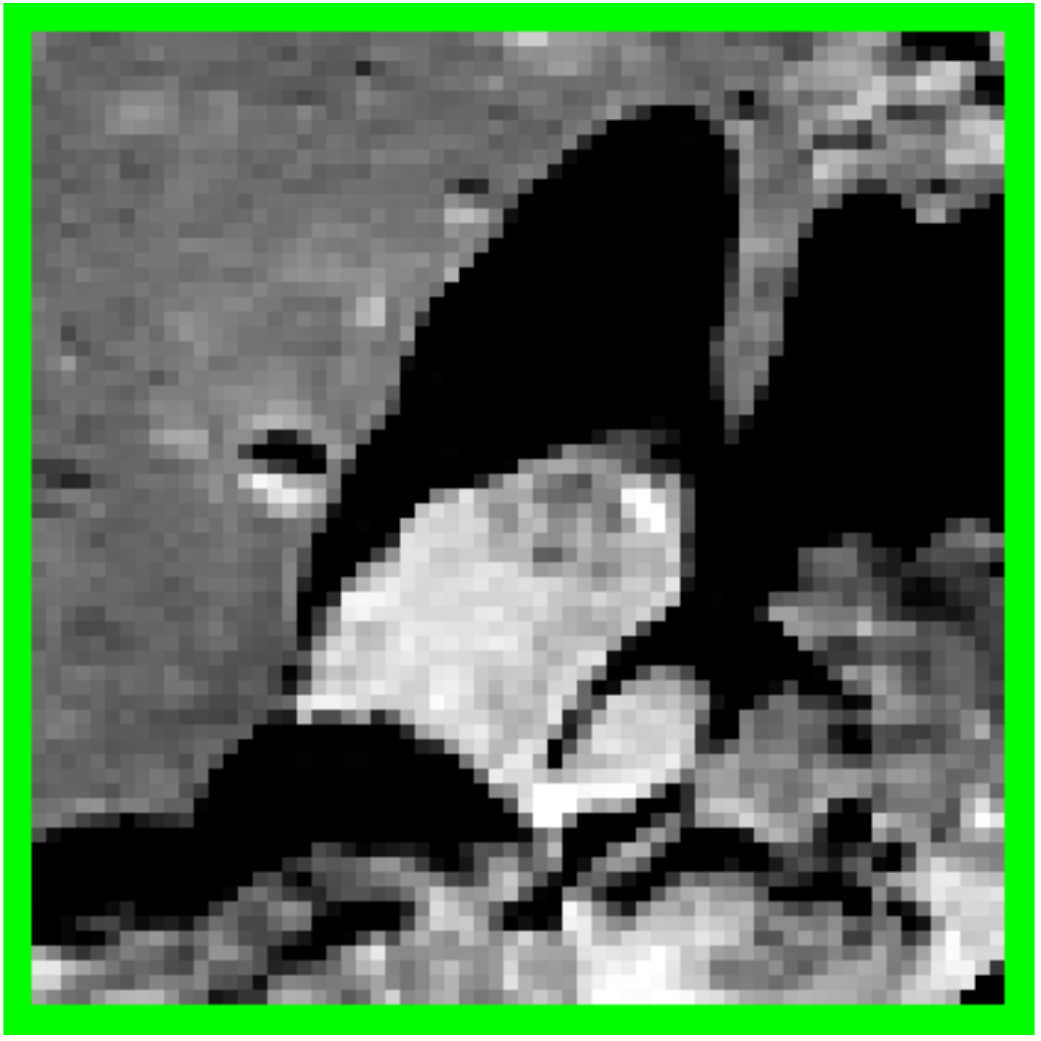}}}}\label{fig_LROC_MaxPol_category_001_image_number_148_Scale_25_Gaussian}}
\subfigure[Original]{\includegraphics[height=4cm]{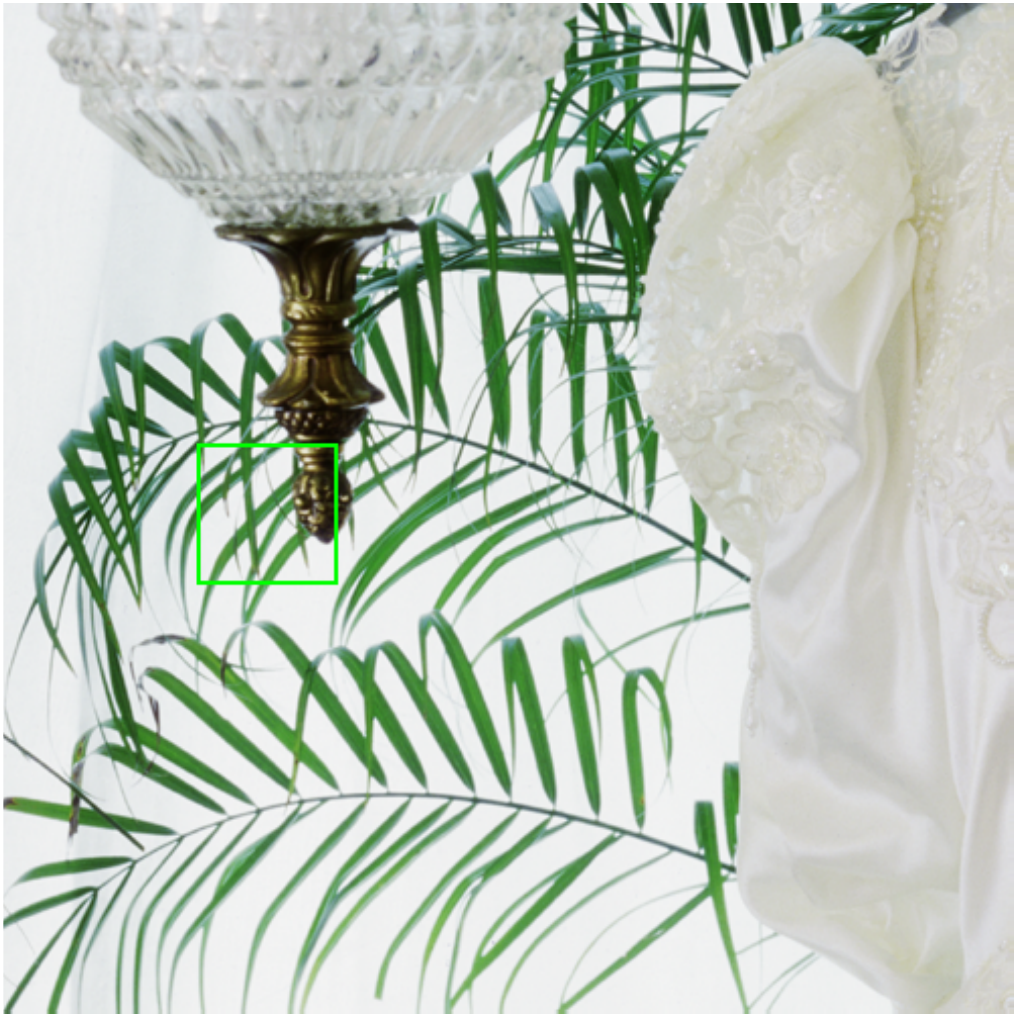}\llap{\makebox[\wd1][r]{\raisebox{0cm}{\includegraphics[height=2cm]{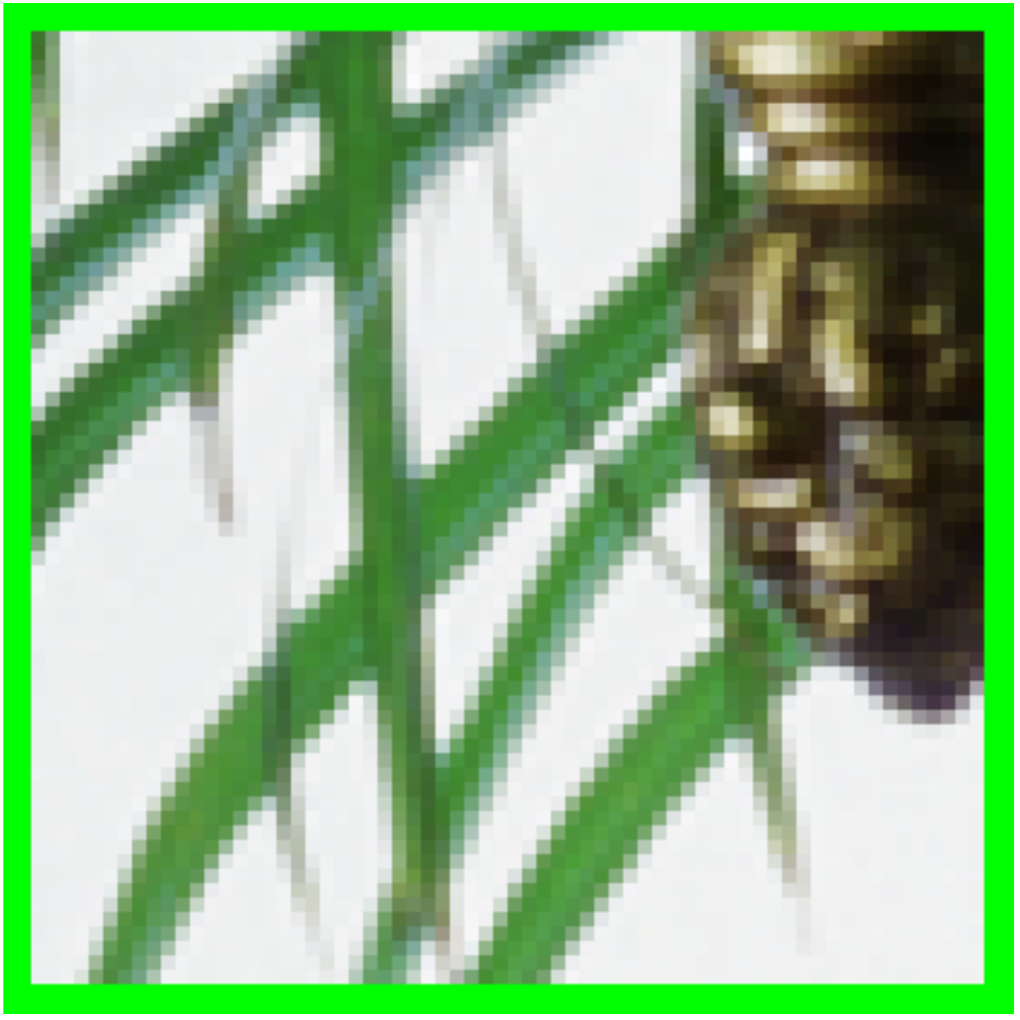}}}}\label{fig_LROC_raw_category_002_image_number_013_Scale_25_Gaussian}}
\subfigure[Krishnan \cite{krishnan2009fast,krishnan2011blind}]{\includegraphics[height=4cm]{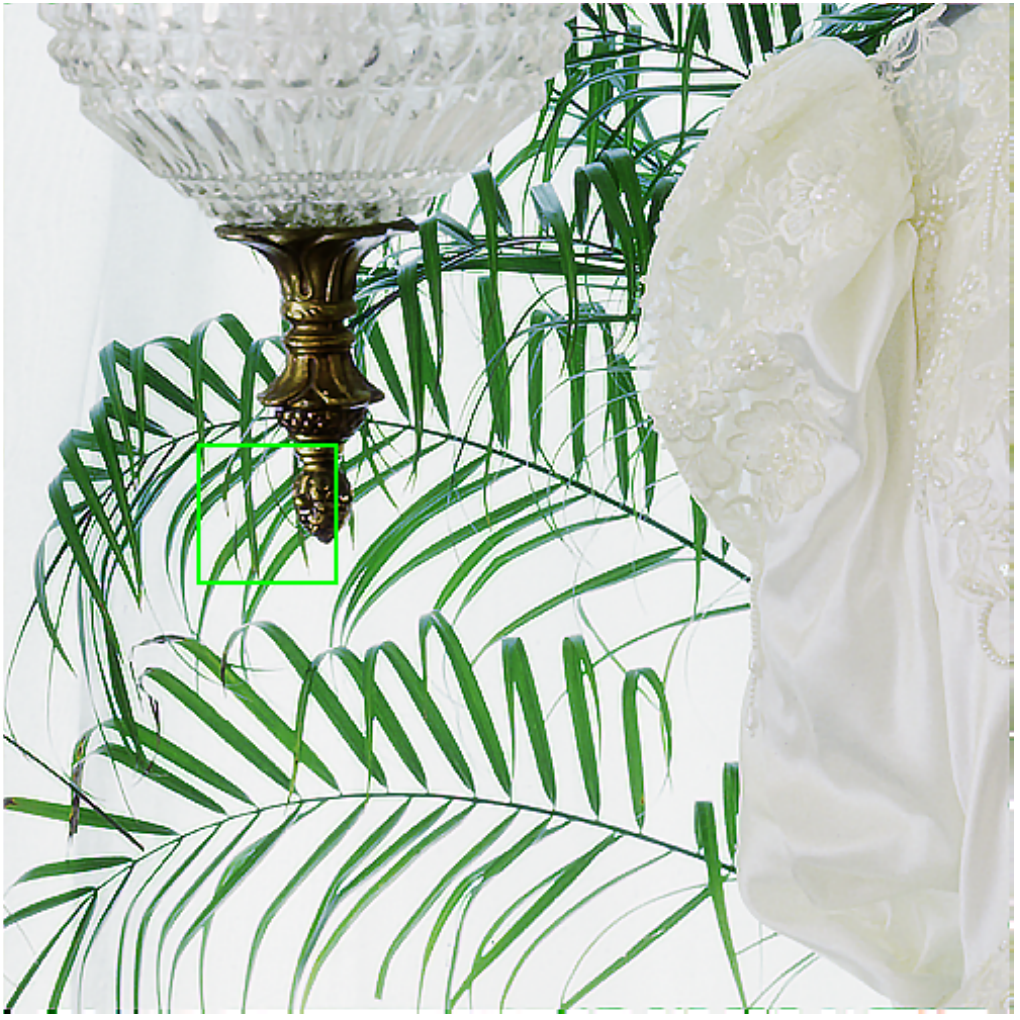}\llap{\makebox[\wd1][r]{\raisebox{0cm}{\includegraphics[height=2cm]{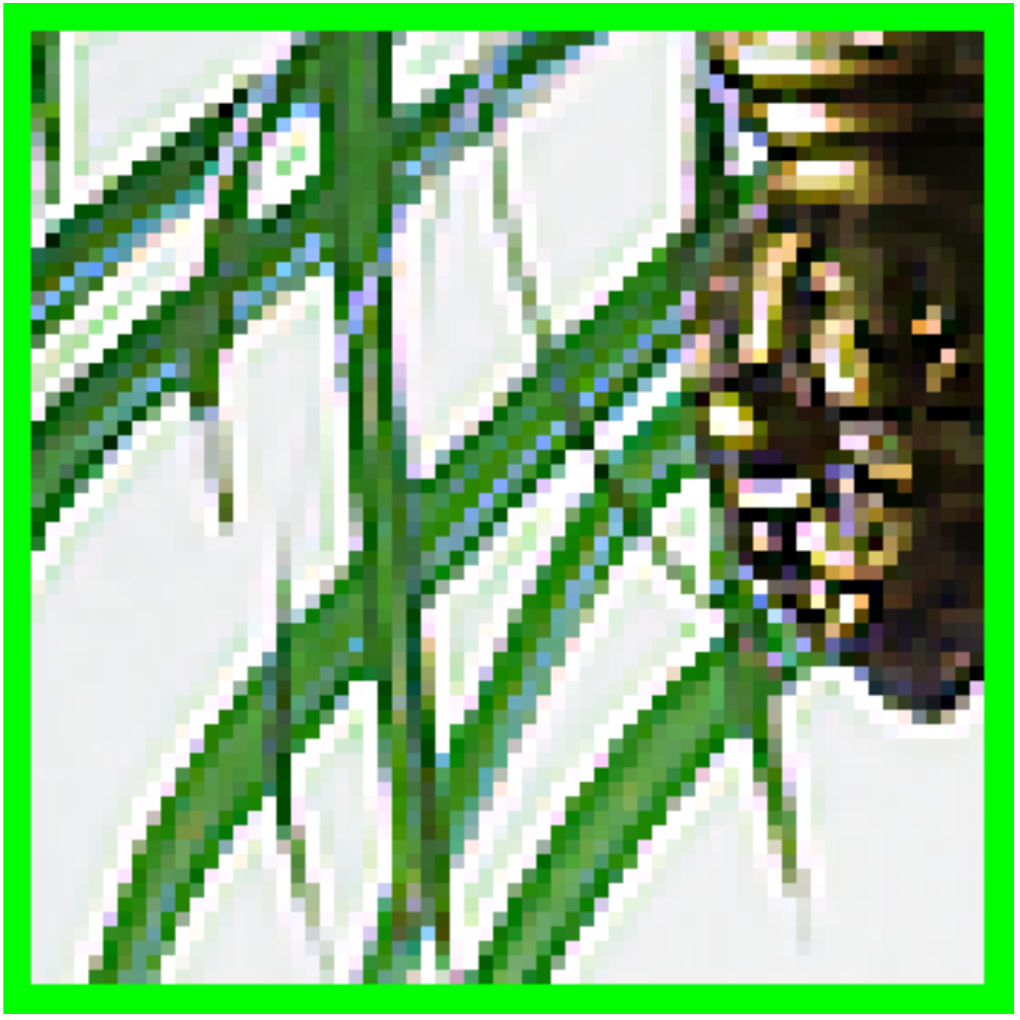}}}}\label{fig_LROC_Krishnan_category_002_image_number_013_Scale_25_Gaussian}}
\subfigure[Chan-PlugPlay \cite{chan2017plug}]{\includegraphics[height=4cm]{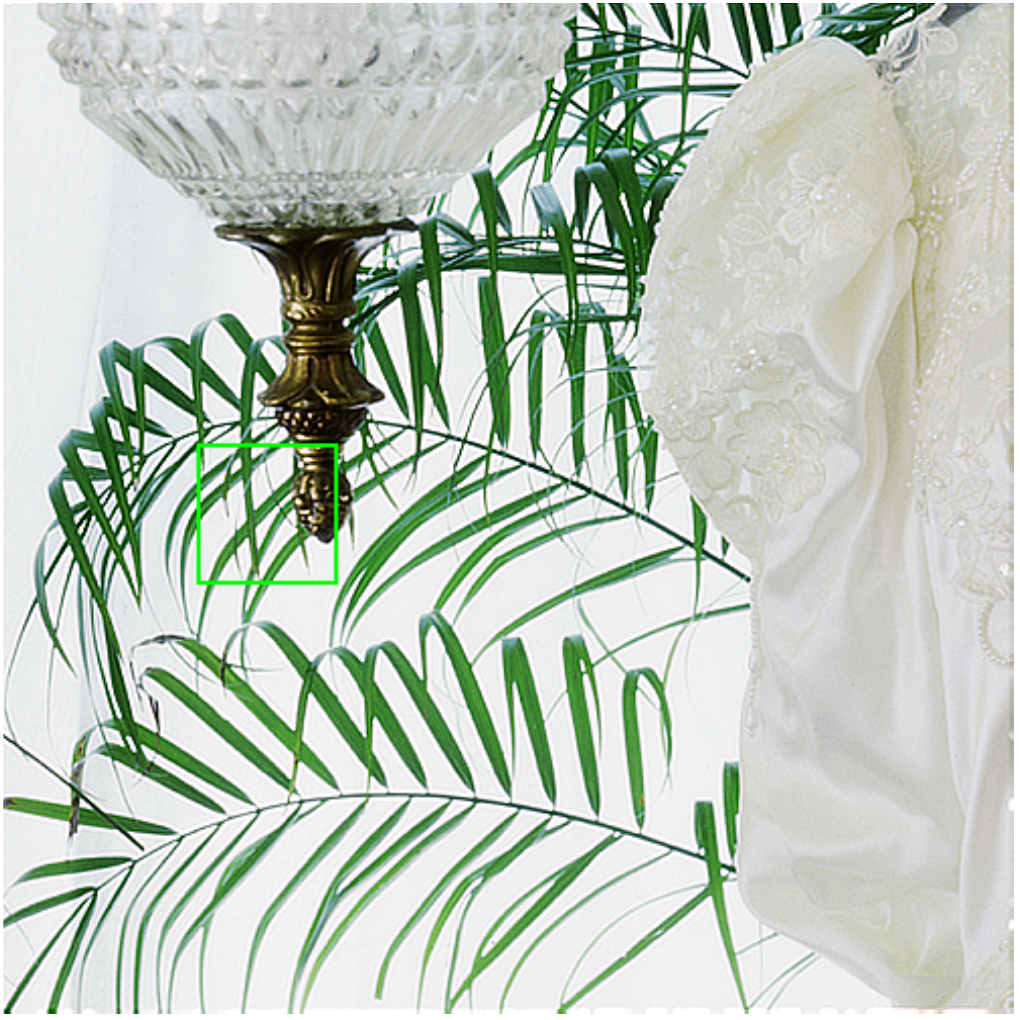}\llap{\makebox[\wd1][r]{\raisebox{0cm}{\includegraphics[height=2cm]{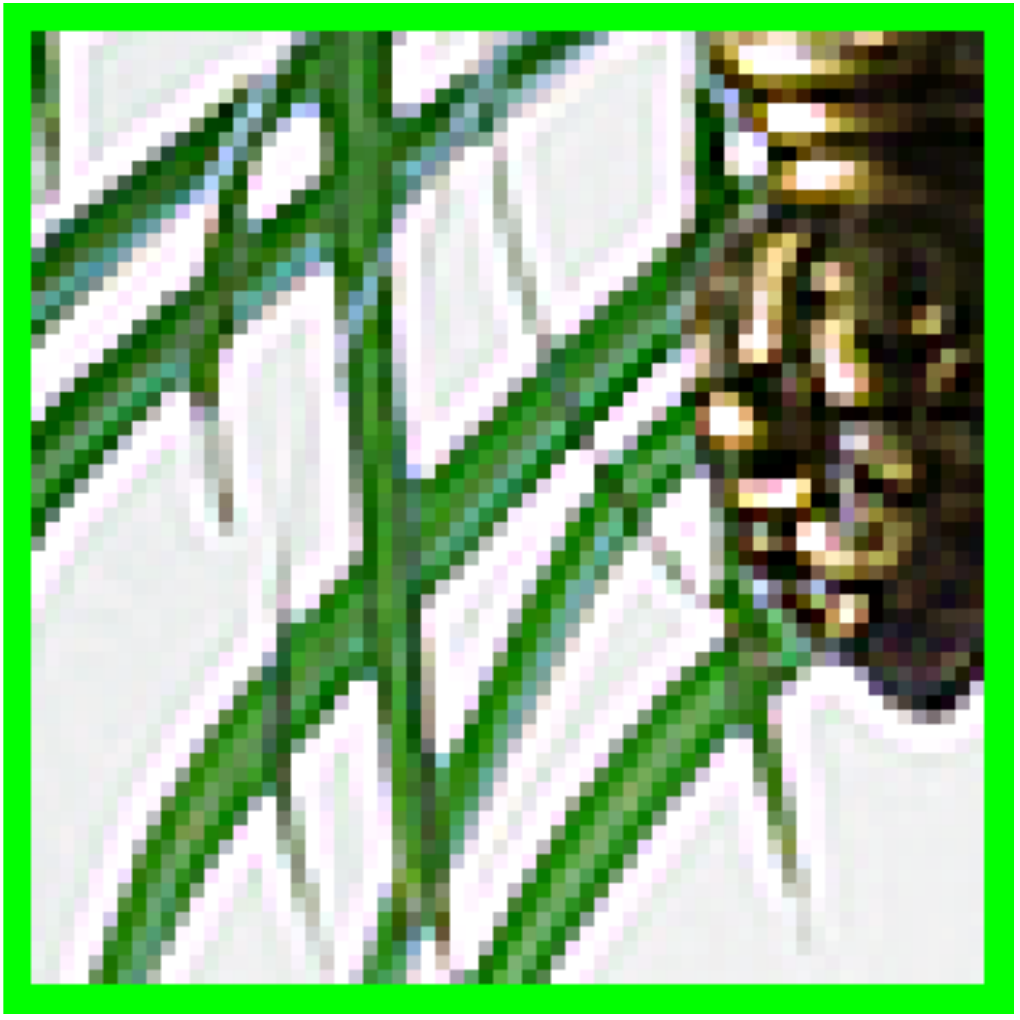}}}}\label{fig_LROC_Chan_PlugPlay_category_002_image_number_013_Scale_25_Gaussian}}
}\vspace{-.05in}
\centerline{
\subfigure[MLP \cite{schuler2013machine}]{\includegraphics[height=4cm]{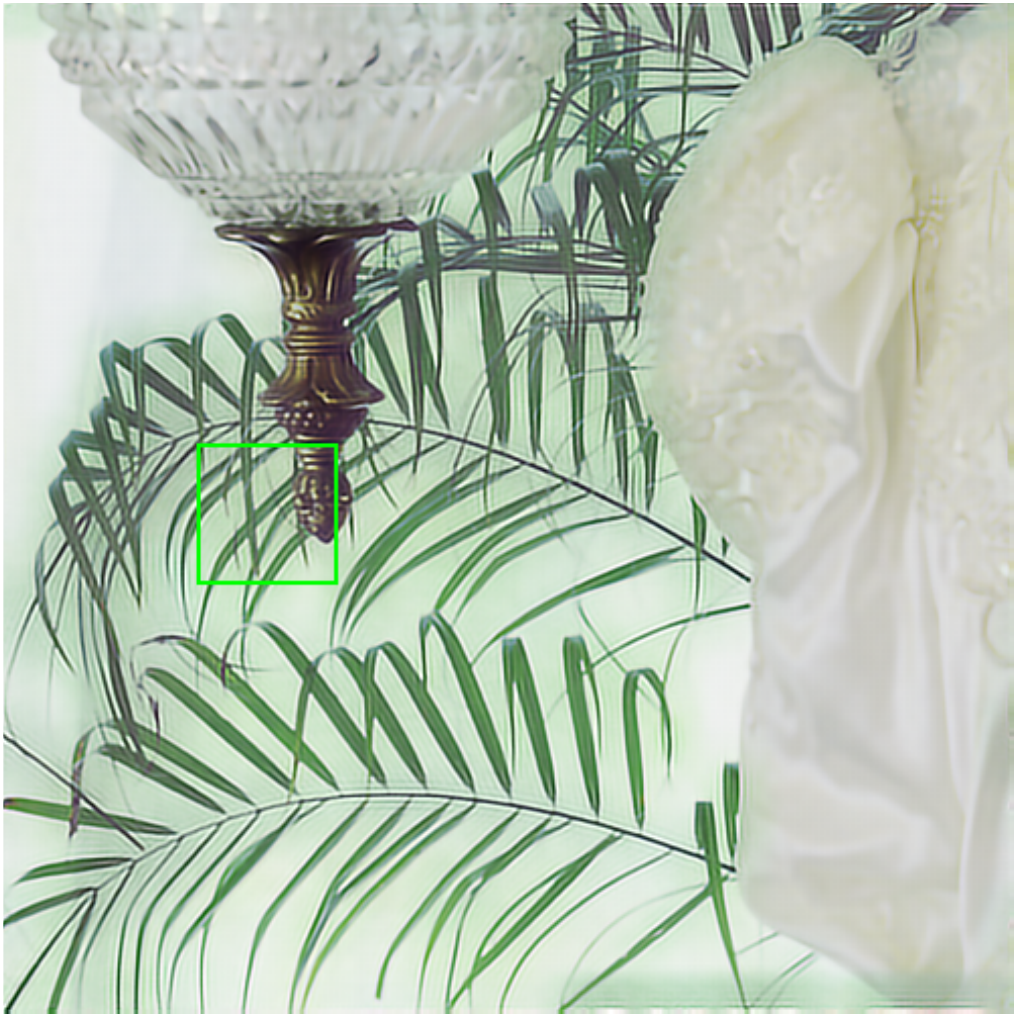}\llap{\makebox[\wd1][r]{\raisebox{0cm}{\includegraphics[height=2cm]{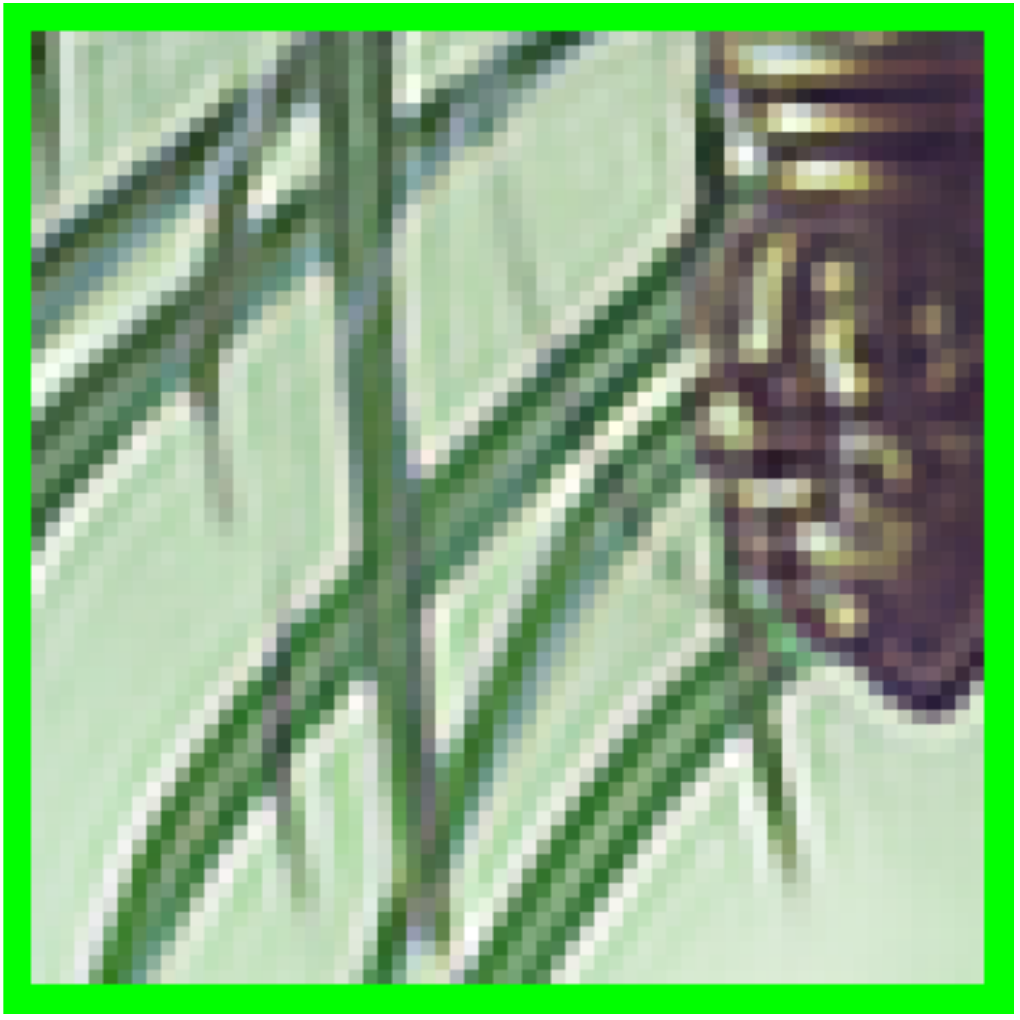}}}}\label{fig_LROC_MLP_category_002_image_number_013_Scale_25_Gaussian}}
\subfigure[Chan-DeconvTV \cite{chan2011augmented}]{\includegraphics[height=4cm]{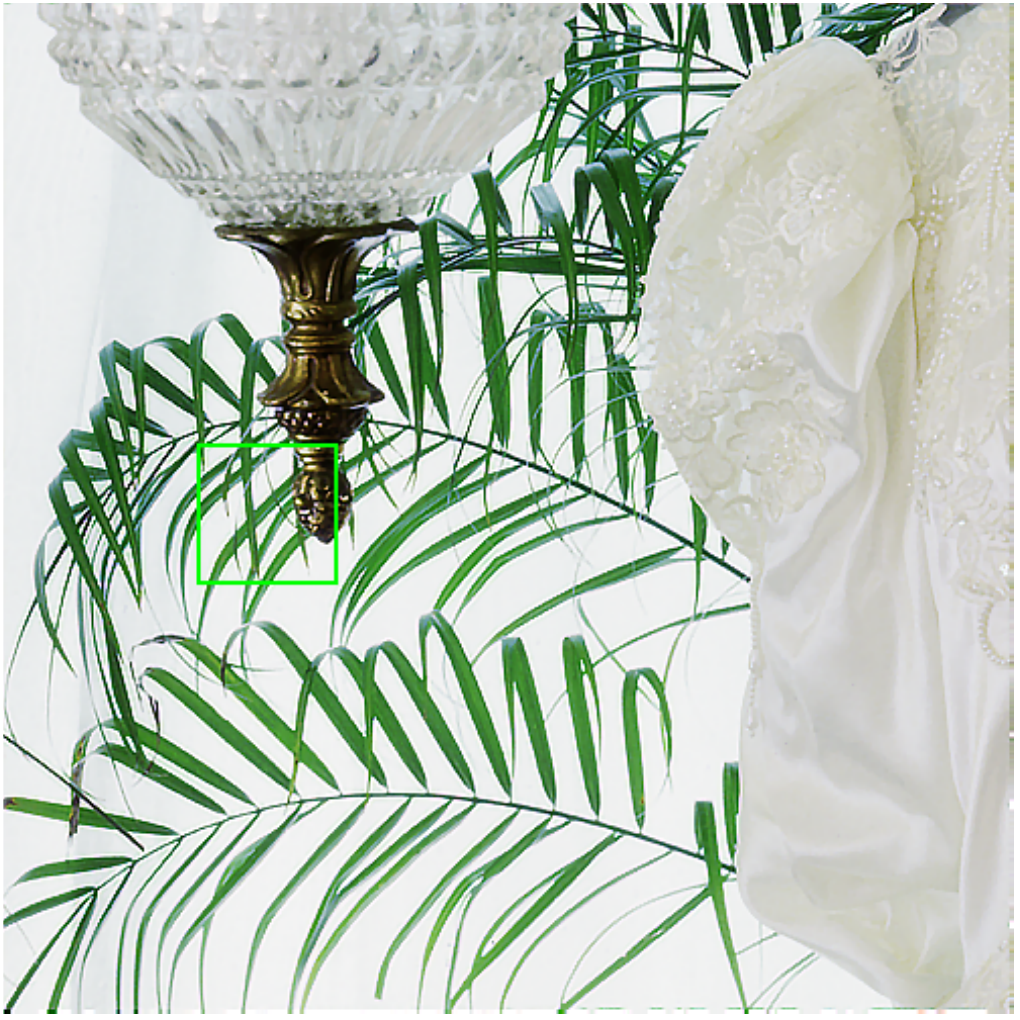}\llap{\makebox[\wd1][r]{\raisebox{0cm}{\includegraphics[height=2cm]{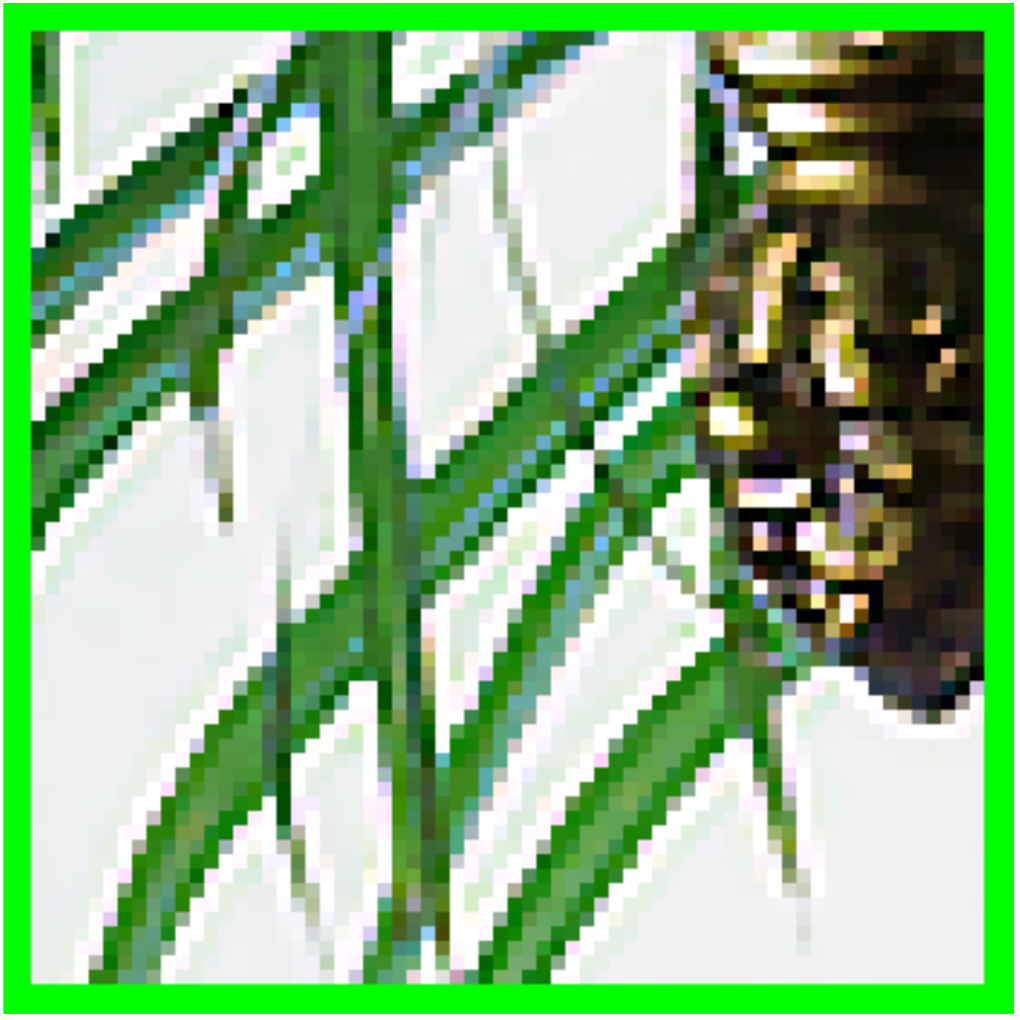}}}}\label{fig_Chan_DeconvTV_category_002_image_number_013_Scale_25_Gaussian}}\setbox1=\hbox{\includegraphics[height=4cm]{image_scan_blurry_category_002_image_number_013_patch}}
\subfigure[IRCNN \cite{zhang2017learning}]{\includegraphics[height=4cm]{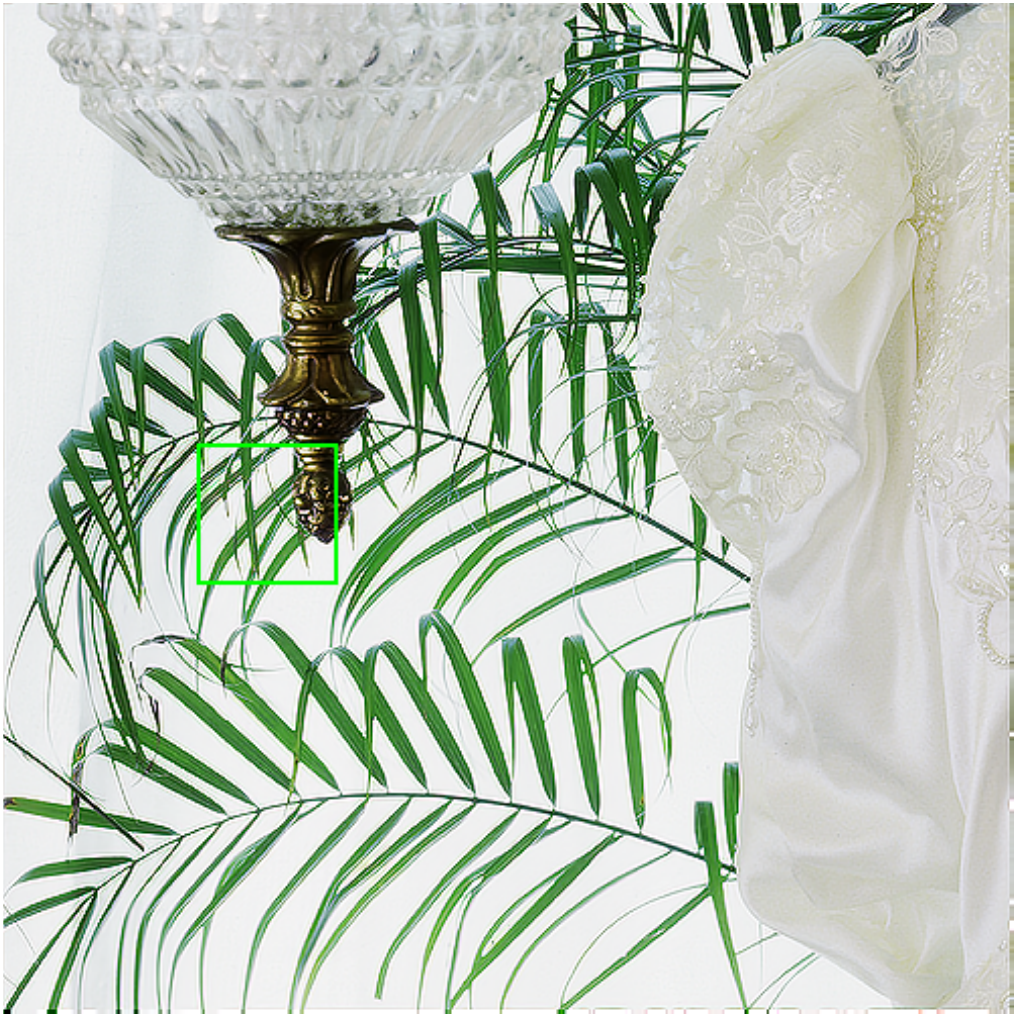}\llap{\makebox[\wd1][r]{\raisebox{0cm}{\includegraphics[height=2cm]{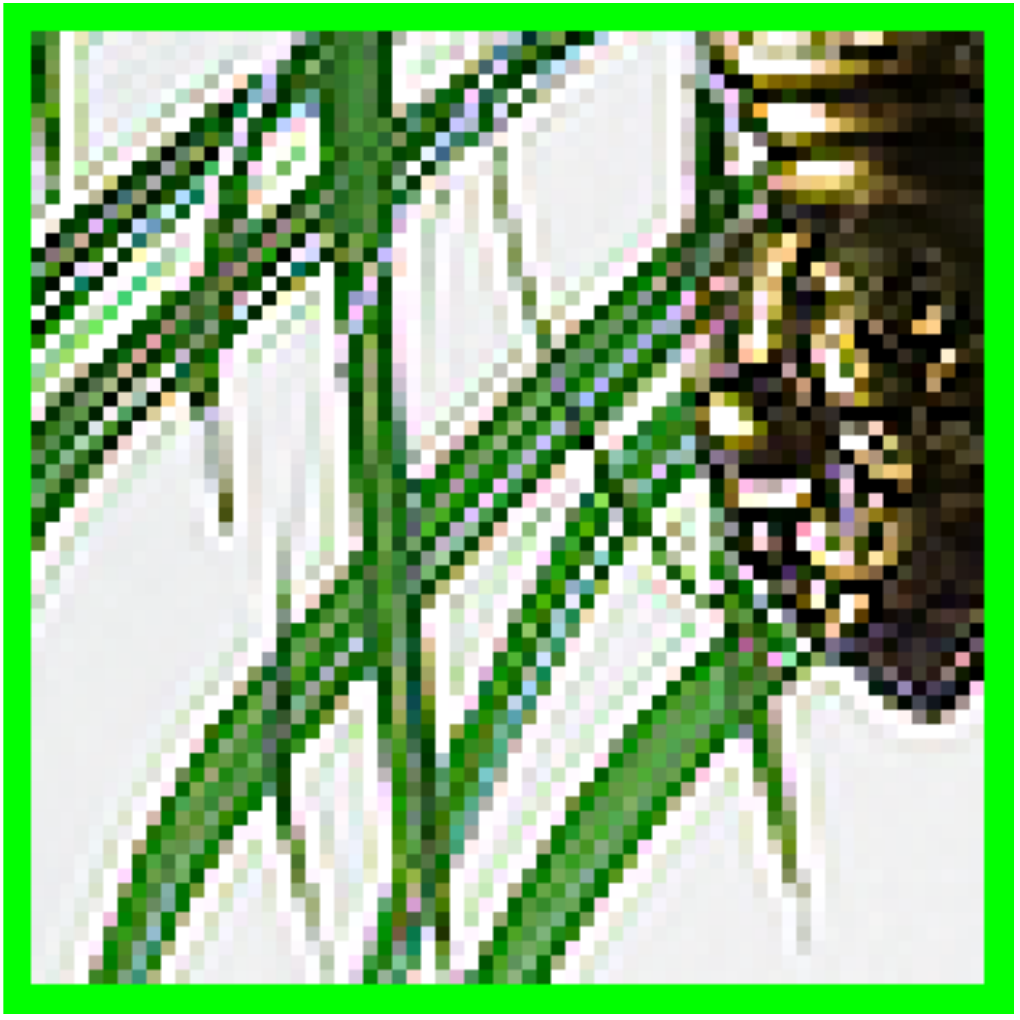}}}}\label{fig_LROC_IRCNN_category_002_image_number_013_Scale_25_Gaussian}}
\subfigure[Simoes \cite{simoes2016framework}]{\includegraphics[height=4cm]{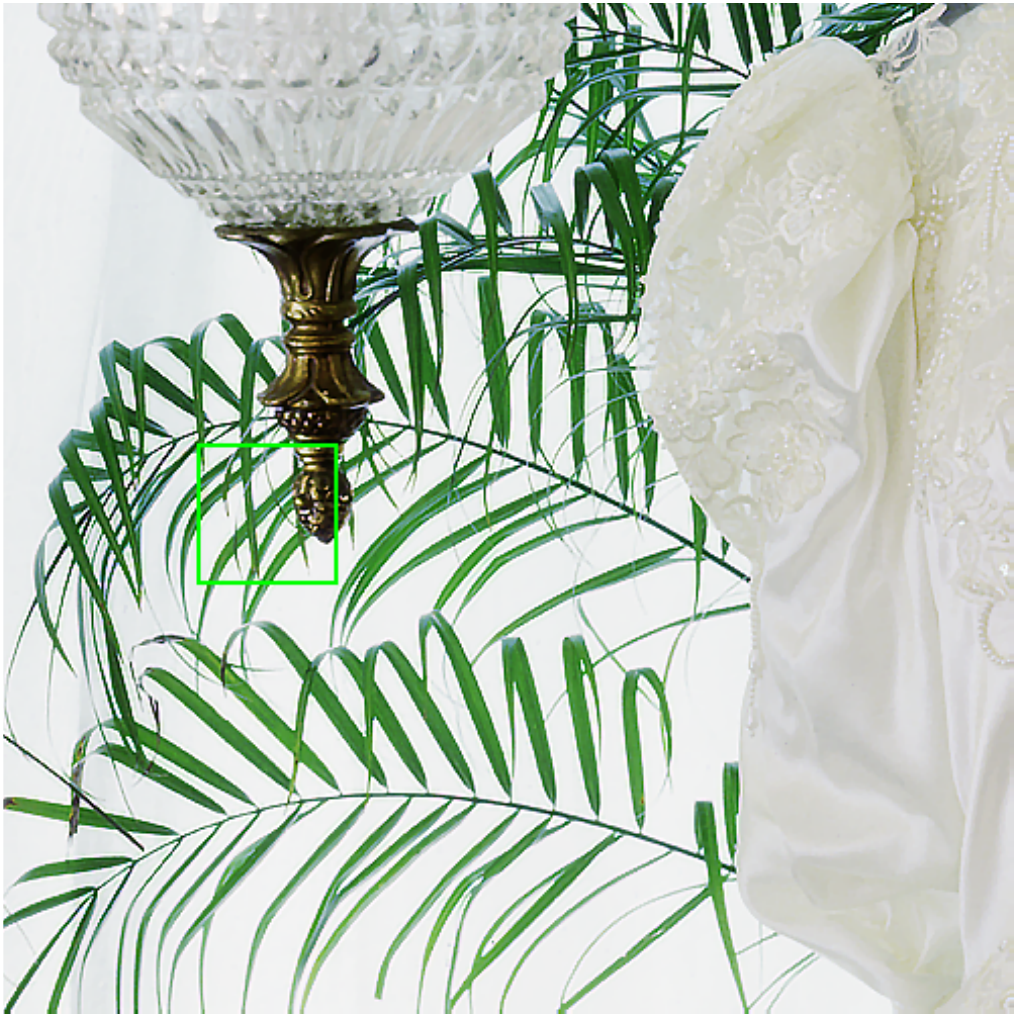}\llap{\makebox[\wd1][r]{\raisebox{0cm}{\includegraphics[height=2cm]{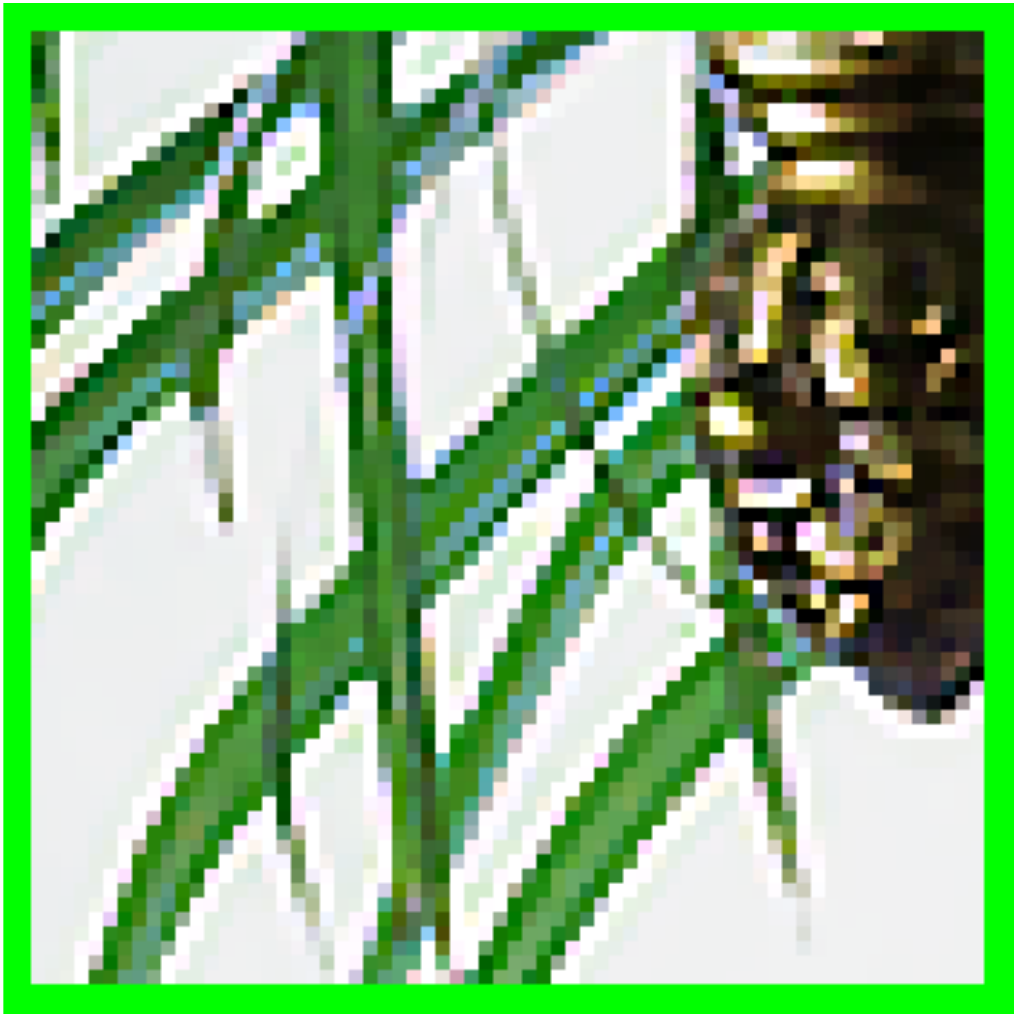}}}}\label{fig_LROC_Simoes_category_002_image_number_013_Scale_25_Gaussian}}
}\vspace{-.05in}
\centerline{
\subfigure[IDD-BM3D \cite{danielyan2012bm3d}]{\includegraphics[height=4cm]{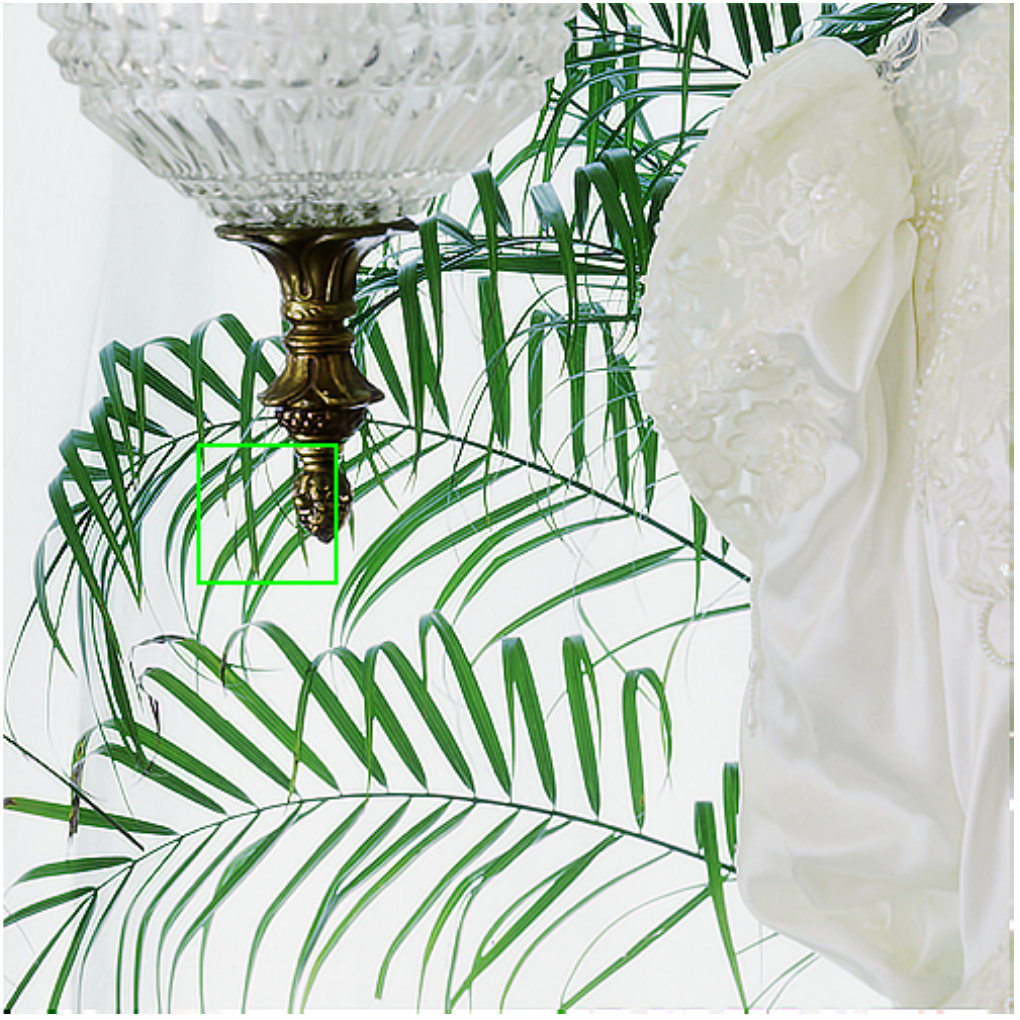}\llap{\makebox[\wd1][r]{\raisebox{0cm}{\includegraphics[height=2cm]{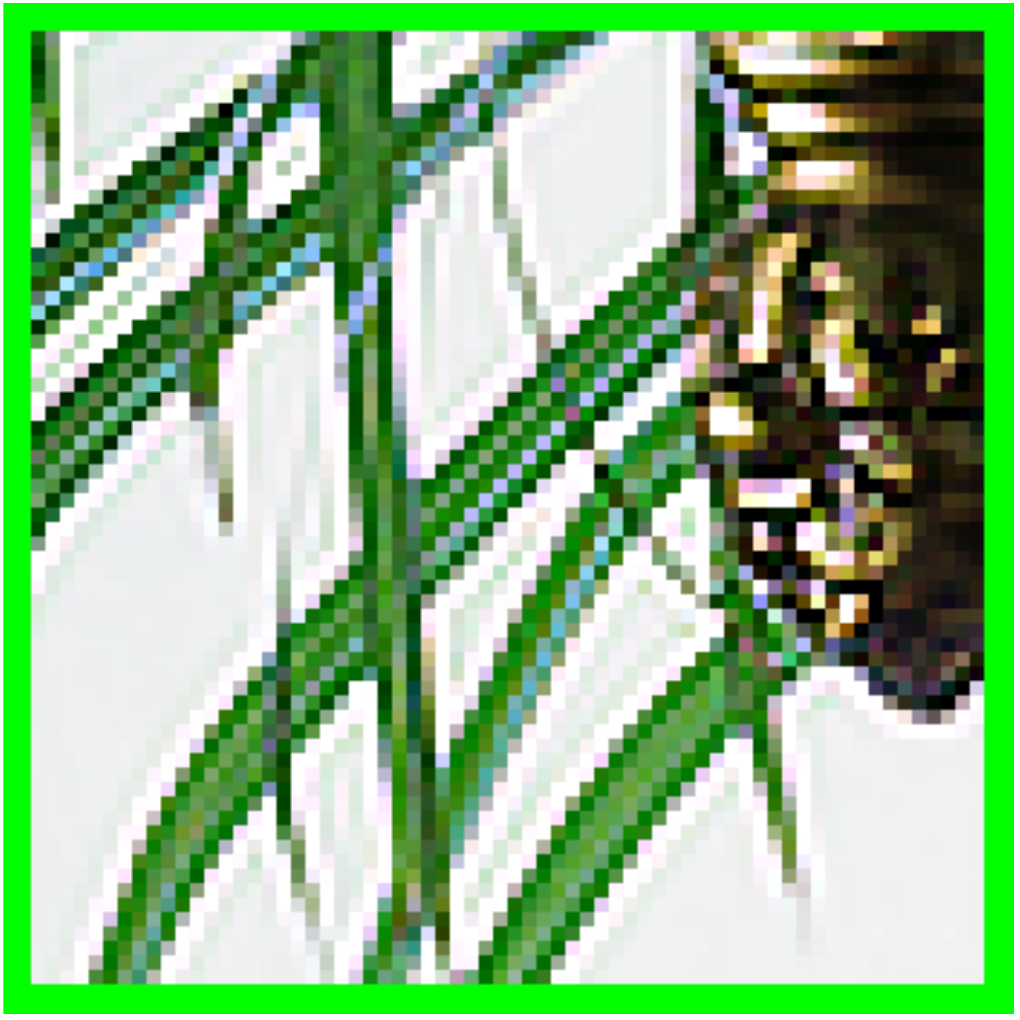}}}}\label{fig_LROC_IDD_BM3D_category_002_image_number_013_Scale_25_Gaussian}}
\subfigure[1Shot-MaxPol]{\includegraphics[height=4cm]{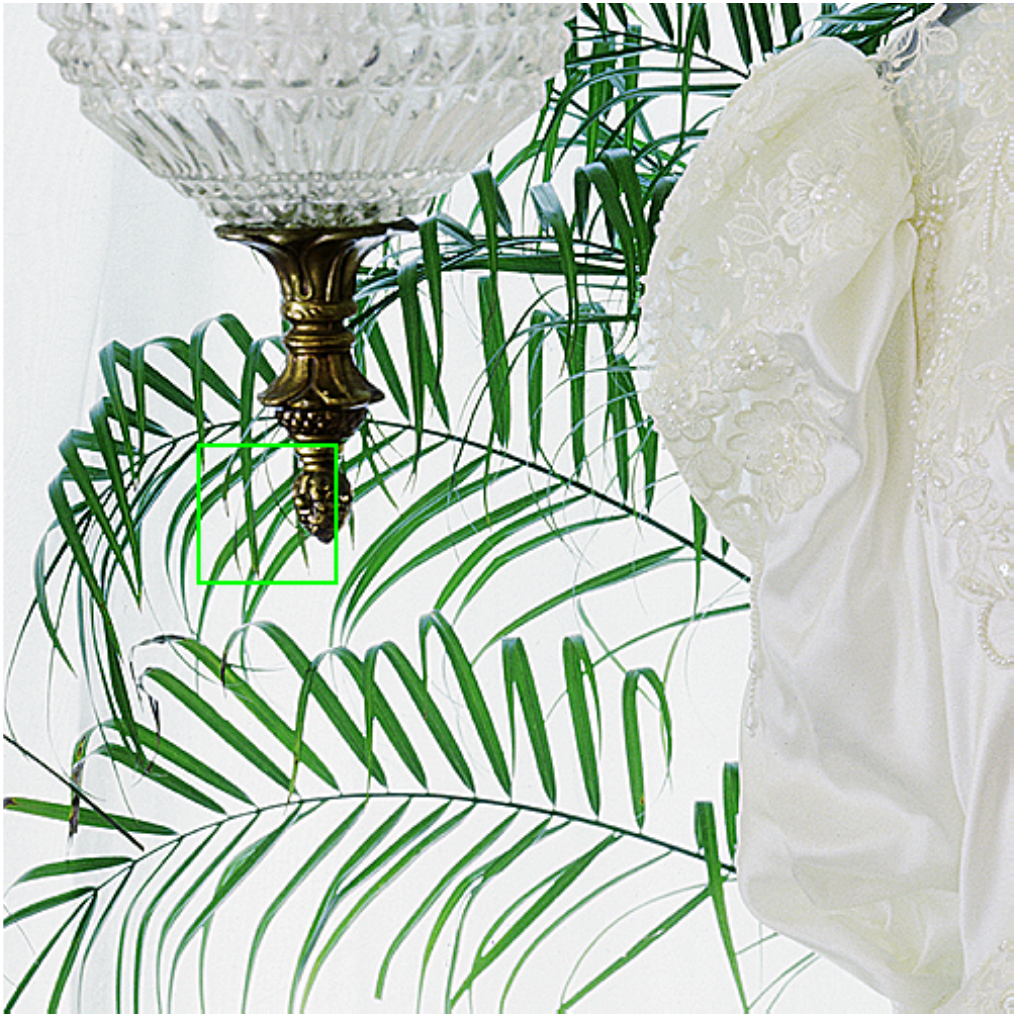}\llap{\makebox[\wd1][r]{\raisebox{0cm}{\includegraphics[height=2cm]{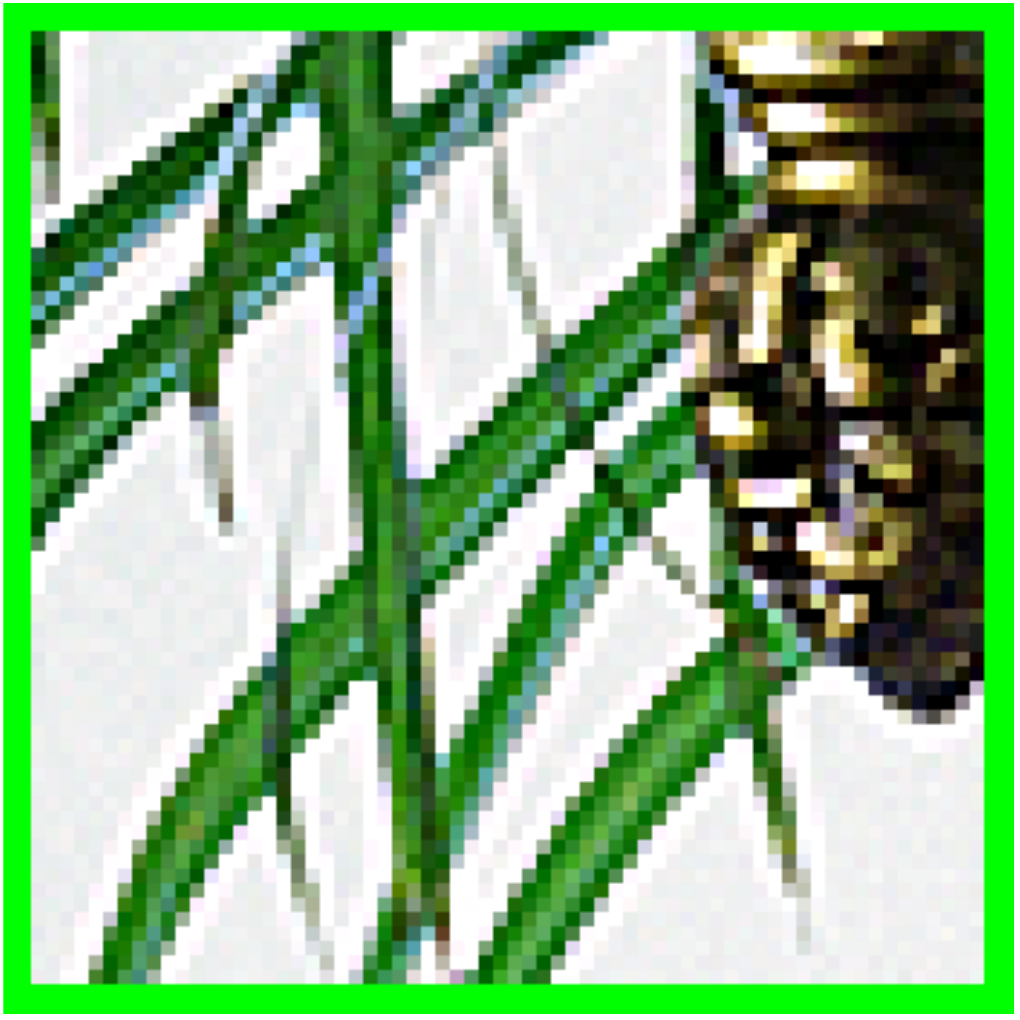}}}}\label{fig_LROC_MaxPol_category_002_image_number_013_Scale_25_Gaussian}}}\caption{Deblurring examples of LROC and McMaster images using different methods. For better visual comparison, please turn off image-smoothing option from Adobe Acrobat view software.}
\label{fig_McMaster_deblurring_examples_category_002_image_number_013_Scale_25_Gaussian}
\end{figure*}

\subsection{Performance Analysis}
Figures \ref{fig_NR_FQA_boxplot} demonstrates the NR-FQA analysis for each deblurring method using Gaussian and Laplacian blind PSF estimation, respectively. The associated PSFs are blindly estimated for both Gaussian and Laplacian models with scale factor $s=2$. The statistical distribution of the scores are shown in box-plots corresponding to $25$ and $75$ percentiles to exclude the outliers (gray circles) and the median scores are shown as red lines overlaid on the box-plot. The figure also includes the original image scores performed on six different categories of the databases explained in Section \ref{sec_natural_databases}. The median NR-FQA for the original LROC, NIR, and Hypersepctral databases are relatively low (high blur) compared to the RGB-Scene, McMaster, and Haze databases. The overall performance of all methods is also demonstrated in Table \ref{table_overall_performance} across different PSF models, where the top three performing methods are shown in bold numbers. Note that 1Shot-MaxPol and IDD-BM3D outrank the other methods for all four different blur models, where IRCNN and Chan-PlugPlay rank in third place for the Gaussian and  Laplacian, respectively. The majority of methods provide reasonable performance on Gaussian models, but provide inferior performances to 1Shot-MaxPol using Laplacian blur. This is simply because the majority of the deblurring methods do not generalize for diverse blurring models for recovery, e.g. IRCNN is trained using only Gaussian blur. In fact, this validates the error analysis on inverse deblurring kernel for 1Shot-MaxPol shown in Figure \ref{fig_GG_feasibility_range}, where the proposed method can be generalized into different blur shapes for deconvolution.

Figures \ref{fig_McMaster_deblurring_examples_category_002_image_number_013_Scale_25_Gaussian} demonstrate the deblurring examples on LROC and McMaster images. For more deblurring examples, please refer to the ``Deblurring Image Examples'' Section in the supplementary materials this paper. Our general observation across many image examples is that detailed recovery using 1Shot-MaxPol is much wider than with other techniques. In particular, this is more noticeable on smooth image areas where methods such as Krishnan, Chan-DeconvTV, IDD-BM3D, Simoes, and IRCNN washout the image details and provide an artificial image look on smoothed regions. Methods such as Chan-PlugPlay, MLP,  IDD-BM3D, and IRCNN also introduce ringing artifacts on contrasting edges. Overall, 1Shot-MaxPol avoids such deficiencies and recovers sharp details with a more natural look.

\subsection{Computational Complexity Analysis}
\begin{figure}[htp]
\centerline{
\subfigure[Computation Speed]{\includegraphics[scale=.6]{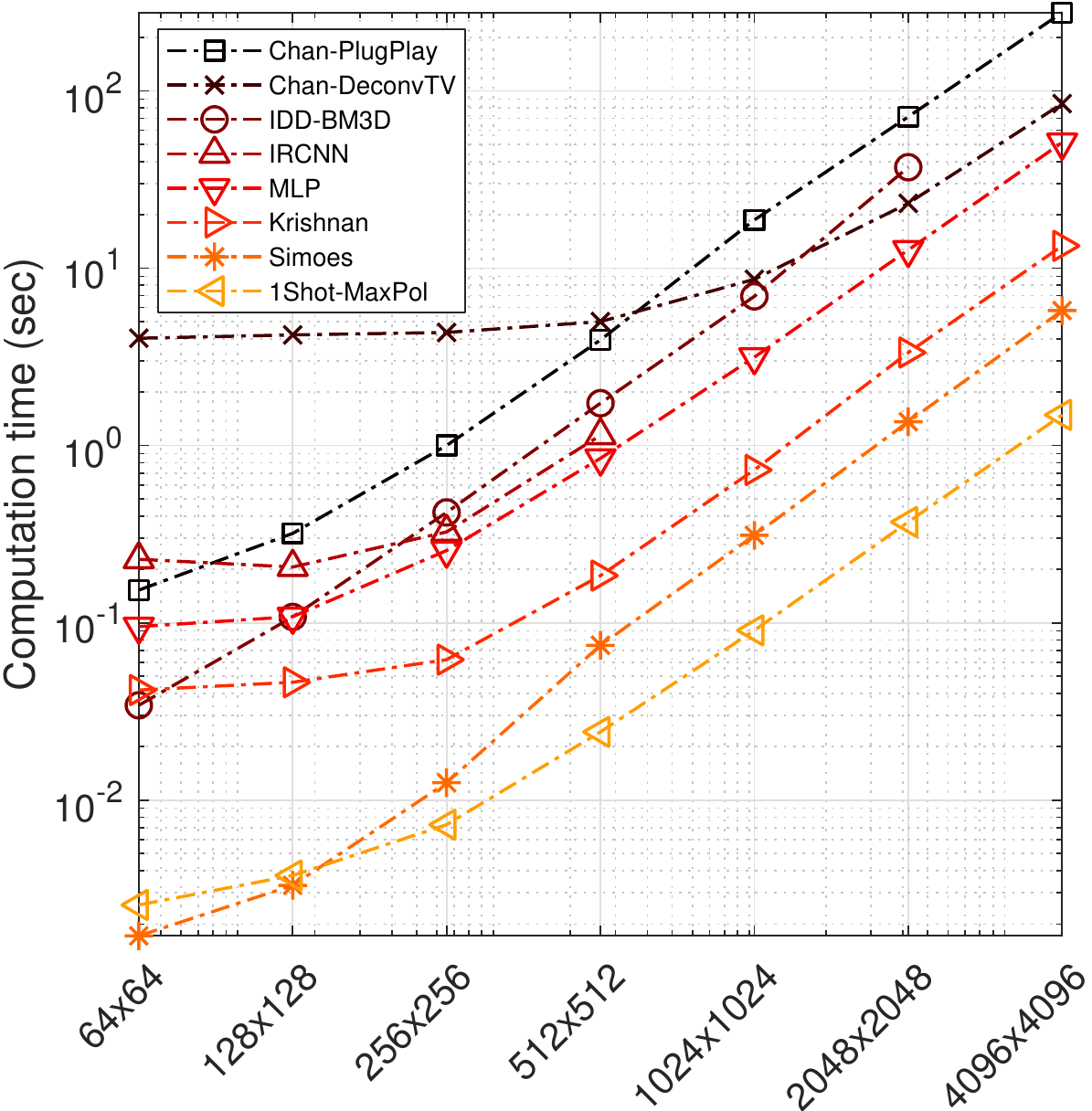}\label{fig_computational_speed}}
}
\centerline{
\subfigure[Performance vs computation]{\includegraphics[scale=.6]{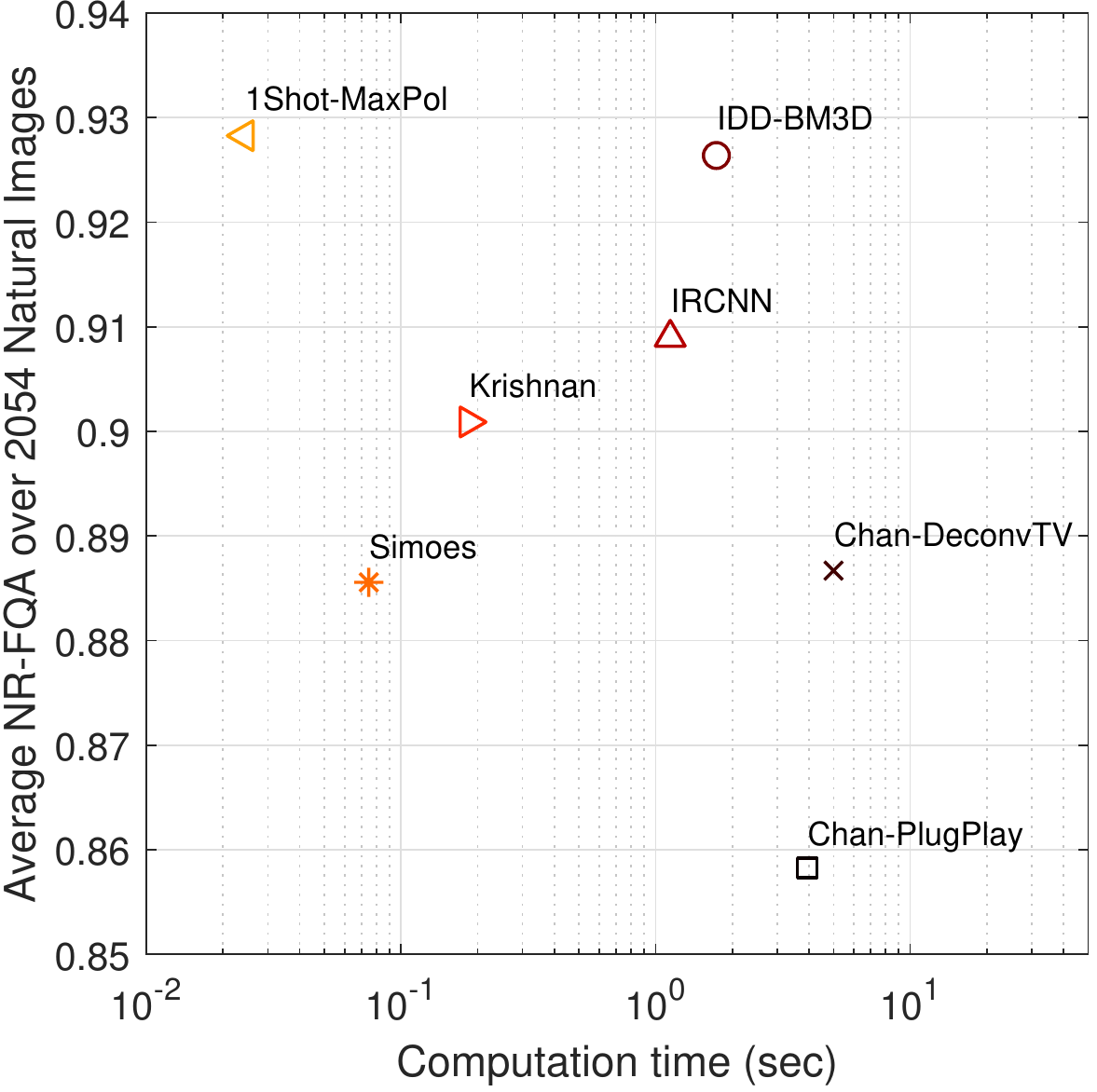}\label{fig_computational_acc_speed}}
}
\caption{Computational complexity analysis of all comparison methods: (a) progression of speed by increasing image size, and (b) NRF-QA comparison vs computation time using an image size of $512\times 512$.}
\label{fig_computational_complexity}
\end{figure}

As one of the main objectives of this paper was to develop a fast deblurring method (while maintaining good performance accuracy), we are keen to analyze the computational complexity of different deblurring methods used in this paper for comparison. We design two sets of experiments to investigate this. First, we analyze the CPU time versus different image sizes for reconstruction. For the CPU time measure, all the experiments were conducted on a Windows station with an AMD FX-8370E 8-Core CPU 3.30 GHz. Figure \ref{fig_computational_speed} demonstrates this complexity, where 1Shot-MaxPol outranks the second and the third top methods, i.e. Simoes and Krishnan, respectively. For instance, 1Shot-MaxPol is $3.43$ and $8.03$ times faster than the second and the third methods on recovering an image tile of size $1024\times 1024$. We perform the second type of assessment by analyzing the computation speed versus average NR-FQA of different methods. This is shown in Figure \ref{fig_computational_acc_speed} where a large y-axis value indicates a high accuracy and a small x-axis value indicates low time consumption. Thus, an ideal method should be located at the top-left corner of the plot. Despite the fact that both 1Shot-MaxPol and IDD-BM3D provide the highest accuracy reconstruction, it worth noting that 1Shot-MaxPol is $71$ times faster than IDD-BM3D. This easily places our proposed method as one of the leading algorithms for optical deblurring in digital archiving applications.

\section{Conclusion}\label{sec_conclusion}
In this paper, we have introduced a novel deblurring method for natural images to correct optical blur (in symmetric form) caused by aberrations, defocus, and turbid medium. The method is called ``1Shot-MaxPol'' and cast as a one-shot convolution filter for blur correction. The merit of our design is the decoupling of the deblurring and denoising problems and the addressing of them individually. For the case of deblurring, we first blindly estimated the PSF statistics for blur modeling using the novel approach of scale-space analysis of the image blur in the Fourier domain. We then constructed an FIR filter kernel for natural image deblurring by casting its dual representation into the Fourier domain for inverse approximation. For the case of denoising, we offered two optional designs for cut-off frequency regulation for inverse deblurring kernel design and Gaussian filter to mitigate the noise effect on deblurred edges. We have gathered $2054$ natural images from six different databases available online that contained optical blur. Experimental results show that our deblurring method significantly outperforms the existing state-of-the-art methods in terms of no-reference focus quality assessment, visual perception error, and computational complexity.

\ifCLASSOPTIONcaptionsoff
  \newpage
\fi

\bibliographystyle{IEEEbib}
\bibliography{references}

\begin{thebibliography}{10}

\bibitem{mahajan1998optical}
Virendra~N Mahajan,
\newblock {\em Optical Imaging and Aberrations: Ray Geometrical Optics},
  vol.~45,
\newblock SPIE press, 1998.

\bibitem{joshi2008psf}
Neel Joshi, Richard Szeliski, and David~J Kriegman,
\newblock ``Psf estimation using sharp edge prediction,''
\newblock in {\em Computer Vision and Pattern Recognition, 2008. CVPR 2008.
  IEEE Conference on}. IEEE, 2008, pp. 1--8.

\bibitem{keller2006objective}
H~Ernst Keller,
\newblock ``Objective lenses for confocal microscopy,''
\newblock in {\em Handbook of biological confocal microscopy}, pp. 145--161.
  Springer, 2006.

\bibitem{bishop2012light}
Tom~E Bishop and Paolo Favaro,
\newblock ``The light field camera: Extended depth of field, aliasing, and
  superresolution,''
\newblock {\em IEEE Transactions on Pattern Analysis and Machine Intelligence},
  vol. 34, no. 5, pp. 972--986, 2012.

\bibitem{hufnagel1964modulation}
RE~Hufnagel and NR~Stanley,
\newblock ``Modulation transfer function associated with image transmission
  through turbulent media,''
\newblock {\em JOSA}, vol. 54, no. 1, pp. 52--61, 1964.

\bibitem{fried1966optical}
David~L Fried,
\newblock ``Optical resolution through a randomly inhomogeneous medium for very
  long and very short exposures,''
\newblock {\em JOSA}, vol. 56, no. 10, pp. 1372--1379, 1966.

\bibitem{narasimhan2002vision}
Srinivasa~G Narasimhan and Shree~K Nayar,
\newblock ``Vision and the atmosphere,''
\newblock {\em International Journal of Computer Vision}, vol. 48, no. 3, pp.
  233--254, 2002.

\bibitem{bonettini2008scaled}
Silvia Bonettini, Riccardo Zanella, and Luca Zanni,
\newblock ``A scaled gradient projection method for constrained image
  deblurring,''
\newblock {\em Inverse problems}, vol. 25, no. 1, pp. 015002, 2008.

\bibitem{bertero2009image}
Mario Bertero, Patrizia Boccacci, Gabriele Desider{\`a}, and G~Vicidomini,
\newblock ``Image deblurring with poisson data: from cells to galaxies,''
\newblock {\em Inverse Problems}, vol. 25, no. 12, pp. 123006, 2009.

\bibitem{zhu2013removing}
Xiang Zhu and Peyman Milanfar,
\newblock ``Removing atmospheric turbulence via space-invariant
  deconvolution,''
\newblock {\em IEEE transactions on pattern analysis and machine intelligence},
  vol. 35, no. 1, pp. 157--170, 2013.

\bibitem{zhu2015fast}
Qingsong Zhu, Jiaming Mai, Ling Shao, et~al.,
\newblock ``A fast single image haze removal algorithm using color attenuation
  prior.,''
\newblock {\em IEEE Trans. Image Processing}, vol. 24, no. 11, pp. 3522--3533,
  2015.

\bibitem{cannon1976blind}
Michael Cannon,
\newblock ``Blind deconvolution of spatially invariant image blurs with
  phase,''
\newblock {\em IEEE Transactions on Acoustics, Speech, and Signal Processing},
  vol. 24, no. 1, pp. 58--63, 1976.

\bibitem{savakis1993blur}
Andreas~E Savakis and H~Joel Trussell,
\newblock ``Blur identification by residual spectral matching,''
\newblock {\em IEEE Transactions on image processing}, vol. 2, no. 2, pp.
  141--151, 1993.

\bibitem{elder1998local}
James~H Elder and Steven~W Zucker,
\newblock ``Local scale control for edge detection and blur estimation,''
\newblock {\em IEEE Transactions on pattern analysis and machine intelligence},
  vol. 20, no. 7, pp. 699--716, 1998.

\bibitem{fergus2006removing}
Rob Fergus, Barun Singh, Aaron Hertzmann, Sam~T Roweis, and William~T Freeman,
\newblock ``Removing camera shake from a single photograph,''
\newblock in {\em ACM transactions on graphics (TOG)}. ACM, 2006, vol.~25, pp.
  787--794.

\bibitem{cho2009fast}
Sunghyun Cho and Seungyong Lee,
\newblock ``Fast motion deblurring,''
\newblock in {\em ACM Transactions on Graphics (TOG)}. ACM, 2009, vol.~28, p.
  145.

\bibitem{whyte2014deblurring}
Oliver Whyte, Josef Sivic, and Andrew Zisserman,
\newblock ``Deblurring shaken and partially saturated images,''
\newblock {\em International journal of computer vision}, vol. 110, no. 2, pp.
  185--201, 2014.

\bibitem{sun2015learning}
Jian Sun, Wenfei Cao, Zongben Xu, Jean Ponce, et~al.,
\newblock ``Learning a convolutional neural network for non-uniform motion blur
  removal.,''
\newblock in {\em CVPR}, 2015, pp. 769--777.

\bibitem{polesel2000image}
Andrea Polesel, Giovanni Ramponi, and V~John Mathews,
\newblock ``Image enhancement via adaptive unsharp masking,''
\newblock {\em IEEE transactions on image processing}, vol. 9, no. 3, pp.
  505--510, 2000.

\bibitem{deng2011generalized}
Guang Deng,
\newblock ``A generalized unsharp masking algorithm,''
\newblock {\em IEEE transactions on Image Processing}, vol. 20, no. 5, pp.
  1249--1261, 2011.

\bibitem{tao2018scale}
Xin Tao, Hongyun Gao, Yi~Wang, Xiaoyong Shen, Jue Wang, and Jiaya Jia,
\newblock ``Scale-recurrent network for deep image deblurring,''
\newblock {\em arXiv preprint arXiv:1802.01770}, 2018.

\bibitem{wang2018training}
Ruxin Wang and Dacheng Tao,
\newblock ``Training very deep cnns for general non-blind deconvolution,''
\newblock {\em IEEE Transactions on Image Processing}, 2018.

\bibitem{schuler2013machine}
Christian~J Schuler, Harold Christopher~Burger, Stefan Harmeling, and Bernhard
  Scholkopf,
\newblock ``A machine learning approach for non-blind image deconvolution,''
\newblock in {\em Proceedings of the IEEE Conference on Computer Vision and
  Pattern Recognition}, 2013, pp. 1067--1074.

\bibitem{schuler2016learning}
Christian~J Schuler, Michael Hirsch, Stefan Harmeling, and Bernhard
  Sch{\"o}lkopf,
\newblock ``Learning to deblur,''
\newblock {\em IEEE transactions on pattern analysis and machine intelligence},
  vol. 38, no. 7, pp. 1439--1451, 2016.

\bibitem{hradivs2015convolutional}
Michal Hradi{\v{s}}, Jan Kotera, Pavel Zemc{\'\i}k, and Filip {\v{S}}roubek,
\newblock ``Convolutional neural networks for direct text deblurring,''
\newblock in {\em Proceedings of BMVC}, 2015, vol.~10.

\bibitem{xu2014deep}
Li~Xu, Jimmy~SJ Ren, Ce~Liu, and Jiaya Jia,
\newblock ``Deep convolutional neural network for image deconvolution,''
\newblock in {\em Advances in Neural Information Processing Systems}, 2014, pp.
  1790--1798.

\bibitem{zhang2017learning}
Kai Zhang, Wangmeng Zuo, Shuhang Gu, and Lei Zhang,
\newblock ``Learning deep cnn denoiser prior for image restoration,''
\newblock in {\em Proceedings of the IEEE Conference on Computer Vision and
  Pattern Recognition}, 2017, pp. 3929--3938.

\bibitem{chan2017plug}
Stanley~H Chan, Xiran Wang, and Omar~A Elgendy,
\newblock ``Plug-and-play admm for image restoration: Fixed-point convergence
  and applications,''
\newblock {\em IEEE Transactions on Computational Imaging}, vol. 3, no. 1, pp.
  84--98, 2017.

\bibitem{gastal2011domain}
Eduardo~SL Gastal and Manuel~M Oliveira,
\newblock ``Domain transform for edge-aware image and video processing,''
\newblock in {\em ACM Transactions on Graphics (ToG)}. ACM, 2011, vol.~30,
  p.~69.

\bibitem{romano2017little}
Yaniv Romano, Michael Elad, and Peyman Milanfar,
\newblock ``The little engine that could: Regularization by denoising (red),''
\newblock {\em SIAM Journal on Imaging Sciences}, vol. 10, no. 4, pp.
  1804--1844, 2017.

\bibitem{danielyan2012bm3d}
Aram Danielyan, Vladimir Katkovnik, and Karen Egiazarian,
\newblock ``Bm3d frames and variational image deblurring,''
\newblock {\em IEEE Transactions on Image Processing}, vol. 21, no. 4, pp.
  1715--1728, 2012.

\bibitem{zoran2011learning}
Daniel Zoran and Yair Weiss,
\newblock ``From learning models of natural image patches to whole image
  restoration,''
\newblock in {\em Computer Vision (ICCV), 2011 IEEE International Conference
  on}. IEEE, 2011, pp. 479--486.

\bibitem{anwar2018image}
Saeed Anwar, Cong~Phuoc Huynh, and Fatih Porikli,
\newblock ``Image deblurring with a class-specific prior,''
\newblock {\em IEEE transactions on pattern analysis and machine intelligence},
  2018.

\bibitem{li2018learning}
Lerenhan Li, Jinshan Pan, Wei-Sheng Lai, Changxin Gao, Nong Sang, and
  Ming-Hsuan Yang,
\newblock ``Learning a discriminative prior for blind image deblurring,''
\newblock {\em arXiv preprint arXiv:1803.03363}, 2018.

\bibitem{simoes2016framework}
Miguel Sim{\~o}es, Luis~B Almeida, Jose Bioucas-Dias, and Jocelyn Chanussot,
\newblock ``A framework for fast image deconvolution with incomplete
  observations,''
\newblock {\em IEEE Transactions on Image Processing}, vol. 25, no. 11, pp.
  5266--5280, 2016.

\bibitem{kim2015generalized}
Tae~Hyun Kim and Kyoung~Mu Lee,
\newblock ``Generalized video deblurring for dynamic scenes,''
\newblock {\em arXiv preprint arXiv:1507.02438}, 2015.

\bibitem{liu2014blind}
Guangcan Liu, Shiyu Chang, and Yi~Ma,
\newblock ``Blind image deblurring using spectral properties of convolution
  operators,''
\newblock {\em IEEE Transactions on image processing}, vol. 23, no. 12, pp.
  5047--5056, 2014.

\bibitem{mosleh2014image}
Ali Mosleh, JM~Pierre Langlois, and Paul Green,
\newblock ``Image deconvolution ringing artifact detection and removal via psf
  frequency analysis,''
\newblock in {\em European Conference on Computer Vision}. Springer, 2014, pp.
  247--262.

\bibitem{pan2014deblurring}
Jinshan Pan, Zhe Hu, Zhixun Su, and Ming-Hsuan Yang,
\newblock ``Deblurring text images via l0-regularized intensity and gradient
  prior,''
\newblock in {\em Computer Vision and Pattern Recognition (CVPR), 2014 IEEE
  Conference on}. IEEE, 2014, pp. 2901--2908.

\bibitem{pan2013fast}
Jinshan Pan and Zhixun Su,
\newblock ``Fast $\ell^0$-regularized kernel estimation for robust motion
  deblurring,''
\newblock {\em IEEE Signal Processing Letters}, vol. 20, no. 9, pp. 841--844,
  2013.

\bibitem{kim2013dynamic}
Tae~Hyun Kim, Byeongjoo Ahn, and Kyoung~Mu Lee,
\newblock ``Dynamic scene deblurring,''
\newblock in {\em Computer Vision (ICCV), 2013 IEEE International Conference
  on}. IEEE, 2013, pp. 3160--3167.

\bibitem{shen2012spatially}
Chih-Tsung Shen, Wen-Liang Hwang, and Soo-Chang Pei,
\newblock ``Spatially-varying out-of-focus image deblurring with l1-2
  optimization and a guided blur map,''
\newblock in {\em Acoustics, Speech and Signal Processing (ICASSP), 2012 IEEE
  International Conference on}. IEEE, 2012, pp. 1069--1072.

\bibitem{sroubek2012robust}
Filip Sroubek and Peyman Milanfar,
\newblock ``Robust multichannel blind deconvolution via fast alternating
  minimization,''
\newblock {\em IEEE Transactions on Image processing}, vol. 21, no. 4, pp.
  1687--1700, 2012.

\bibitem{dong2011image}
Weisheng Dong, Lei Zhang, Guangming Shi, and Xiaolin Wu,
\newblock ``Image deblurring and super-resolution by adaptive sparse domain
  selection and adaptive regularization,''
\newblock {\em IEEE Transactions on Image Processing}, vol. 20, no. 7, pp.
  1838--1857, 2011.

\bibitem{zhang2011sparse}
Haichao Zhang, Jianchao Yang, Yanning Zhang, and Thomas~S Huang,
\newblock ``Sparse representation based blind image deblurring,''
\newblock in {\em Multimedia and Expo (ICME), 2011 IEEE International
  Conference on}. IEEE, 2011, pp. 1--6.

\bibitem{zhang2011close}
Haichao Zhang, Jianchao Yang, Yanning Zhang, Nasser~M Nasrabadi, and Thomas~S
  Huang,
\newblock ``Close the loop: Joint blind image restoration and recognition with
  sparse representation prior,''
\newblock in {\em Computer Vision (ICCV), 2011 IEEE International Conference
  on}. IEEE, 2011, pp. 770--777.

\bibitem{bai2018graph}
Yuanchao Bai, Gene Cheung, Xianming Liu, and Wen Gao,
\newblock ``Graph-based blind image deblurring from a single photograph,''
\newblock {\em arXiv preprint arXiv:1802.07929}, 2018.

\bibitem{lou2015weighted}
Yifei Lou, Tieyong Zeng, Stanley Osher, and Jack Xin,
\newblock ``A weighted difference of anisotropic and isotropic total variation
  model for image processing,''
\newblock {\em SIAM Journal on Imaging Sciences}, vol. 8, no. 3, pp.
  1798--1823, 2015.

\bibitem{zhang2014image}
Jian Zhang, Debin Zhao, Ruiqin Xiong, Siwei Ma, and Wen Gao,
\newblock ``Image restoration using joint statistical modeling in a
  space-transform domain,''
\newblock {\em IEEE Transactions on Circuits and Systems for Video Technology},
  vol. 24, no. 6, pp. 915--928, 2014.

\bibitem{xu2012depth}
Li~Xu and Jiaya Jia,
\newblock ``Depth-aware motion deblurring,''
\newblock in {\em Computational Photography (ICCP), 2012 IEEE International
  Conference on}. IEEE, 2012, pp. 1--8.

\bibitem{chan2011augmented}
Stanley~H Chan, Ramsin Khoshabeh, Kristofor~B Gibson, Philip~E Gill, and
  Truong~Q Nguyen,
\newblock ``An augmented lagrangian method for total variation video
  restoration,''
\newblock {\em IEEE Transactions on Image Processing}, vol. 20, no. 11, pp.
  3097--3111, 2011.

\bibitem{afonso2010fast}
Manya~V Afonso, Jos{\'e}~M Bioucas-Dias, and M{\'a}rio~AT Figueiredo,
\newblock ``Fast image recovery using variable splitting and constrained
  optimization,''
\newblock {\em IEEE transactions on image processing}, vol. 19, no. 9, pp.
  2345--2356, 2010.

\bibitem{li2018pure}
Jizhou Li, Florian Luisier, and Thierry Blu,
\newblock ``Pure-let image deconvolution,''
\newblock {\em IEEE Transactions on Image Processing}, vol. 27, no. 1, pp.
  92--105, 2018.

\bibitem{bertero2010discrepancy}
Mario Bertero, Patrizia Boccacci, Giorgio Talenti, Riccardo Zanella, and Luca
  Zanni,
\newblock ``A discrepancy principle for poisson data,''
\newblock {\em Inverse problems}, vol. 26, no. 10, pp. 105004, 2010.

\bibitem{wiener1949extrapolation}
Norbert Wiener, Norbert Wiener, Cyberneticist Mathematician, Norbert Wiener,
  Norbert Wiener, and Cybern{\'e}ticien Math{\'e}maticien,
\newblock ``Extrapolation, interpolation, and smoothing of stationary time
  series: with engineering applications,''
\newblock 1949.

\bibitem{gonzalez2012digital}
Rafael~C Gonzalez and Richard~E Woods,
\newblock ``Digital image processing,'' 2012.

\bibitem{xiao2016learning}
Lei Xiao, Jue Wang, Wolfgang Heidrich, and Michael Hirsch,
\newblock ``Learning high-order filters for efficient blind deconvolution of
  document photographs,''
\newblock in {\em European Conference on Computer Vision}. Springer, 2016, pp.
  734--749.

\bibitem{zuo2013generalized}
Wangmeng Zuo, Deyu Meng, Lei Zhang, Xiangchu Feng, and David Zhang,
\newblock ``A generalized iterated shrinkage algorithm for non-convex sparse
  coding,''
\newblock in {\em Computer Vision (ICCV), 2013 IEEE International Conference
  on}. IEEE, 2013, pp. 217--224.

\bibitem{krishnan2011blind}
Dilip Krishnan, Terence Tay, and Rob Fergus,
\newblock ``Blind deconvolution using a normalized sparsity measure,''
\newblock in {\em Computer Vision and Pattern Recognition (CVPR), 2011 IEEE
  Conference on}. IEEE, 2011, pp. 233--240.

\bibitem{dabov2008image}
Kostadin Dabov, Alessandro Foi, Vladimir Katkovnik, and Karen Egiazarian,
\newblock ``Image restoration by sparse 3d transform-domain collaborative
  filtering,''
\newblock in {\em Image Processing: Algorithms and Systems VI}. International
  Society for Optics and Photonics, 2008, vol. 6812, p. 681207.

\bibitem{neelamani2004forward}
Ramesh Neelamani, Hyeokho Choi, and Richard Baraniuk,
\newblock ``Forward: Fourier-wavelet regularized deconvolution for
  ill-conditioned systems,''
\newblock {\em IEEE Transactions on signal processing}, vol. 52, no. 2, pp.
  418--433, 2004.

\bibitem{schmidt2013discriminative}
Uwe Schmidt, Carsten Rother, Sebastian Nowozin, Jeremy Jancsary, and Stefan
  Roth,
\newblock ``Discriminative non-blind deblurring,''
\newblock in {\em Computer Vision and Pattern Recognition (CVPR), 2013 IEEE
  Conference on}. IEEE, 2013, pp. 604--611.

\bibitem{dong2013nonlocally}
Weisheng Dong, Lei Zhang, Guangming Shi, and Xin Li,
\newblock ``Nonlocally centralized sparse representation for image
  restoration,''
\newblock {\em IEEE Transactions on Image Processing}, vol. 22, no. 4, pp.
  1620--1630, 2013.

\bibitem{sun2013edge}
Libin Sun, Sunghyun Cho, Jue Wang, and James Hays,
\newblock ``Edge-based blur kernel estimation using patch priors,''
\newblock in {\em Computational photography (ICCP), 2013 IEEE international
  conference on}. IEEE, 2013, pp. 1--8.

\bibitem{whyte2012non}
Oliver Whyte, Josef Sivic, Andrew Zisserman, and Jean Ponce,
\newblock ``Non-uniform deblurring for shaken images,''
\newblock {\em International journal of computer vision}, vol. 98, no. 2, pp.
  168--186, 2012.

\bibitem{amizic2012sparse}
Bruno Amizic, Rafael Molina, and Aggelos~K Katsaggelos,
\newblock ``Sparse bayesian blind image deconvolution with parameter
  estimation,''
\newblock {\em EURASIP Journal on Image and Video Processing}, vol. 2012, no.
  1, pp. 20, 2012.

\bibitem{levin2011efficient}
Anat Levin, Yair Weiss, Fredo Durand, and William~T Freeman,
\newblock ``Efficient marginal likelihood optimization in blind
  deconvolution,''
\newblock in {\em Computer Vision and Pattern Recognition (CVPR), 2011 IEEE
  Conference on}. IEEE, 2011, pp. 2657--2664.

\bibitem{levin2011understanding}
Anat Levin, Yair Weiss, Fredo Durand, and William~T Freeman,
\newblock ``Understanding blind deconvolution algorithms,''
\newblock {\em IEEE transactions on pattern analysis and machine intelligence},
  vol. 33, no. 12, pp. 2354--2367, 2011.

\bibitem{krishnan2009fast}
Dilip Krishnan and Rob Fergus,
\newblock ``Fast image deconvolution using hyper-laplacian priors,''
\newblock in {\em Advances in Neural Information Processing Systems}, 2009, pp.
  1033--1041.

\bibitem{shan2008high}
Qi~Shan, Jiaya Jia, and Aseem Agarwala,
\newblock ``High-quality motion deblurring from a single image,''
\newblock in {\em Acm transactions on graphics (tog)}. ACM, 2008, vol.~27,
  p.~73.

\bibitem{yuan2007image}
Lu~Yuan, Jian Sun, Long Quan, and Heung-Yeung Shum,
\newblock ``Image deblurring with blurred/noisy image pairs,''
\newblock in {\em ACM Transactions on Graphics (TOG)}. ACM, 2007, vol.~26,
  p.~1.

\bibitem{biggs1997acceleration}
David~SC Biggs and Mark Andrews,
\newblock ``Acceleration of iterative image restoration algorithms,''
\newblock {\em Applied optics}, vol. 36, no. 8, pp. 1766--1775, 1997.

\bibitem{richardson1972bayesian}
William~Hadley Richardson,
\newblock ``Bayesian-based iterative method of image restoration,''
\newblock {\em JOSA}, vol. 62, no. 1, pp. 55--59, 1972.

\bibitem{lucy1974iterative}
Leon~B Lucy,
\newblock ``An iterative technique for the rectification of observed
  distributions,''
\newblock {\em The astronomical journal}, vol. 79, pp. 745, 1974.

\bibitem{geman1987stochastic}
Stuart Geman and Donald Geman,
\newblock ``Stochastic relaxation, gibbs distributions, and the bayesian
  restoration of images,''
\newblock in {\em Readings in Computer Vision}, pp. 564--584. Elsevier, 1987.

\bibitem{kundur1996blind}
Deepa Kundur and Dimitrios Hatzinakos,
\newblock ``Blind image deconvolution,''
\newblock {\em IEEE signal processing magazine}, vol. 13, no. 3, pp. 43--64,
  1996.

\bibitem{yuan2008progressive}
Lu~Yuan, Jian Sun, Long Quan, and Heung-Yeung Shum,
\newblock ``Progressive inter-scale and intra-scale non-blind image
  deconvolution,''
\newblock {\em ACM transactions on graphics (TOG)}, vol. 27, no. 3, pp. 74,
  2008.

\bibitem{dabov2007image}
Kostadin Dabov, Alessandro Foi, Vladimir Katkovnik, and Karen Egiazarian,
\newblock ``Image denoising by sparse 3-d transform-domain collaborative
  filtering,''
\newblock {\em IEEE Transactions on image processing}, vol. 16, no. 8, pp.
  2080--2095, 2007.

\bibitem{HosseiniPltaniotis_MaxPol_TIP_2017}
M.~S. Hosseini and K.~N. Plataniotis,
\newblock ``Derivative kernels: Numerics and applications,''
\newblock {\em IEEE Transactions on Image Processing}, vol. 26, no. 10, pp.
  4596--4611, Oct 2017.

\bibitem{HosseiniPltaniotis_MaxPol_SIAM_2017}
Mahdi~S. Hosseini and Konstantinos~N. Plataniotis,
\newblock ``Finite differences in forward and inverse imaging problems: Maxpol
  design,''
\newblock {\em SIAM Journal on Imaging Sciences}, vol. 10, no. 4, pp.
  1963--1996, 2017.

\bibitem{kelley1999iterative}
Carl~T Kelley,
\newblock {\em Iterative methods for optimization}, vol.~18,
\newblock Siam, 1999.

\bibitem{subbotin1923law}
M~Th Subbotin,
\newblock ``On the law of frequency of error,''
\newblock {\em MATEM. c6.}, vol. 31, no. 2, pp. 296--301, 1923.

\bibitem{kotz2012laplace}
Samuel Kotz, Tomasz Kozubowski, and Krzystof Podgorski,
\newblock {\em The Laplace distribution and generalizations: a revisit with
  applications to communications, economics, engineering, and finance},
\newblock Springer Science \& Business Media, 2012.

\bibitem{metari2007new}
Samy Metari and Fran{\c{c}}ois Deschenes,
\newblock ``A new convolution kernel for atmospheric point spread function
  applied to computer vision,''
\newblock in {\em Computer Vision, 2007. ICCV 2007. IEEE 11th International
  Conference on}. IEEE, 2007, pp. 1--8.

\bibitem{stallinga2010accuracy}
Sjoerd Stallinga and Bernd Rieger,
\newblock ``Accuracy of the gaussian point spread function model in 2d
  localization microscopy,''
\newblock {\em Optics express}, vol. 18, no. 24, pp. 24461--24476, 2010.

\bibitem{field1987relations}
David~J Field,
\newblock ``Relations between the statistics of natural images and the response
  properties of cortical cells,''
\newblock {\em Josa a}, vol. 4, no. 12, pp. 2379--2394, 1987.

\bibitem{brady1995s}
Nuala Brady and David~J Field,
\newblock ``What's constant in contrast constancy? the effects of scaling on
  the perceived contrast of bandpass patterns,''
\newblock {\em Vision research}, vol. 35, no. 6, pp. 739--756, 1995.

\bibitem{field1997visual}
David~J Field and Nuala Brady,
\newblock ``Visual sensitivity, blur and the sources of variability in the
  amplitude spectra of natural scenes,''
\newblock {\em Vision research}, vol. 37, no. 23, pp. 3367--3383, 1997.

\bibitem{dubeau2011fourier}
F~Dubeau and S~El~Mashoubi,
\newblock ``The fourier transform of the multidimentional generalized gaussian
  distribution,''
\newblock {\em International Journal of Pure and Applied Mathematics}, vol. 67,
  no. 4, pp. 443--454, 2011.

\bibitem{bahrami2014fast}
Khosro Bahrami and Alex~C. Kot,
\newblock ``A fast approach for no-reference image sharpness assessment based
  on maximum local variation,''
\newblock {\em IEEE Signal Processing Letters}, vol. 21, no. 6, pp. 751--755,
  2014.

\bibitem{robinson2010lunar}
MS~Robinson, SM~Brylow, M~Tschimmel, D~Humm, SJ~Lawrence, PC~Thomas, BW~Denevi,
  E~Bowman-Cisneros, J~Zerr, MA~Ravine, et~al.,
\newblock ``Lunar reconnaissance orbiter camera (lroc) instrument overview,''
\newblock {\em Space science reviews}, vol. 150, no. 1-4, pp. 81--124, 2010.

\bibitem{zhang2011color}
Lei Zhang, Xiaolin Wu, Antoni Buades, and Xin Li,
\newblock ``Color demosaicking by local directional interpolation and nonlocal
  adaptive thresholding,''
\newblock {\em Journal of Electronic imaging}, vol. 20, no. 2, pp. 023016,
  2011.

\bibitem{nascimento2016spatial}
S{\'e}rgio~MC Nascimento, Kinjiro Amano, and David~H Foster,
\newblock ``Spatial distributions of local illumination color in natural
  scenes,''
\newblock {\em Vision research}, vol. 120, pp. 39--44, 2016.

\bibitem{BS11}
M.~Brown and S.~S\"usstrunk,
\newblock ``Multispectral {SIFT} for scene category recognition,''
\newblock in {\em Computer Vision and Pattern Recognition (CVPR11)}, 2011, pp.
  177--184.

\bibitem{ren2016single}
Wenqi Ren, Si~Liu, Hua Zhang, Jinshan Pan, Xiaochun Cao, and Ming-Hsuan Yang,
\newblock ``Single image dehazing via multi-scale convolutional neural
  networks,''
\newblock in {\em European conference on computer vision}. Springer, 2016, pp.
  154--169.

\bibitem{humm2016flight}
DC~Humm, M~Tschimmel, SM~Brylow, P~Mahanti, TN~Tran, SE~Braden, S~Wiseman,
  J~Danton, EM~Eliason, and MS~Robinson,
\newblock ``Flight calibration of the lroc narrow angle camera,''
\newblock {\em Space Science Reviews}, vol. 200, no. 1-4, pp. 431--473, 2016.

\bibitem{mahanti2016inflight}
P~Mahanti, DC~Humm, MS~Robinson, AK~Boyd, R~Stelling, H~Sato, BW~Denevi,
  SE~Braden, E~Bowman-Cisneros, SM~Brylow, et~al.,
\newblock ``Inflight calibration of the lunar reconnaissance orbiter camera
  wide angle camera,''
\newblock {\em Space Science Reviews}, vol. 200, no. 1-4, pp. 393--430, 2016.

\bibitem{speyerer2012flight}
EJ~Speyerer, RV~Wagner, MS~Robinson, DC~Humm, K~Becker, J~Anderson, and
  P~Thomas,
\newblock ``In-flight geometric calibration of the lunar reconnaissance orbiter
  camera,''
\newblock in {\em 22nd Congress of the International Society for Photogrammetry
  and Remote Sensing, ISPRS 2012}. International Society for Photogrammetry and
  Remote Sensing, 2012.

\end{thebibliography}

\begin{IEEEbiography}[{\includegraphics[width=1in,height=1.25in,clip,keepaspectratio]{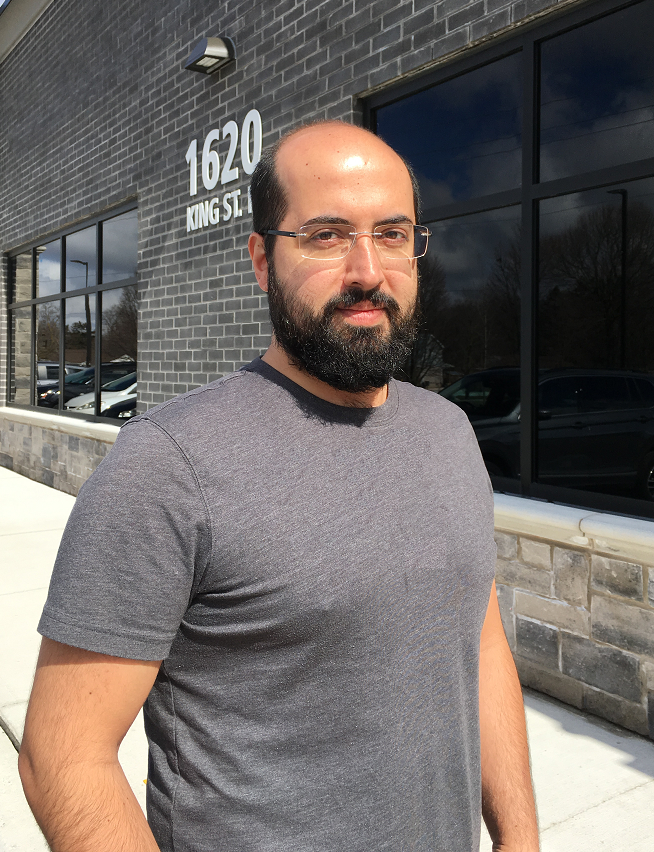}}]{Mahdi S. Hosseini}
received the B.Sc. ``cum laude'', the M.Sc., M.A.Sc., and Ph.D. degrees in electrical engineering from University of Tabriz, University of Tehran, University of Waterloo, and University of Toronto in $2004$, $2007$, $2010$, and $2016$, respectively. He is a post-doctoral fellow at the University of Toronto (UofT) and research scientist at Huron Digital Pathology in Waterloo, Ontario, Canada. His research is primarily focused on understanding and developing robust computational imaging and machine learning techniques for analyzing images from medical imaging platforms in digital pathology both from applied and theoretical perspectives. He leads a research team of UofT students for R\&D development of the image analysis pipelines required for digital and computational pathology. Dr. Hosseini has served as a reviewer for the \textsc{IEEE transaction on Image Processing}, the \textsc{IEEE transaction on Signal Processing} and \textsc{IEEE Transactions on Circuits and Systems for Video Technology}.
\end{IEEEbiography}

\begin{IEEEbiography}[{\includegraphics[width=1in,height=1.25in,clip,keepaspectratio]{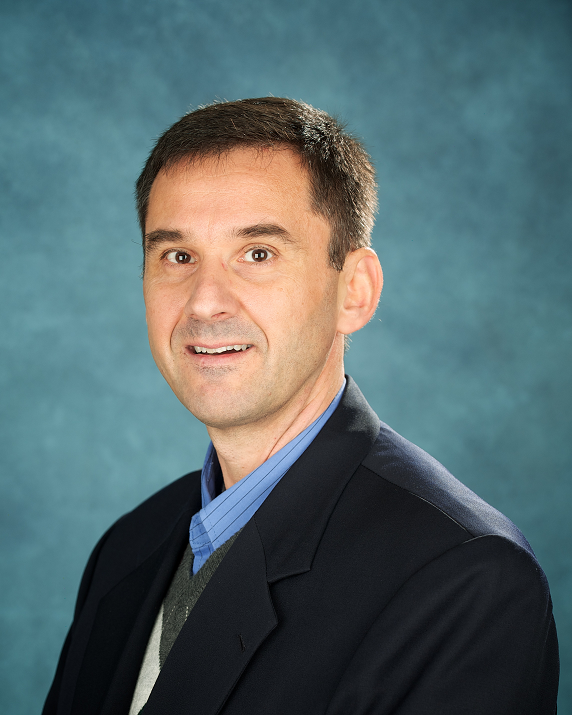}}]{Konstantinos N. Plataniotis}
(S’$93$-M’$95$-SM’$03$- F’$12$) is a  Professor  and  the  Bell  Canada  Chair in Multimedia with the ECE Department at the University of Toronto. He is the founder and inau- gural Director-Research for the Identity, Privacy and Security Institute (IPSI) at the University of Toronto and he has served as the Director for the Knowledge Media Design Institute (KMDI) at the University of Toronto from January 2010 to July 2012. His research interests are: knowledge and digital media design, multimedia systems, biometrics, image \& signal processing, communications systems and pattern recognition. Among his publications in these fields are the recent books WLAN positioning systems ($2012$) and Multi-linear subspace learning: Reduction of multidimensional data ($2013$).

Dr. Plataniotis is a registered professional engineer in Ontario, Fellow of the IEEE and Fellow of the Engineering Institute of Canada. He has served as the Editor-in-Chief of the IEEE Signal Processing Letters, and as Technical Co-Chair of the IEEE $2013$ International Conference in Acoustics, Speech and Signal Processing. He was the IEEE Signal Processing Society Vice President for Membership ($2014$ -$2016$). He is the General Co-Chair for  $2017$ IEEE GlobalSIP,   the $2018$ IEEE International Conference on Image Processing (ICIP $2018$), and the $2021$ IEEE International Conference on Acoustics, Speech and Signal Processing (ICASSP $2021$) .
\end{IEEEbiography}

\end{document}